\documentclass[fleqn,usenatbib]{mnras}

\usepackage[T1]{fontenc}

\DeclareRobustCommand{\VAN}[3]{#2}
\let\VANthebibliography\thebibliography
\def\thebibliography{\DeclareRobustCommand{\VAN}[3]{##3}\VANthebibliography}

\usepackage{graphicx}	
\usepackage{amsmath}	
\usepackage{amssymb}	

\usepackage{color}

\usepackage{url}

\title[Three triply eclipsing \textit{TESS} triples]{Triply eclipsing triple stars in the northern \textit{TESS} fields: TICs 193993801, 388459317 and 52041148}

\author[Borkovits et al.]{
T. Borkovits$^{1,2,3}$\thanks{E-mail: borko@electra.bajaobs.hu},
T. Mitnyan$^1$,
S. A. Rappaport$^4$,
T. Pribulla$^5$,
B. P. Powell$^6$,
\newauthor
V. B. Kostov$^{6,7}$,
I. B. B\'\i r\'o$^1$, 
I. Cs\'anyi$^1$,
Z. Garai$^{3,5,8,9}$, 
B. L. Gary$^{10}$, 
T. G. Kaye$^{11}$, 
\newauthor
R. Kom\v{z}\'{\i}k$^5$,
I.~Terentev$^{12}$,
M.\,Omohundro$^{12}$,
R.~Gagliano$^{13}$,
T.\,Jacobs$^{14}$,
\newauthor
M.\,H.\,Kristiansen$^{15,16}$,
D.\,LaCourse$^{17}$,
H.\,M.\,Schwengeler$^{12}$,
D.\,Czavalinga$^1$, 
B.\,Seli$^{2,18}$,
\newauthor
C.\,X.\,Huang$^{4}$,
A.\,P\'al$^{2,4}$,
A.\,Vanderburg$^{19}$,
J.\,E.\,Rodriguez$^{20}$,
D.\,J.\,Stevens$^{21,22}$ \\
$^{1}$Baja Astronomical Observatory of University of Szeged, H-6500 Baja, Szegedi \'ut, Kt. 766, Hungary\\
$^{2}$Konkoly Observatory, Research Centre for Astronomy and Earth Sciences,  H-1121 Budapest, Konkoly Thege Mikl\'os \'ut 15-17, Hungary\\
$^3$ ELTE Gothard Astrophysical Observatory, H-9700 Szombathely, Szent Imre h. u. 112, Hungary \\
$^4$ Department of Physics, Kavli Institute for Astrophysics and Space Research, M.I.T., Cambridge, MA 02139, USA\\
$^5$ Astronomical Institute of the Slovak Academy of Sciences, Tatransk\'a Lomnica, Slovakia \\
$^6$ NASA Goddard Space Flight Center, 8800 Greenbelt Road, Greenbelt, MD 20771, USA \\
$^7$ SETI Institute, 189 Bernardo Avenue, Suite 200, Mountain View, CA 94043, USA\\
$^8$ MTA-ELTE Exoplanet Research Group, H-9700 Szombathely, Szent Imre h. u. 112, Hungary \\
$^9$ MTA-ELTE Lend\"ulet Milky Way Research Group, H-9700 Szombathely, Szent Imre h. u. 112, Hungary \\
$^{10}$ Amateur Astronomer, Hereford Arizona Observatory, AZ \\
$^{11}$ Amateur Astronomer, Patterson Observatory, AZ\\
$^{12}$ Citizen Scientist, c/o Zooniverse, Department of Physics, University of Oxford, Denys Wilkinson Building, Keble Road, Oxford, OX1 3RH, UK \\
$^{13}$ Amateur Astronomer, Glendale, AZ 85308 \\
$^{14}$ Amateur Astronomer, 12812 SE 69th Place Bellevue, WA 98006, USA \\
$^{15}$ Brorfelde Observatory, Observator Gyldenkernes Vej 7, DK-4340 T\o ll\o se, Denmark \\
$^{16}$ DTU Space, National Space Institute, Technical University of Denmark, Elektrovej 327, DK-2800 Lyngby, Denmark \\
$^{17}$ Amateur Astronomer, 7507 52nd Place NE Marysville, WA 98270, USA \\
$^{18}$ E\"otv\"os University, Department of Astronomy, Pf. 32, 1518 Budapest, Hungary \\
$^{19}$ Department of Astronomy, The University of Wisconsin-Madison, 475 N.\,Charter St., Madison, WI 53706, USA \\
$^{20}$ Department of Physics and Astronomy, Michigan State University, East Lansing, MI 48824, USA \\
$^{21}$ Department of Astronomy \& Astrophysics, The Pennsylvania State University, 525 Davey Lab, University Park, PA 16802, USA \\
$^{22}$ Center for Exoplanets and Habitable Worlds, The Pennsylvania State University, 525 Davey Lab, University Park, PA 16802, USA }
\date{Accepted XXX. Received YYY; in original form ZZZ}

\pubyear{2021}

\begin{document}
\label{firstpage}
\pagerange{\pageref{firstpage}--\pageref{lastpage}}
\maketitle

\begin{abstract}
 In this work we report the discovery and analysis of three new triply eclipsing triple star systems found with the {\em TESS} mission during its observations of the northern skies: TICs 193993801, 388459317, and 52041148.  We utilized the {\em TESS} precision photometry of the binary eclipses and third-body eclipsing events, ground-based archival and follow-up photometric data, eclipse timing variations, archival spectral energy distributions, as well as theoretical evolution tracks in a joint photodynamical analysis to deduce the system masses and orbital parameters of both the inner and outer orbits.  In one case (TIC 193993801) we also obtained radial velocity measurements of all three stars.  This enabled us to `calibrate' our analysis approach with and without `truth' (i.e., RV) data.  We find that the masses are good to 1-3\% accuracy with RV data and 3-10\% without the use of RV data.  In all three systems we were able to find the outer orbital period before doing any detailed analysis by searching for a longer-term periodicity in the ASAS-SN archival photometry data---just a few thousand ASAS-SN points enabled us to find the outer periods of 49.28 d, 89.86 d, and 177.0 d, respectively. From our full photodynamical analysis we find that all three systems are coplanar to within $1^\circ - 3^\circ$.  The outer eccentricities of the three systems are 0.003, 0.10, and 0.62, respectively (i.e., spanning a factor of 200).  The masses of the three stars \{Aa, Ab, and B\} in the three systems are: \{1.31, 1.19, 1.34\}, \{1.82, 1.73, 2.19\}, and \{1.62, 1.48, 2.74\} M$_\odot$, respectively.
\end{abstract}




\section{Introduction}
\label{sect_intro}

After binary stars, triple-star systems are the simplest multi-stellar configurations to study and understand.  There is only one basic stable configuration for triples with an inner binary being orbited by an outer third star in a hierarchical configuration.  There is no hard and fast rule about the shortest ratio of outer to inner (i.e., binary) period ratio, $P_{\rm out}/P_{\rm bin}$ that is allowed, but various numerical studies (see, e.g., \citealt{eggleton_kiseleva95}, Eqn.~1; \citealt{mikkola08}, Eqn.~10, and references therein) have led us to a relatively simple expression that we have found convenient and reasonably robust:
\begin{eqnarray}
\frac{a_{\rm out}}{ a_{\rm in}} & \gtrsim & 2.8 \left(\frac{M_{\rm trip}}{M_{\rm bin}}\right)^{2/5} \frac{(1+e_{\rm out})^{2/5}}{(1-e_{\rm out})^{6/5}} \,(1+e_{\rm in})  \\
\frac{P_{\rm out}}{P_{\rm in}} & \gtrsim & 4.7 \left(\frac{M_{\rm trip}}{M_{\rm bin}}\right)^{1/10} \frac{(1+e_{\rm out})^{3/5}}{(1-e_{\rm out})^{9/5}} \,(1+e_{\rm in})^{3/2}  \nonumber
\end{eqnarray}
where $e_{\rm out}$ and $e_{\rm in}$ refer to the orbital eccentricity of the outer and inner (i.e., binary) orbits, respectively, $M_{\rm trip}$ and $M_{\rm bin}$ refer to the masses of the entire triple system and the inner binary, respectively, and the $a$'s are the semi-major axes of the orbits.  These expressions are summarized in Eqn.~(16) of \citet{rappaport13} and are here modified with multiplicative factors of $(1 + e_{\rm in})^{2/2,3/2}$ which we adopt from the Eggleton-Kiseleva stability requirement (\citealt{eggleton_kiseleva95}, Eqn.~1). These hold best for nearly co-planar systems, but are probably not too far off for highly non-coplanar systems as well.  

Thus, as a general rule, if the orbital eccentricities are small to modest in size, the outer orbital period of the triple system must be greater than $\sim$5 times the period of the inner binary, and more typically a factor of 10 times the period of the inner binary for a common value of the outer eccentricity of $e_{\rm out} \simeq  0.3$.  The shortest known ratio of $P_{\rm out}/P_{\rm bin}$ in a triple system is about 5.4 for LHS~1070 \citep{xiaetal19}, which consists of three very low mass red dwarfs with inner and outer periods of 18.2 and 99 years, respectively. Due to their long periods, however, even the short period perturbations (i.e., those that are comparable with the periods of the inner and outer binary periods) are not easily measurable on a human timescale. Moreover, the outer period is known only with a large uncertainty and, therefore, the period ratio is uncertain as well \citep[see][for further discussion]{tokovinin21}. Amongst compact stellar triples, where both periods are known with much higher accuracy, the record-holder is KIC~7668648 with $P_{\rm out}/P_{\rm bin}\sim7$, which ultimately turned out to be a triply eclipsing triple system \citep{borkovitsetal15,orosz15}.\footnote{We also note that the majority of the known circumbinary planets are also located close to the dynamical stability limit.}  

Outer periods in triple systems can range all the way from millions of years in some visual systems \citep{tokovinin18}, all the way down to 33.03 days in the case of $\lambda$~Tau \citep{ebbighausenstruve56}. Interestingly, the second and third shortest known outer period systems have outer periods that are only a few percent longer, namely,  33.92 days in the case of KOI~126 \citep{carter11} and 33.95 days for HD~144548 \citep{alonso15}.  However, in principle, there is no dynamical reason for triple stars not to have stable outer orbital periods as short as a couple of days with, e.g., $P_{\rm bin} \simeq 0.25$ days and $P_{\rm out} \simeq 2$ days.

Compact triple star systems are quite fascinating objects to study because all their interesting timescales for dynamical interaction can happen over just a few observing seasons, or even on times as short as months and weeks.  Another feature of compact triples is that the a priori probability for third-body (or `outer') eclipses\footnote{In this case the binary stars can eclipse the third star or vice versa.}  to appear grows with decreasing outer orbital period roughly as $\propto P_{\rm out}^{-2/3}$.  There are two advantages of finding third-body eclipses in triple systems.  First, it immediately reveals the triple nature of the system without the need for radial velocity or eclipse timing variation studies.  Second, it carries a considerable amount of important information for the analysis of these systems, in particular for determining the properties of the outer orbit \citep[see, e.~g.][]{carter11,borkovitsetal13,masudaetal15,orosz15,alonso15}.  

\begin{figure}
\begin{center}
\includegraphics[width=1.02 \columnwidth]{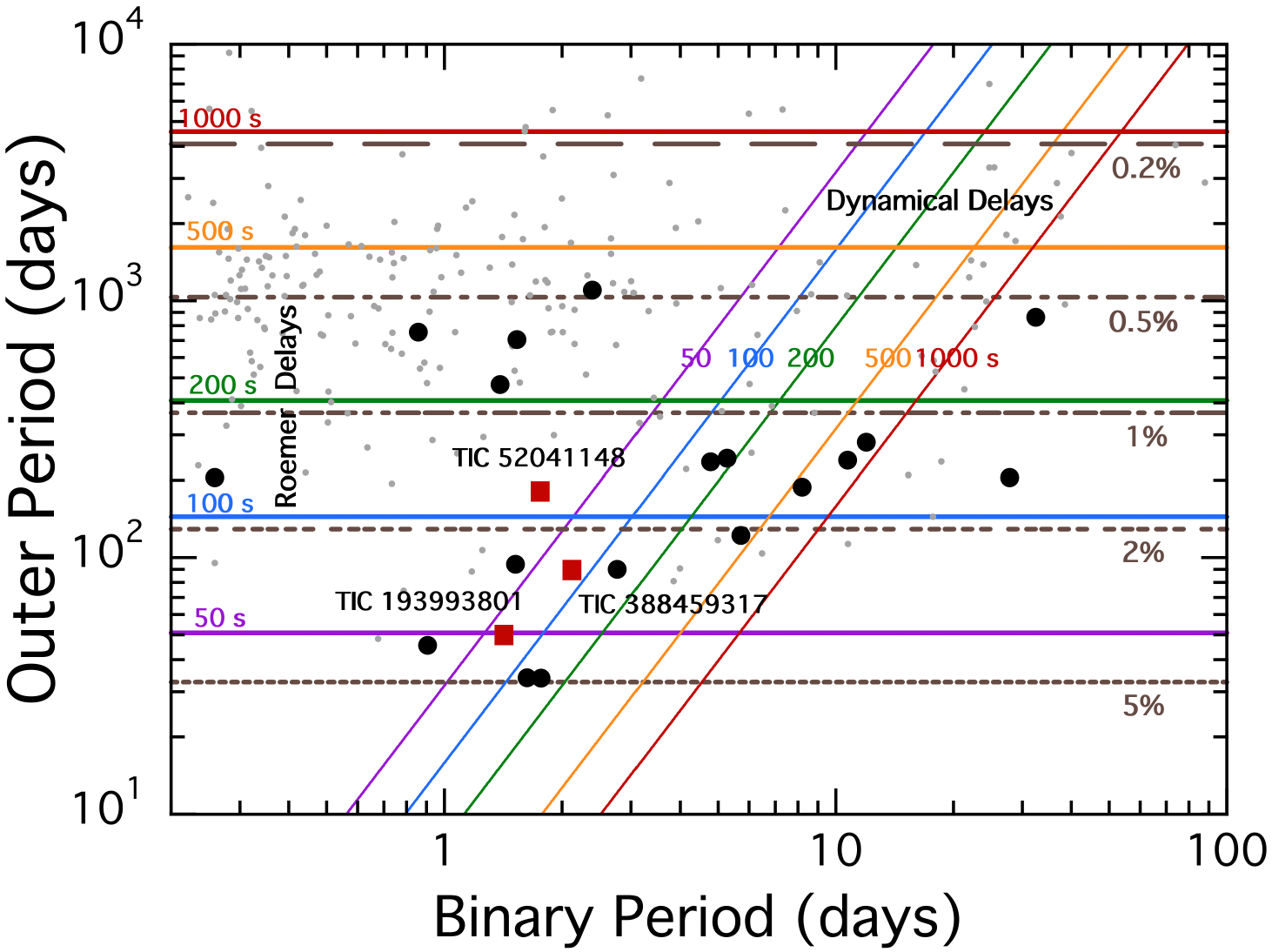}
\caption{Summary of compact triple systems updated from the work of \citet{borkovitsetal20b}. The results are plotted in the plane of the outer third body orbital period versus the inner binary period.  We also overlay illustrative lines of constant Roemer delays (coloured horizontal lines) and dynamical delays (coloured diagonal lines) that would show up in eclipse timing variations of binary stars with third-body companions (see, e.~g., \citealt{borkovitsetal15}). Horizontal dashed lines represent rough estimates of the eclipse probability for a third body orbiting the binary. The 17 filled black circles are known triply eclipsing triple systems, while the three red squares represent the new triply eclipsing triples reported in this work. The fainter dots are some 200 other triple systems found with Kepler that have measured or estimated outer orbital periods (see \citealt{borkovitsetal16}).}
\label{fig:triples}
\end{center}
\end{figure}

We show in Figure \ref{fig:triples} an updated summary of the known compact triples first presented in \citet{borkovitsetal20b}.  The information is plotted in the plane of $P_{\rm out}$ vs.~$P_{\rm bin}$.  All but the triply eclipsing triples shown in this diagram were found by eclipse timing variations (ETVs), and the $\approx 200$ found in this way with {\em Kepler} are shown as grey dots \citep{borkovitsetal16}.  Superposed on the plot are lines of constant Roemer delays as well as dynamical delays where the period of inner binary is lengthened by the presence of the third body (see, e.g., \citealt{rappaport13}; \citealt{borkovitsetal15}; \citealt{borkovitsetal16}). The filled black circles are the known triple systems which exhibit third-body eclipses.  The three red squares are the newly discovered systems reported in this work. 

Here we present the discovery and detailed analyses of three new compact triply eclipsing triple star systems.  In Section \ref{sec:obs} we present the {\em TESS} observations (Sect.~\ref{sec:TESS}) which led to the discovery of these objects; the use of archival photometric data to help determine the outer orbital period via third-body eclipses (Sect.~\ref{sec:archival}); follow-up ground-based observations to gather more timing data for the binary and third-body eclipses (Sect.~\ref{sec:follow_up}); and radial velocity measurements for one of the objects (Sect.~\ref{sec:RVs}).  The detailed photodynamical model by which we analyze jointly the photometric lightcurves, eclipse timing variations, spectral energy distributions, and radial velocities (where available) is reviewed in Section \ref{sec:dyn_mod}.  The net results are in the form of three comprehensive tables, one for each triple system, of detailed extracted parameters for the masses, radii, and effective temperatures, as well as the orbital parameters for both the inner and outer orbits.  We summarize our results and discuss a few of the salient findings from our study in Section \ref{sec:discussion}.

\begin{table*}
\centering
\caption{Main properties of the three systems from different catalogs}
\begin{tabular}{lccc}
\hline
\hline
Parameter & 193993801 & 388459317 & 52041148 \\
\hline
RA (J2000) & $15:25:15.011$ & $22:39:51.187$ & $00:56:39.011$ \\  
Dec (J2000)& $+50:35:27.64$ & $+54:58:48.22$ & $+65:26:57.88$ \\  
$T^a$ & $10.9412\pm0.0069$  & $14.0809\pm0.0292$ & $12.6336\pm0.0165$ \\
$G^b$ & $11.3150\pm0.0011$  & $14.6024\pm0.0004$ & $13.4641\pm0.0005$ \\
$G_{\rm BP}^b$ & $11.5597\pm0.0018$ & $15.0356\pm0.0024$ & $14.2360\pm0.0014$ \\
$G_{\rm RP}^b$ & $10.8716\pm0.0016$ & $13.9857\pm0.0011$ & $12.5838\pm0.0017$ \\
B$^a$ & $11.940 \pm 0.193$ & $15.898\pm0.395$ & $15.201\pm0.031$ \\
V$^c$ & $11.269 \pm 0.013$ & $14.894\pm0.103$ & $14.097\pm0.092$ \\
g$'^c$& $11.443 \pm 0.113$ & $15.210\pm0.015$ & $14.561\pm0.015$ \\
r$'^c$& $11.357 \pm 0.104$ & $14.644\pm0.033$ & $13.512\pm0.113$ \\
i$'^c$& $11.257 \pm 0.083$ & $14.351\pm0.054$ & $12.926\pm0.083$ \\
J$^d$ & $10.388 \pm 0.023$ & $13.130\pm0.024$ & $11.320\pm0.023$ \\
H$^d$ & $10.167 \pm 0.015$ & $12.955\pm0.032$ & $11.002\pm0.024$ \\
K$^d$ & $10.114 \pm 0.016$ & $12.814\pm0.034$ & $10.810\pm0.019$ \\
W1$^e$ & $10.130\pm0.024$  & $12.778\pm0.023$ & $10.684\pm0.023$ \\
W2$^e$ & $10.154\pm0.022$  & $12.625\pm0.023$ & $10.664\pm0.022$ \\
W3$^e$ & $10.077\pm0.041$  & $12.266\pm0.328$ & $10.303\pm0.093$ \\
W4$^e$ & $ 9.614$          & $8.793\pm0.376$  & $7.148\pm0.112$ \\
$T_{\rm eff}$ (K)$^a$ & $6239\pm99$ & $7630\pm123$ & $6994\pm126$ \\
Distance (pc)$^f$ & $676\pm26$ & $4042\pm268$ & $5931\pm407$ \\ 
$[M/H]^a$ & $0.028 \pm 0.017$ & $-$ & $-$ \\ 
$E(B-V)^a$ & $0.020 \pm 0.004$ & $0.567\pm0.038$ & $0.908\pm0.021$ \\
$\mu_\alpha$ (mas ~${\rm yr}^{-1}$)$^b$ & $-10.85\pm0.06$ & $-3.01\pm0.02$ & $-1.65\pm0.01$ \\ 
$\mu_\delta$ (mas ~${\rm yr}^{-1}$)$^b$ &  $-4.87\pm0.07$ & $-2.43\pm0.02$ & $0.07\pm0.01$ \\ 
\hline
\label{tbl:mags}  
\end{tabular}

\textit{Notes.}  (a) TESS Input Catalog (TIC v8.2) \citep{TIC8}. (b) Gaia EDR3 \citep{GaiaEDR3}. (c) AAVSO Photometric All Sky Survey (APASS) DR9, \citep{APASS}, \url{http://vizier.u-strasbg.fr/viz-bin/VizieR?-source=II/336/apass9}. (d) 2MASS catalog \citep{2MASS}.  (e) WISE point source catalog \citep{WISE}. (f) \citet{bailer-jonesetal21}. \\
Note also, that for the SED analysis in Sect.~\ref{sec:dyn_mod} the uncertainties of the passband magnitudes were set to $\sigma_\mathrm{mag}=\mathrm{max}(\sigma_\mathrm{catalog},0.030)$ to avoid the strong overdominance of the extremely accurate Gaia magnitudes over the other measurements.
\end{table*}

\section{Observations}
\label{sec:obs}

\subsection{TESS light curves}
\label{sec:TESS}

Our `Visual Surveyors Group' (VSG) continues to search for multistellar systems in \textit{TESS} lightcurves.  We estimate that, thus far, we have visually inspected some 3 million lightcurves from {\em TESS}.  Such visual searches are a complement to more automated ones using machine learning algorithms (see, e.g., \citealt{powell21}).  The lightcurves are displayed with Allan Schmitt's {\tt LcTools} and {\tt LcViewer} software \citep{schmitt19}, which allows for an inspection of a typical lightcurve in just a few seconds.  For {\em TESS}'s first excursion into the northern hemisphere, we largely made use of the lightcurve pipeline {\tt QLP} \citep{huang20}.  

The first signatures that are looked for in terms of identifying triples, quadruples, and higher order stellar systems are: (i) an eclipsing binary (EB) lightcurve with an additional strangely shaped extra eclipse or rapid succession of isolated eclipses; and (ii) two (or possibly more) sets of EB lightcurves with different periods.  Many of the latter detections ($\gtrsim 80\%$) turn out to be false positives due to the large {\em TESS} pixel size where two completely independent EBs may accidentally land near each other on the sky.  Most false positives are identified using the {\tt Lightkurve} software \citep{lightkurve18}. We normally generate a $15 \times 15$ pixel mask and use the software's interactive aperture feature in order to either (i) rapidly locate the individual EBs when their positions are sufficiently separated or (ii) compare how the eclipse depths scale with  aperture location. For those cases where the eclipses do not scale equally with aperture position, we {\em reject} the candidates as possible bound systems that are likely to exhibit interesting dynamical interactions within a human lifetime (i.e., $\sim$100 year). Sources that pass this check still require further vetting (e.g., with studies of eclipse timing variations, radial velocities, or high-resolution imaging) to ascertain more robustly whether the EBs are physically associated, and if there are, or will be, detectable dynamical interactions on interesting timescales. 

One gratifying aspect of finding triply eclipsing triples is that they are in a sense `self-vetting'.  In particular, there is no way for a single binary, or sets of independent stars or binaries to produce such `extra' eclipsing events.  Therefore, additional vetting becomes largely unnecessary.

While searching through the lightcurves obtained from the first northerly {\em TESS} observations we have found more than $\sim$30 of these triply eclipsing triples.  We report on three of them in this work: TIC 193993801, TIC 388459317, and TIC 52041148. (See Table~\ref{tbl:mags} for the main catalog data of the targets.)

\begin{figure}
\includegraphics[width=0.5\textwidth]{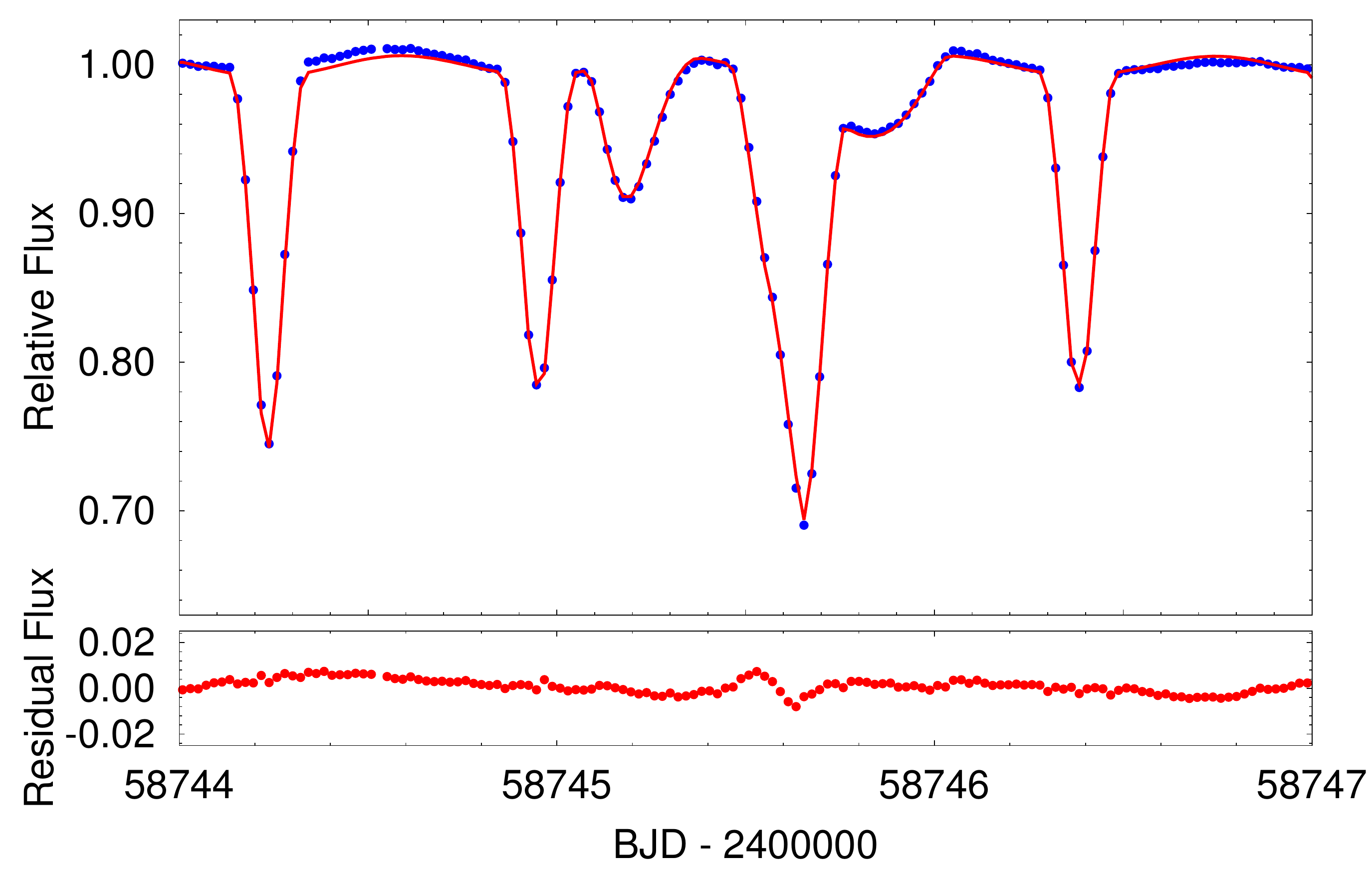}
\includegraphics[width=0.5\textwidth]{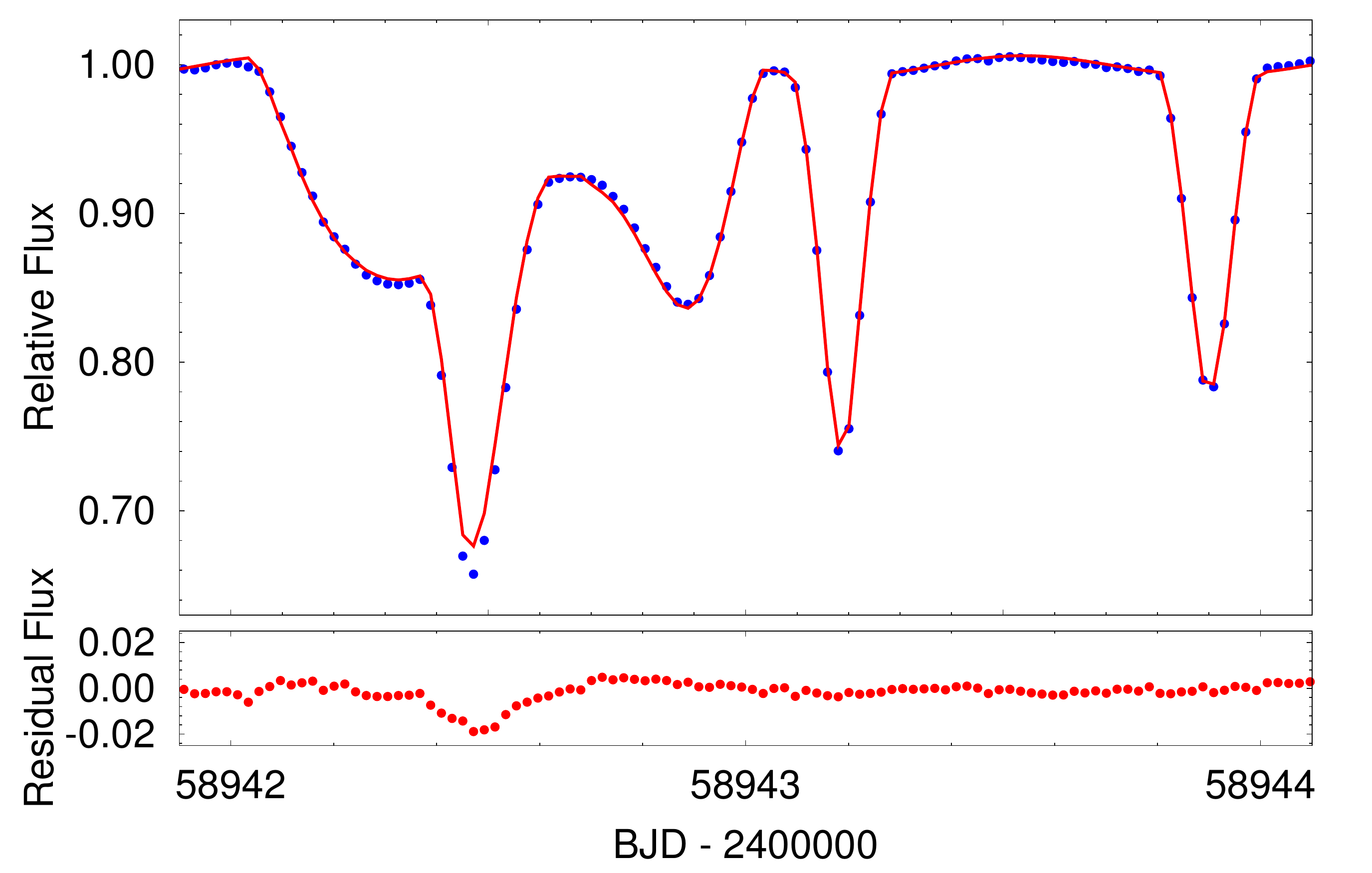} 
\includegraphics[width=0.5\textwidth]{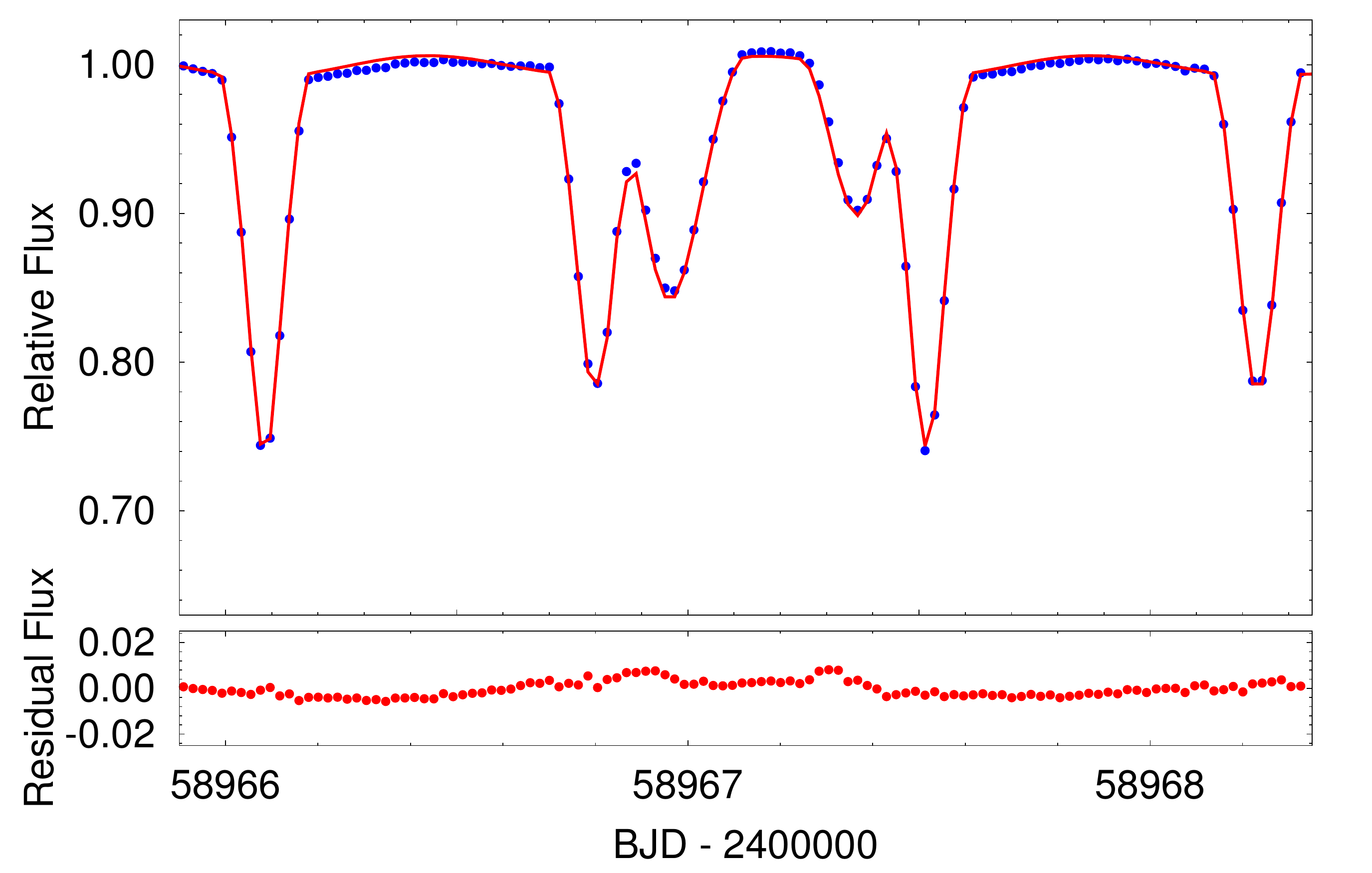}  
 \caption{Three third-body, or outer eclipses of TIC~193993801 observed with \textit{TESS} in sectors 16, 23 and 24, respectively. Blue dots represent the observed {\em TESS} full frame image (FFI) fluxes, while red lines represent the bestfitting spectro-photodynamical model. In the case of the first two third-body eclipses (upper and middle panels) the distant, third star eclipsed the members of the inner, close pair, while in the bottom panel the third star was eclipsed separately by the two inner stars.}
\label{fig:T193993801lcswithfit}
\end{figure}

\begin{figure}
\includegraphics[width=0.5\textwidth]{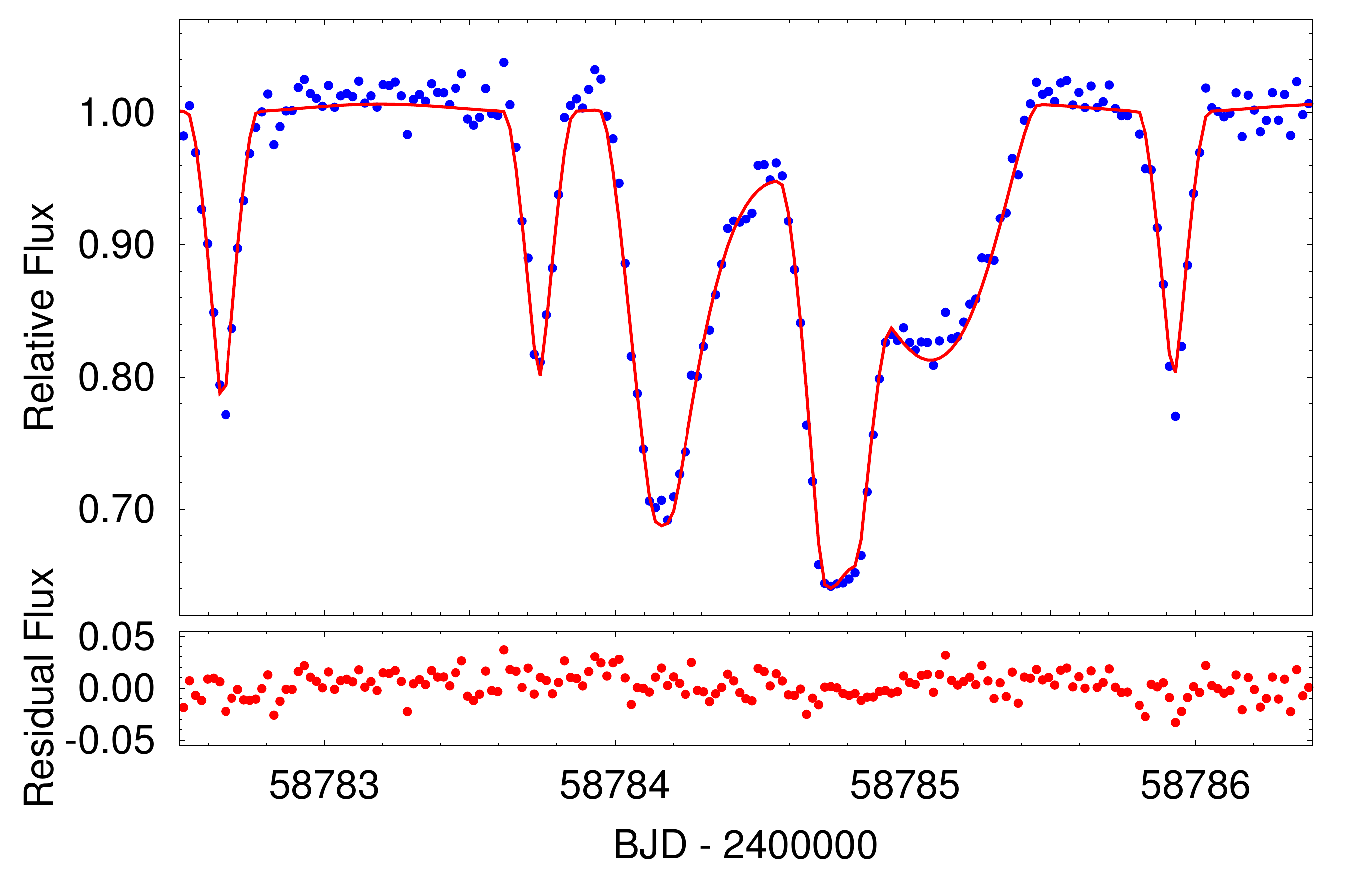}
 \caption{The sole thid-body eclipse of TIC\,388459317 observed with \textit{TESS} in sector 17. As above, blue dots represent the observed FFI fluxes, while the red curve represents the bestfitting spectro-photodynamical model. During this event the third star was eclipsed by the two inner binary stars.}
\label{fig:T388459317lcswithfit}
\end{figure}

\begin{figure}
\includegraphics[width=0.5\textwidth]{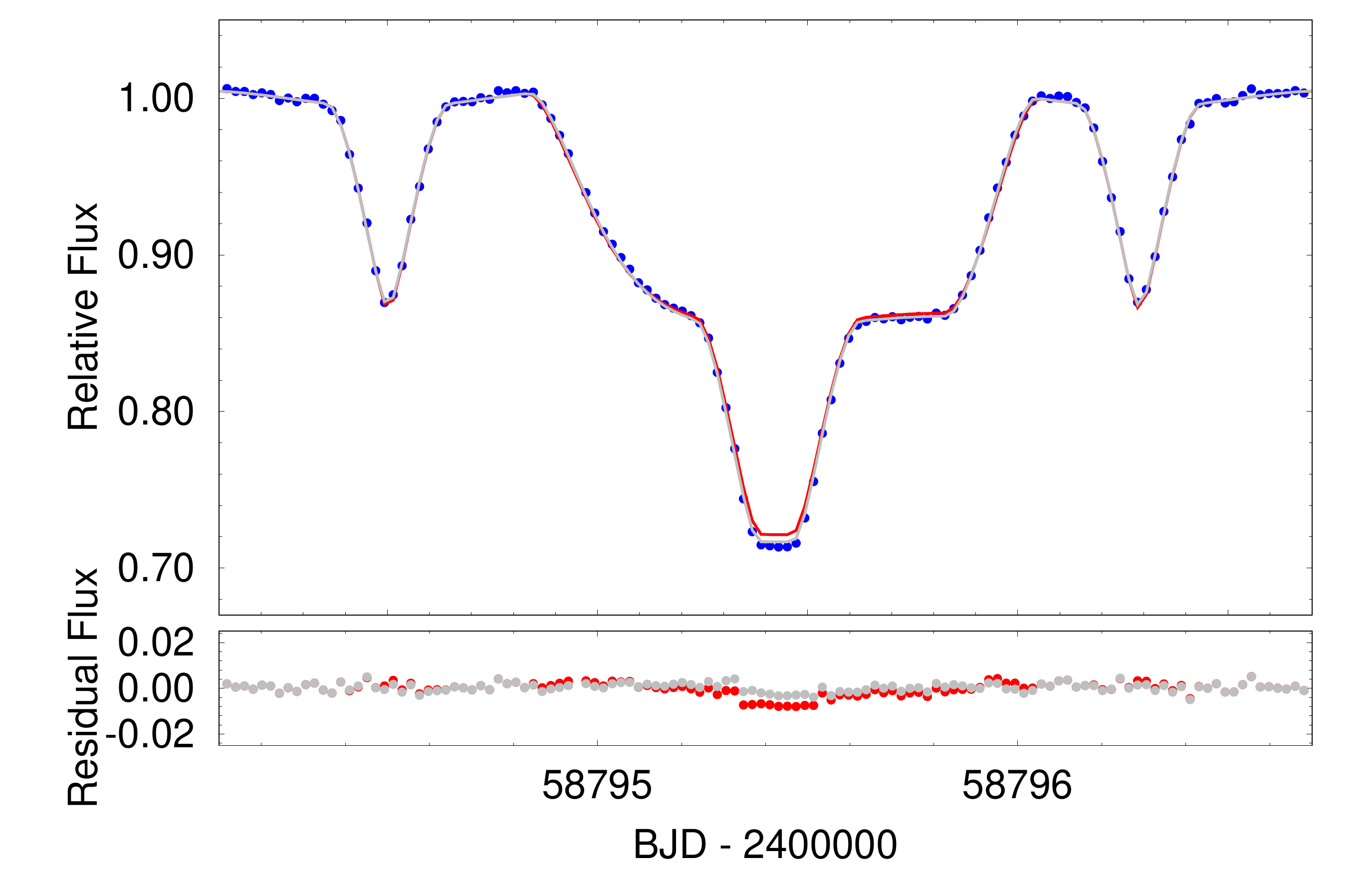}
\includegraphics[width=0.5\textwidth]{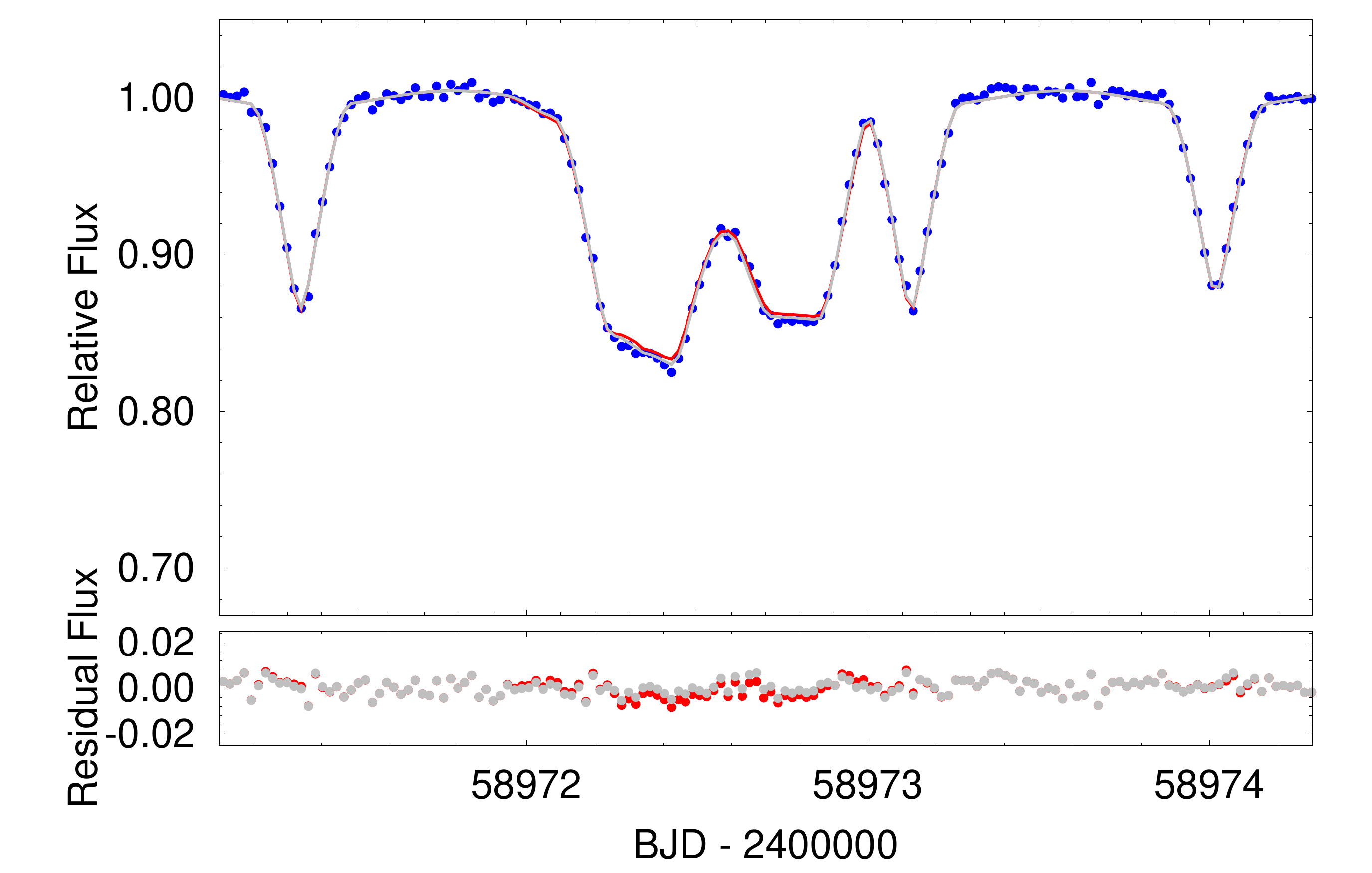} 
 \caption{Two third-body, or outer eclipses of TIC~52041148 observed with \textit{TESS} in Sectors 18 and 24, respectively. Blue dots represent the observed FFI fluxes, while red and gray curves stand for the bestfitting photodynamical models with and without additional fits using SED and \texttt{PARSEC} isochrones constraints, respectively. In the case of both outer eclipses, the distant third star eclipsed the members of the inner, close binary pair.}
\label{fig:T052041148lcswithfit}
\end{figure}

All three targets were measured in {\em TESS}'s 30-min cadence mode. TIC~193993801 was observed in Sectors 16, 23 and 24. A set of extra eclipses was detected in each sector (see Fig.~\ref{fig:T193993801lcswithfit}), which made the identification of the target as a potential triply eclipsing triple system candidate quite trivial.  TIC~388459317 was observed in Sectors 16 and 17. We have identified one complex, $\sim$$1.5$-day-long extra eclipsing event in Sector 17 (see Fig.~\ref{fig:T388459317lcswithfit}). Finally, TIC~52041148 was observed by \textit{TESS} in Sectors 18, 24 and 25. This target exhibited an extra eclipsing event in both Sectors 18 and 24 (see Fig.~\ref{fig:T052041148lcswithfit}). 

In the following steps of the analysis, more sophisticated photometric data processing was carried out by employing convolution-based differential image analysis implemented using the tasks of the {\sc FITSH} package \citep{2012MNRAS.421.1825P}. The processing steps are based on the ones used in the extraction of fluxes from moving objects in the Solar System \citep{paletal20} as well as other types of pulsating variables \citep{plachyetal21}.

\begin{figure}
\center
\includegraphics[width=0.5\textwidth]{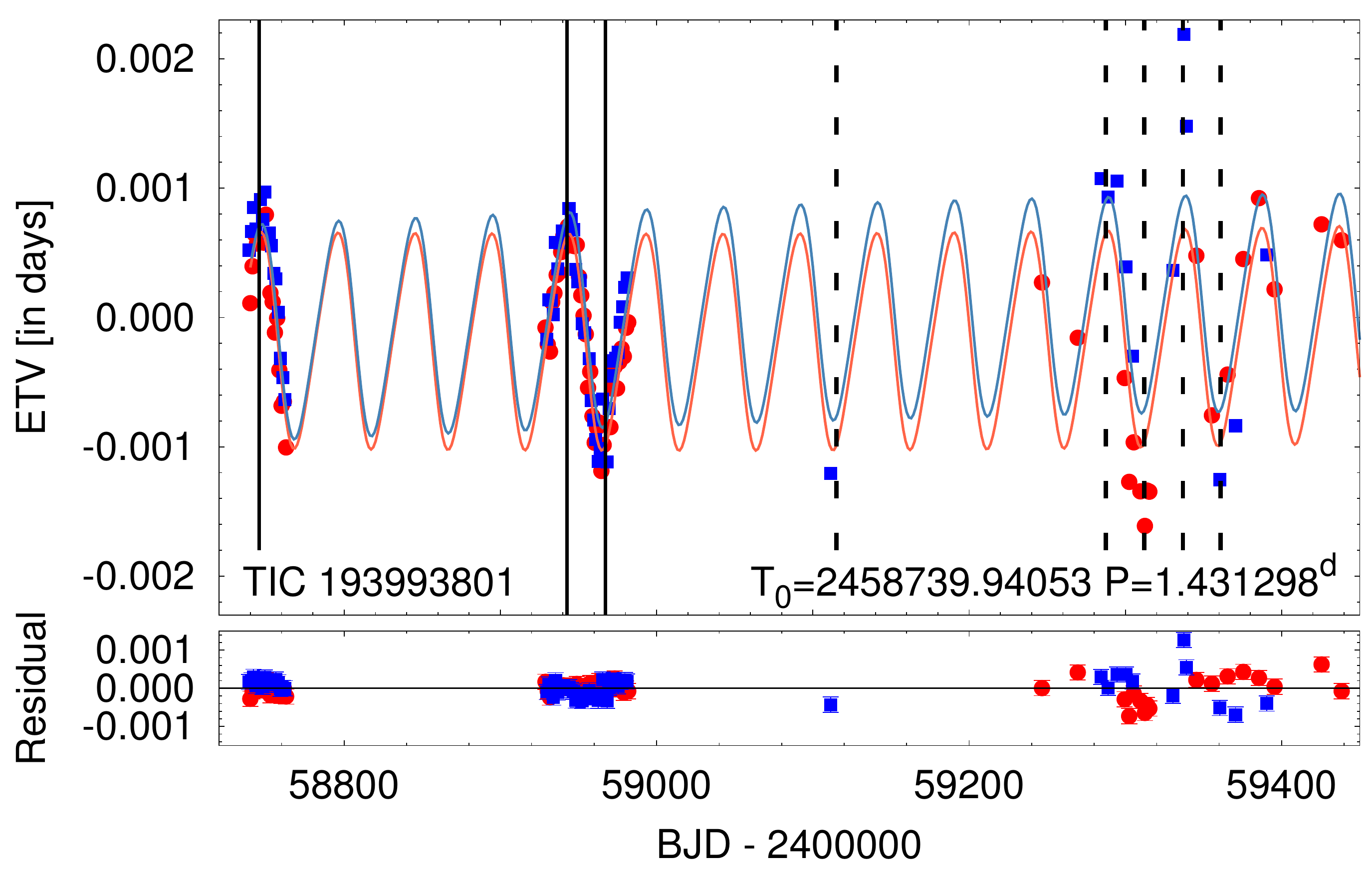} 
\includegraphics[width=0.5\textwidth]{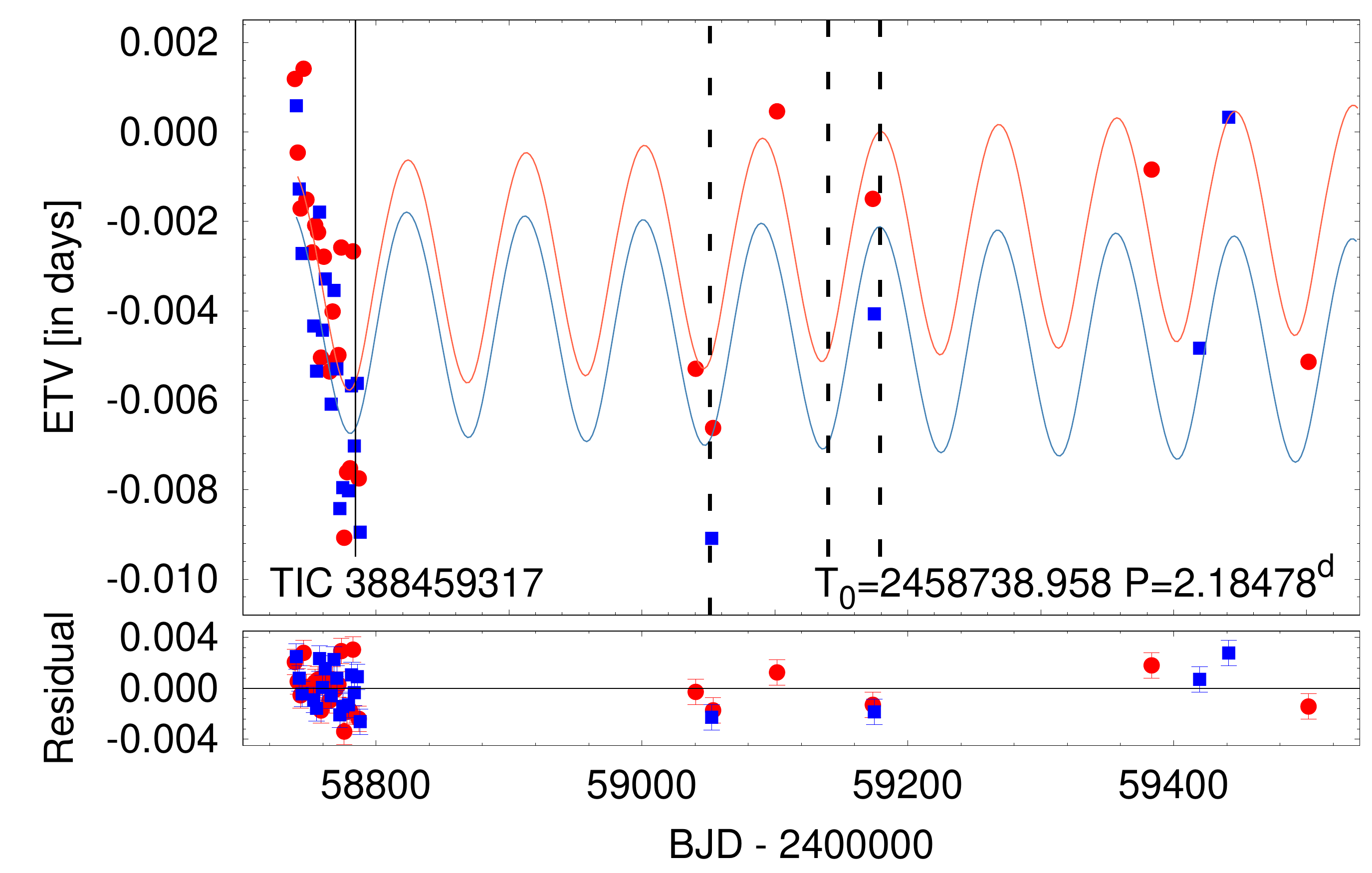} 
\includegraphics[width=0.5\textwidth]{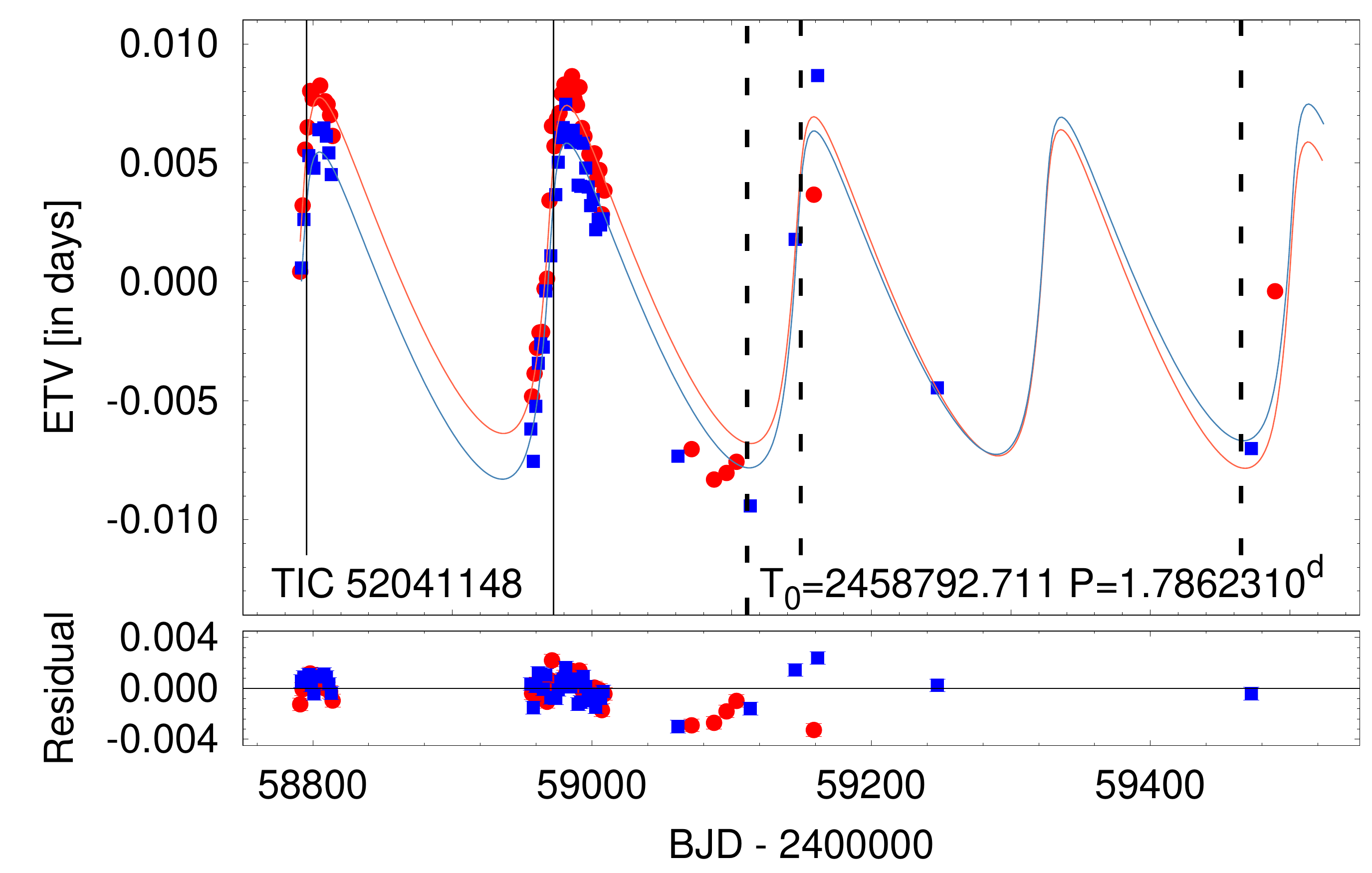} 
\caption{Eclipse timing variations of TICs 193993801 (upper), 388459317 (middle) and 52041148 (bottom). Red and blue dots represent the primary and secondary binary eclipses, respectively, while the correspondingly colored curves connect the best-fitting photodynamical model ETV points. Solid and dashed vertical black lines indicate the times of the third-body (i.e., outer) eclipses detected by \textit{TESS} and during ground-based follow-up observations, respectively. The lower panel shows the residuals.}
\label{fig:ETVswithfit}
\end{figure}

Our next step in analyzing systems with detected third-body eclipsing events is to search for eclipse timing variations (ETVs).  Such triply eclipsing triple systems typically have outer periods in the range of 40-200 days or else the probability for an outer eclipse diminishes rapidly with increasing size of the outer orbit.  Therefore, we anticipate the likelihood of also finding detectable ETVs.  The expected dynamical delays for a coplanar system are proportional to $(3e_{\rm out}/2 \pi) \zeta P_{\rm bin}^2/P_{\rm out}$, where the `out' subscript refers to the outer orbit, and $\zeta$ is the mass ratio of the third component to the total triple-star system.  Thus for typical parameters of $P_{\rm bin} \sim 3$ d, $P_{\rm out} \simeq 100$ d, and $e_{\rm out} \sim 0.3$ dynamical delays of $\sim$0.004 d (5.5 min) are expected.  For combinations of shorter-period inner binaries and longer outer periods we expect the classical light travel time delays (LTTE) to dominate the ETVs.
To measure the ETVs we use our standard procedure as described in \citet{borkovitsetal16}. We list the times of minima of the three targets in Tables~\ref{tab:T193993801ToM} -- \ref{tab:T052041148ToM} and plot the ETV curves in Fig.~\ref{fig:ETVswithfit}. These ETV data carry a number of valuable pieces of information for the further analyses of these systems.   

For example, in the case of TIC 193993801 the ETV curve itself, even without the third-body eclipses, reveals a $\sim$49-day-period third companion. From the nearly perfectly sinusoidal shape of the ETV curve, one can assume that the outer orbit may be nearly circular. Furthermore, the first and second \textit{TESS}-observed third-body eclipses are located close to the upper extrema of the ETV curve (i.e. close to the maximum delay), while the third one is near the lower extremum.  Thus, insofar as the ETV is dominated by the geometrical light-travel time effect, the first and second extra events occurred during the inferior conjunction of the distant third star, while the third one was a superior conjunction event.

In the case of TIC 52041148 the similar curvature of the two sections of the ETV curves around the times of the two third-body eclipses suggested that the two \textit{TESS}-observed events belong to the same type of the outer eclipses (i.e., primary vs.~secondary). Hence, we readily got an upper limit for the outer period of this triple, as it cannot be longer than the duration between the two events. 

In contrast to the previous two systems, in the case of TIC 388459317, the \textit{TESS} ETV curve by itself did not carry any readily usable information about the outer orbit.

\begin{table*}
\caption{Binary Mid Eclipse Times for TIC~193993801.}
 \label{tab:T193993801ToM}
\begin{tabular}{@{}lrllrllrl}
\hline
BJD & Cycle  & std. dev. & BJD & Cycle  & std. dev. & BJD & Cycle  & std. dev. \\ 
$-2\,400\,000$ & no. &   \multicolumn{1}{c}{$(d)$} & $-2\,400\,000$ & no. &   \multicolumn{1}{c}{$(d)$} & $-2\,400\,000$ & no. &   \multicolumn{1}{c}{$(d)$} \\ 
\hline
58739.225610 &   -0.5 & 0.000081 & 58936.744183 &  137.5 & 0.000298 & 58972.526035 &  162.5 & 0.000077 \\ 
58739.940505 &    0.0 & 0.000190 & 58937.460156 &  138.0 & 0.000072 & 58973.241730 &  163.0 & 0.000192 \\ 
58740.657011 &    0.5 & 0.000187 & 58938.175788 &  138.5 & 0.000331 & 58973.957329 &  163.5 & 0.000100 \\ 
58741.372040 &    1.0 & 0.000064 & 58938.891582 &  139.0 & 0.000111 & 58974.672985 &  164.0 & 0.000064 \\ 
58742.088529 &    1.5 & 0.000582 & 58939.607086 &  139.5 & 0.000077 & 58975.388704 &  164.5 & 0.000238 \\ 
58742.803597 &    2.0 & 0.000060 & 58940.323043 &  140.0 & 0.000101 & 58976.104447 &  165.0 & 0.000234 \\ 
58743.519648 &    2.5 & 0.000198 & 58943.185550 &  142.0 & 0.000066 & 58976.820321 &  165.5 & 0.000185 \\ 
58744.234845 &    3.0 & 0.000036 & 58943.901111 &  142.5 & 0.000126 & 58977.535837 &  166.0 & 0.000121 \\ 
58744.951027 &    3.5 & 0.000117 & 58944.616921 &  143.0 & 0.000105 & 58978.251723 &  166.5 & 0.000099 \\ 
58746.382386 &    4.5 & 0.000278 & 58945.332291 &  143.5 & 0.000137 & 58978.967050 &  167.0 & 0.000245 \\ 
58747.097436 &    5.0 & 0.000094 & 58946.048108 &  144.0 & 0.000116 & 58979.683172 &  167.5 & 0.000399 \\ 
58747.813581 &    5.5 & 0.000177 & 58946.763538 &  144.5 & 0.000098 & 58980.398548 &  168.0 & 0.000050 \\ 
58748.528738 &    6.0 & 0.000084 & 58947.479382 &  145.0 & 0.000052 & 58981.114617 &  168.5 & 0.000301 \\ 
58749.245059 &    6.5 & 0.000102 & 58948.194596 &  145.5 & 0.000156 & 58981.829822 &  169.0 & 0.000081 \\ 
58749.960311 &    7.0 & 0.000022 & 58948.910704 &  146.0 & 0.000417 & 59111.361155 &  259.5 & 0.000103 \\ 
58752.107291 &    8.5 & 0.000111 & 58949.625744 &  146.5 & 0.000165 & 59246.620293 &  354.0 & 0.000040 \\ 
58752.822358 &    9.0 & 0.000113 & 58950.341732 &  147.0 & 0.000170 & 59269.520633 &  370.0 & 0.000054 \\ 
58753.538450 &    9.5 & 0.000104 & 58951.057035 &  147.5 & 0.000407 & 59284.550493 &  380.5 & 0.000031 \\ 
58754.253626 &   10.0 & 0.000103 & 58951.772834 &  148.0 & 0.000076 & 59288.844246 &  383.5 & 0.000047 \\ 
58754.969482 &   10.5 & 0.000182 & 58952.488017 &  148.5 & 0.000334 & 59294.569560 &  387.5 & 0.000019 \\ 
58755.684584 &   11.0 & 0.000071 & 58953.203937 &  149.0 & 0.000187 & 59299.577580 &  391.0 & 0.000042 \\ 
58756.400821 &   11.5 & 0.000262 & 58953.919211 &  149.5 & 0.000106 & 59300.294088 &  391.5 & 0.000127 \\ 
58757.116049 &   12.0 & 0.000121 & 58954.635190 &  150.0 & 0.000112 & 59302.439373 &  393.0 & 0.000053 \\ 
58757.831766 &   12.5 & 0.000067 & 58956.066102 &  151.0 & 0.000052 & 59304.587293 &  394.5 & 0.000025 \\ 
58758.546987 &   13.0 & 0.000225 & 58956.781622 &  151.5 & 0.000369 & 59305.302275 &  395.0 & 0.000065 \\ 
58759.262719 &   13.5 & 0.000170 & 58957.497457 &  152.0 & 0.000093 & 59309.595790 &  398.0 & 0.000018 \\ 
58759.977997 &   14.0 & 0.000027 & 58958.212582 &  152.5 & 0.000298 & 59312.458118 &  400.0 & 0.000032 \\ 
58760.693828 &   14.5 & 0.000499 & 58958.928468 &  153.0 & 0.000177 & 59313.889689 &  401.0 & 0.000046 \\ 
58761.409279 &   15.0 & 0.000160 & 58959.643762 &  153.5 & 0.000078 & 59315.320979 &  402.0 & 0.000097 \\ 
58762.125004 &   15.5 & 0.000109 & 58960.359632 &  154.0 & 0.000108 & 59330.351320 &  412.5 & 0.000041 \\ 
58762.840198 &   16.0 & 0.000088 & 58961.074991 &  154.5 & 0.000733 & 59337.509634 &  417.5 & 0.000153 \\ 
58928.155362 &  131.5 & 0.000184 & 58961.791063 &  155.0 & 0.000053 & 59338.940221 &  418.5 & 0.000083 \\ 
58928.871754 &  132.0 & 0.000122 & 58962.506158 &  155.5 & 0.000211 & 59345.380062 &  423.0 & 0.000042 \\ 
58929.587171 &  132.5 & 0.000078 & 58963.222398 &  156.0 & 0.000143 & 59355.397913 &  430.0 & 0.000060 \\ 
58930.302948 &  133.0 & 0.000054 & 58963.937513 &  156.5 & 0.000136 & 59360.406959 &  433.5 & 0.000065 \\ 
58931.018721 &  133.5 & 0.000149 & 58964.653370 &  157.0 & 0.000117 & 59365.417315 &  437.0 & 0.000041 \\ 
58931.734335 &  134.0 & 0.000054 & 58965.369235 &  157.5 & 0.000222 & 59370.426462 &  440.5 & 0.000080 \\ 
58932.450115 &  134.5 & 0.000290 & 58966.084774 &  158.0 & 0.000048 & 59375.437294 &  444.0 & 0.000038 \\ 
58933.165904 &  135.0 & 0.000072 & 58968.231445 &  159.5 & 0.000155 & 59385.456851 &  451.0 & 0.000055 \\ 
58933.881325 &  135.5 & 0.000096 & 58969.662954 &  160.5 & 0.000276 & 59390.465954 &  454.5 & 0.000095 \\ 
58934.597304 &  136.0 & 0.000333 & 58970.378793 &  161.0 & 0.000190 & 59395.475232 &  458.0 & 0.000078 \\ 
58935.313141 &  136.5 & 0.000156 & 58971.094554 &  161.5 & 0.000259 & 59425.532993 &  479.0 & 0.000122 \\ 
58936.028769 &  137.0 & 0.000150 & 58971.810329 &  162.0 & 0.000064 & 59438.414552 &  488.0 & 0.000041 \\ 
\hline
\end{tabular}

{\it Notes.} Integer and half-integer cycle numbers refer to primary and secondary eclipses, respectively. Most of the eclipses (cycle nos. $-0.5$ to $169.0$) were observed by the \textit{TESS} spacecraft. The last 29 eclipses were observed during our photometric follow-up campaign.
\end{table*}

\begin{table*}
\caption{Binary Mid Eclipse Times for TIC~388459317.}
 \label{tab:T388459317ToM}
\begin{tabular}{@{}lrllrllrl}
\hline
BJD & Cycle  & std. dev. & BJD & Cycle  & std. dev. & BJD & Cycle  & std. dev. \\ 
$-2\,400\,000$ & no. &   \multicolumn{1}{c}{$(d)$} & $-2\,400\,000$ & no. &   \multicolumn{1}{c}{$(d)$} & $-2\,400\,000$ & no. &   \multicolumn{1}{c}{$(d)$} \\ 
\hline
58738.959183 &    0.0 & 0.000999 & 58761.894902 &   10.5 & 0.000778 & 58783.738969 &   20.5 & 0.000753 \\ 
58740.050975 &    0.5 & 0.001334 & 58765.170004 &   12.0 & 0.000850 & 58785.925147 &   21.5 & 0.000720 \\ 
58741.142318 &    1.0 & 0.000401 & 58766.261662 &   12.5 & 0.000449 & 58787.015412 &   22.0 & 0.000822 \\ 
58742.233893 &    1.5 & 0.001226 & 58767.356124 &   13.0 & 0.000342 & 58788.106602 &   22.5 & 0.000798 \\ 
58743.325849 &    2.0 & 0.000283 & 58768.448985 &   13.5 & 0.000271 & 58789.194873$^*$ &   23.0 & 0.000689 \\ 
58744.417232 &    2.5 & 0.000438 & 58769.539806 &   14.0 & 0.001258 & 59040.452345 &  138.0 & 0.000047 \\ 
58745.513752 &    3.0 & 0.001082 & 58770.632011 &   14.5 & 0.000405 & 59052.464842 &  143.5 & 0.000099 \\ 
58747.695604 &    4.0 & 0.000571 & 58771.724709 &   15.0 & 0.000634 & 59053.559698 &  144.0 & 0.000146 \\ 
58752.063987 &    6.0 & 0.000648 & 58772.813668 &   15.5 & 0.000665 & 59101.631939 &  166.0 & 0.000159 \\ 
58753.154730 &    6.5 & 0.000653 & 58773.911894 &   16.0 & 0.000835 & 59140.946668$^*$ &  184.0 & 0.000198 \\ 
58754.249371 &    7.0 & 0.000833 & 58774.998918 &   16.5 & 0.000254 & 59173.727724 &  199.0 & 0.000102 \\ 
58755.338500 &    7.5 & 0.000640 & 58776.090186 &   17.0 & 0.000989 & 59174.817539 &  199.5 & 0.000137 \\ 
58756.433995 &    8.0 & 0.000475 & 58778.276431 &   18.0 & 0.000677 & 59383.467258 &  295.0 & 0.000339 \\ 
58757.526834 &    8.5 & 0.000671 & 58779.368405 &   18.5 & 0.000250 & 59419.512135 &  311.5 & 0.000087 \\ 
58758.615973 &    9.0 & 0.000721 & 58780.461298 &   19.0 & 0.000799 & 59441.365099 &  321.5 & 0.000065 \\ 
58759.708979 &    9.5 & 0.000604 & 58781.555532 &   19.5 & 0.000708 & 59501.441078 &  349.0 & 0.000076 \\ 
58760.803010 &   10.0 & 0.000569 & 58782.650930 &   20.0 & 0.000559 &&& \\ 
\hline
\end{tabular}

{\it Notes.} Integer and half-integer cycle numbers refer to primary and secondary eclipses, respectively. Most of the eclipses (cycle nos. $0.0$ to $23.0$) were observed by the \textit{TESS} spacecraft. The last 11 eclipses were observed during our photometric follow-up campaign, however, the two eclipses denoted by asterics were not used for the analysis.
\end{table*}

\begin{table*}
\caption{Binary Mid Eclipse Times for TIC~52041148.}
 \label{tab:T052041148ToM}
\begin{tabular}{@{}lrllrllrl}
\hline
BJD & Cycle  & std. dev. & BJD & Cycle  & std. dev. & BJD & Cycle  & std. dev. \\ 
$-2\,400\,000$ & no. &   \multicolumn{1}{c}{$(d)$} & $-2\,400\,000$ & no. &   \multicolumn{1}{c}{$(d)$} & $-2\,400\,000$ & no. &   \multicolumn{1}{c}{$(d)$} \\ 
\hline
58790.924998 &   -1.0 & 0.001088 & 58964.187141 &   96.0 & 0.001152 & 58995.453022 &  113.5 & 0.000429 \\       
58791.818404 &   -0.5 & 0.003674 & 58965.079483 &   96.5 & 0.001325 & 58997.238502 &  114.5 & 0.001239 \\ 
58792.714170 &    0.0 & 0.001337 & 58965.975140 &   97.0 & 0.001131 & 58998.133102 &  115.0 & 0.000734 \\ 
58793.606751 &    0.5 & 0.001093 & 58966.868252 &   97.5 & 0.001080 & 58999.023678 &  115.5 & 0.000998 \\ 
58794.502805 &    1.0 & 0.000865 & 58967.761685 &   98.0 & 0.001016 & 58999.918753 &  116.0 & 0.001017 \\ 
58796.289759 &    2.0 & 0.000689 & 58969.551239 &   99.0 & 0.000323 & 59000.810334 &  116.5 & 0.000778 \\ 
58797.181844 &    2.5 & 0.000821 & 58970.441922 &   99.5 & 0.001125 & 59001.705426 &  117.0 & 0.000691 \\ 
58798.077597 &    3.0 & 0.000691 & 58971.340561 &  100.0 & 0.002395 & 59002.595368 &  117.5 & 0.000907 \\ 
58798.967783 &    3.5 & 0.000672 & 58973.126272 &  101.0 & 0.004843 & 59003.490469 &  118.0 & 0.000837 \\ 
58799.863324 &    4.0 & 0.000944 & 58974.016991 &  101.5 & 0.001347 & 59004.382018 &  118.5 & 0.000849 \\ 
58800.753860 &    4.5 & 0.002069 & 58974.913118 &  102.0 & 0.001010 & 59005.277151 &  119.0 & 0.000841 \\ 
58804.327777 &    6.5 & 0.000439 & 58975.805010 &  102.5 & 0.000855 & 59006.168090 &  119.5 & 0.000788 \\ 
58805.222844 &    7.0 & 0.000370 & 58976.700022 &  103.0 & 0.000731 & 59007.061496 &  120.0 & 0.000723 \\ 
58806.113293 &    7.5 & 0.000449 & 58977.591766 &  103.5 & 0.000931 & 59007.954599 &  120.5 & 0.000811 \\ 
58807.008060 &    8.0 & 0.000419 & 58978.486974 &  104.0 & 0.000582 & 59008.848779 &  121.0 & 0.000833 \\ 
58807.900436 &    8.5 & 0.000411 & 58979.378549 &  104.5 & 0.000738 & 59012.429804$^*$ &  123.0 & 0.000673 \\  
58808.794618 &    9.0 & 0.000494 & 58980.273254 &  105.0 & 0.000630 & 59036.530944$^*$ &  136.5 & 0.000531 \\  
58809.686435 &    9.5 & 0.000471 & 58981.165843 &  105.5 & 0.000487 & 59061.531198 &  150.5 & 0.000097 \\      
58810.580651 &   10.0 & 0.000497 & 58982.059389 &  106.0 & 0.000514 & 59071.355924 &  156.0 & 0.000166 \\      
58811.471800 &   10.5 & 0.000603 & 58983.846571 &  107.0 & 0.001082 & 59087.431071 &  165.0 & 0.000056 \\      
58812.366545 &   11.0 & 0.000562 & 58984.736738 &  107.5 & 0.000993 & 59096.362283 &  170.0 & 0.000071 \\      
58813.257156 &   11.5 & 0.000855 & 58985.632532 &  108.0 & 0.000715 & 59103.507631 &  174.0 & 0.000144 \\      
58814.151883 &   12.0 & 0.002758 & 58986.523328 &  108.5 & 0.000642 & 59113.329970 &  179.5 & 0.000071 \\      
58956.144958 &   91.5 & 0.001171 & 58987.417818 &  109.0 & 0.000638 & 59145.493329 &  197.5 & 0.000064 \\      
58957.039530 &   92.0 & 0.002321 & 58988.309225 &  109.5 & 0.000613 & 59158.892302 &  205.0 & 0.000389 \\      
58957.929565 &   92.5 & 0.002986 & 58989.203753 &  110.0 & 0.000623 & 59161.577011 &  206.5 & 0.000078 \\      
58958.826656 &   93.0 & 0.001857 & 58990.093584 &  110.5 & 0.000621 & 59166.927119$^*$ &  209.5 & 0.000261 \\  
58959.718294 &   93.5 & 0.001687 & 58990.990630 &  111.0 & 0.000499 & 59247.302330 &  254.5 & 0.000123 \\      
58960.613720 &   94.0 & 0.001338 & 58991.879830 &  111.5 & 0.000584 & 59472.364889 &  380.5 & 0.000084 \\      
58961.506508 &   94.5 & 0.001298 & 58992.775136 &  112.0 & 0.000484 & 59489.340694 &  390.0 & 0.000093 \\      
58962.400536 &   95.0 & 0.001368 & 58993.667758 &  112.5 & 0.000504 & 59512.573082$^*$ &  403.0 & 0.000069 \\
58963.293110 &   95.5 & 0.001162 & 58994.561239 &  113.0 & 0.000434 && \\ 
\hline
\end{tabular}

{\it Notes.} Integer and half-integer cycle numbers refer to primary and secondary eclipses, respectively. Most of the eclipses (cycle nos. $-1.0$ to $121.0$) were observed by the \textit{TESS} spacecraft. The last 16 eclipses were observed during our photometric follow-up campaign, however, the four eclipses denoted by asterics were not used for the analysis.
\end{table*}



\subsection{Probing the Outer Orbit with Archival Data}
\label{sec:archival}

After discovering a triply eclipsing triple system with \textit{TESS} data, the question immediately arises as to the nature of its outer orbit, e.g., period and eccentricity.  We therefore routinely examine the ASAS-SN (\citealt{shappee14}; \citealt{kochanek17}) and ATLAS (\citealt{tonry18}; \citealt{smith20}) archival data sets.  The former typically have $\sim$1500-3000 photometric measurements of a given source, while ATLAS provides approximately 1500 measurements.  However, the latter set has somewhat better photometric precision, and therefore the two data sets are of comparable statistical precision.  Of course, we also make use of KELT \citep{pepperetal07}, WASP \citep{2006PASP..118.1407P}, and/or HAT \citep{bakosetal02} data when they are available.  

For TIC 193993801, TIC 388459317, and TIC 52041148, there are both ASAS-SN and ATLAS photometry available for all three.  Additionally, there are KELT data available for TIC 193993801.  For this latter source, the ATLAS data are somewhat affected by saturation effects and therefore we do not include them in the analysis of that source.  Our procedure for combining the ATLAS and ASAS-SN data sets is as follows.  We first median normalize the different colors within the ATLAS data set to each other, and the same for the ASAS-SN data set.  We then median normalize the ATLAS to the ASAS-SN data sets.  Outliers to the very bright and very faint side of the median value are eliminated, being careful not to exclude any low points that might conceivably be part of an eclipse. 

We next apply a Box Least Squares transform \citep{kovacs02} to the combined archival data set to determine the long-term (over several years) average sidereal period of the eclipsing binary.  Once we know that period accurately we remove that signal from the data train as follows.  We fit simultaneously for between 30 and 100 harmonics of the EB, depending on how sharp its eclipsing features are.  These best-fit sines and cosines are evaluated at each time of the archival data set, and then that value is subtracted from the flux at that point.  

\begin{figure}
\vspace{0.0cm}
\begin{center}
\includegraphics[width=0.45 \textwidth]{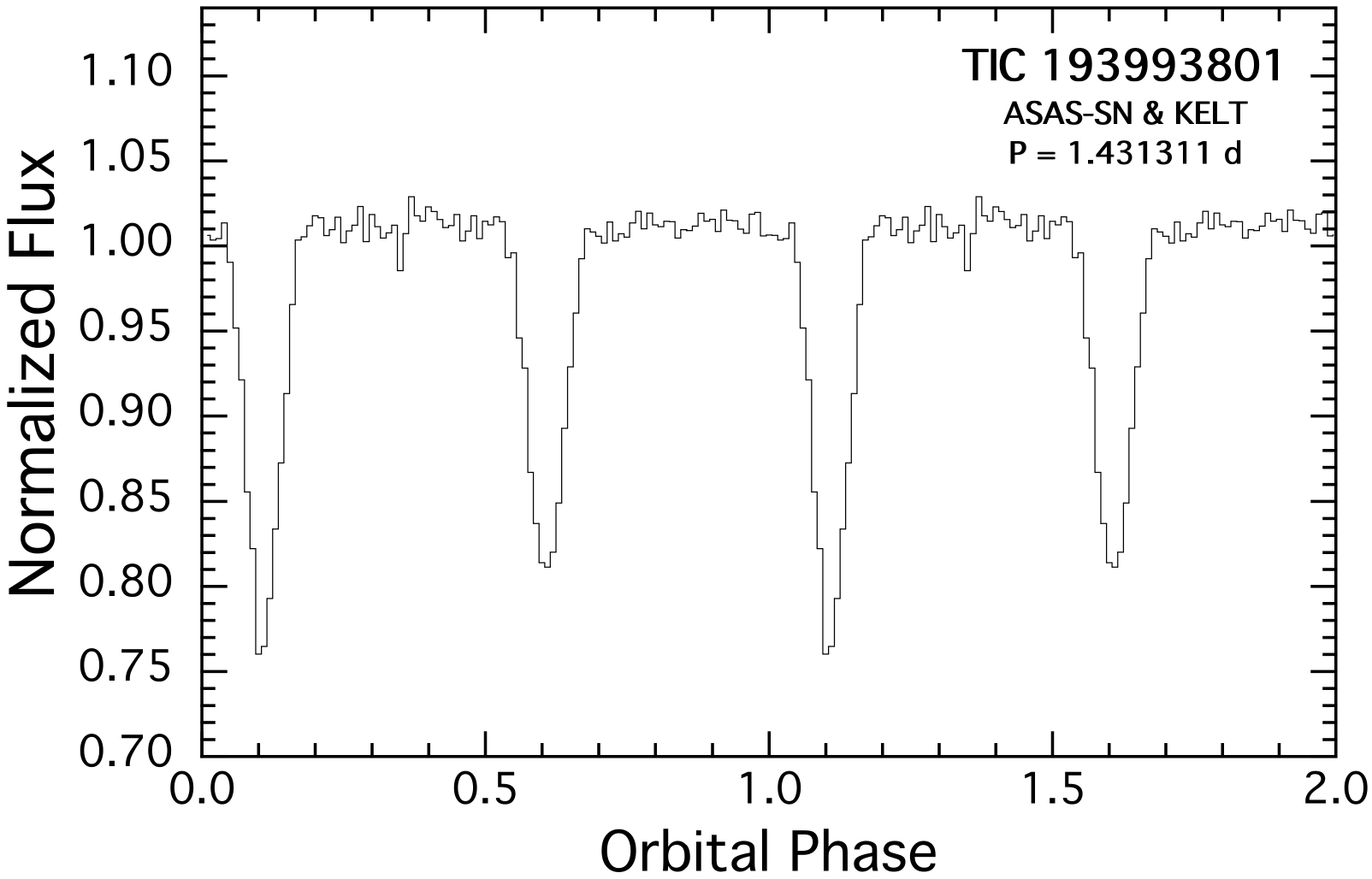} \vglue0.02cm \hglue-0.12cm
\includegraphics[width=0.465 \textwidth]{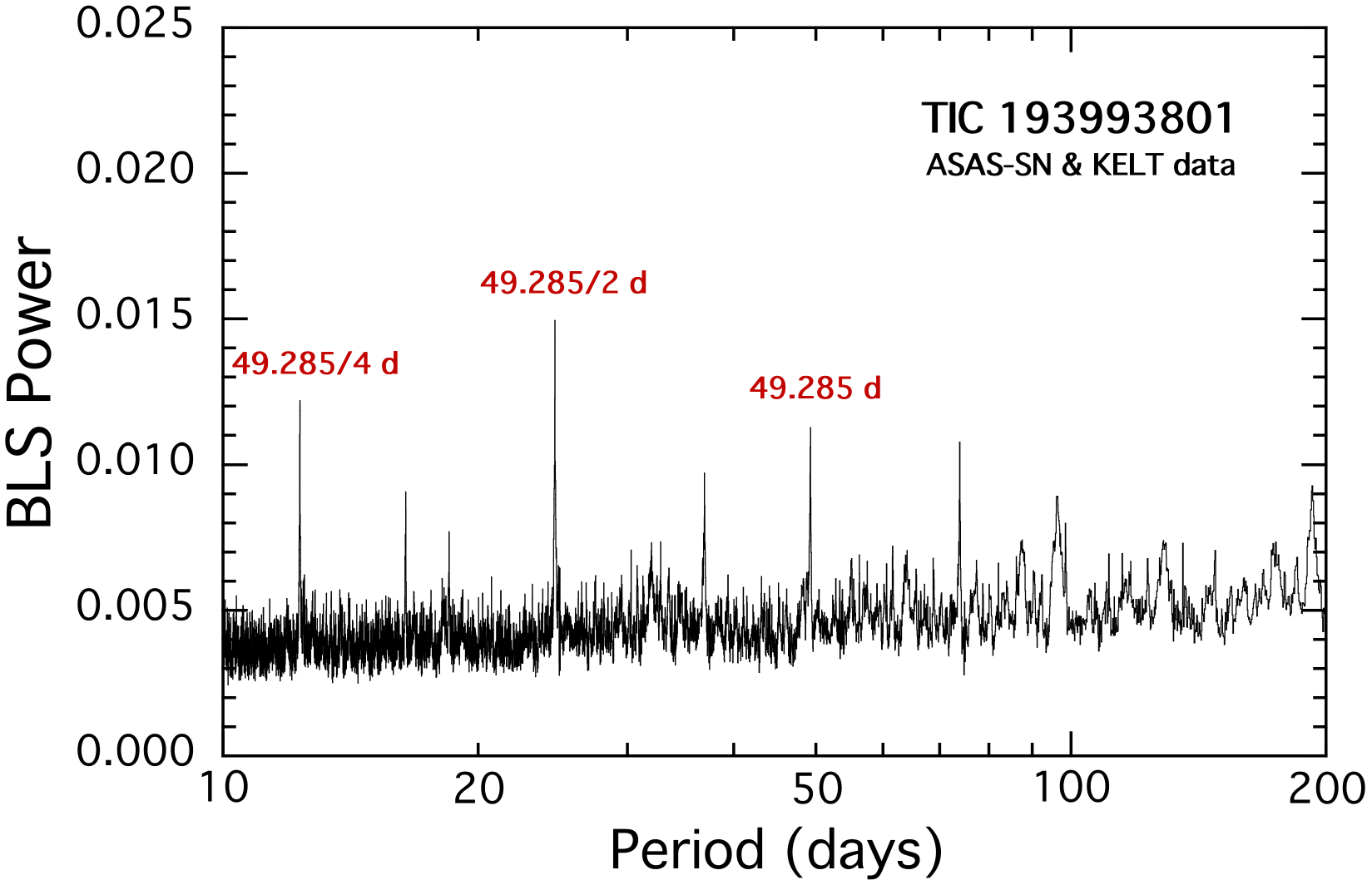} \vglue0.05cm \hglue0.15cm
\includegraphics[width=0.453 \textwidth]{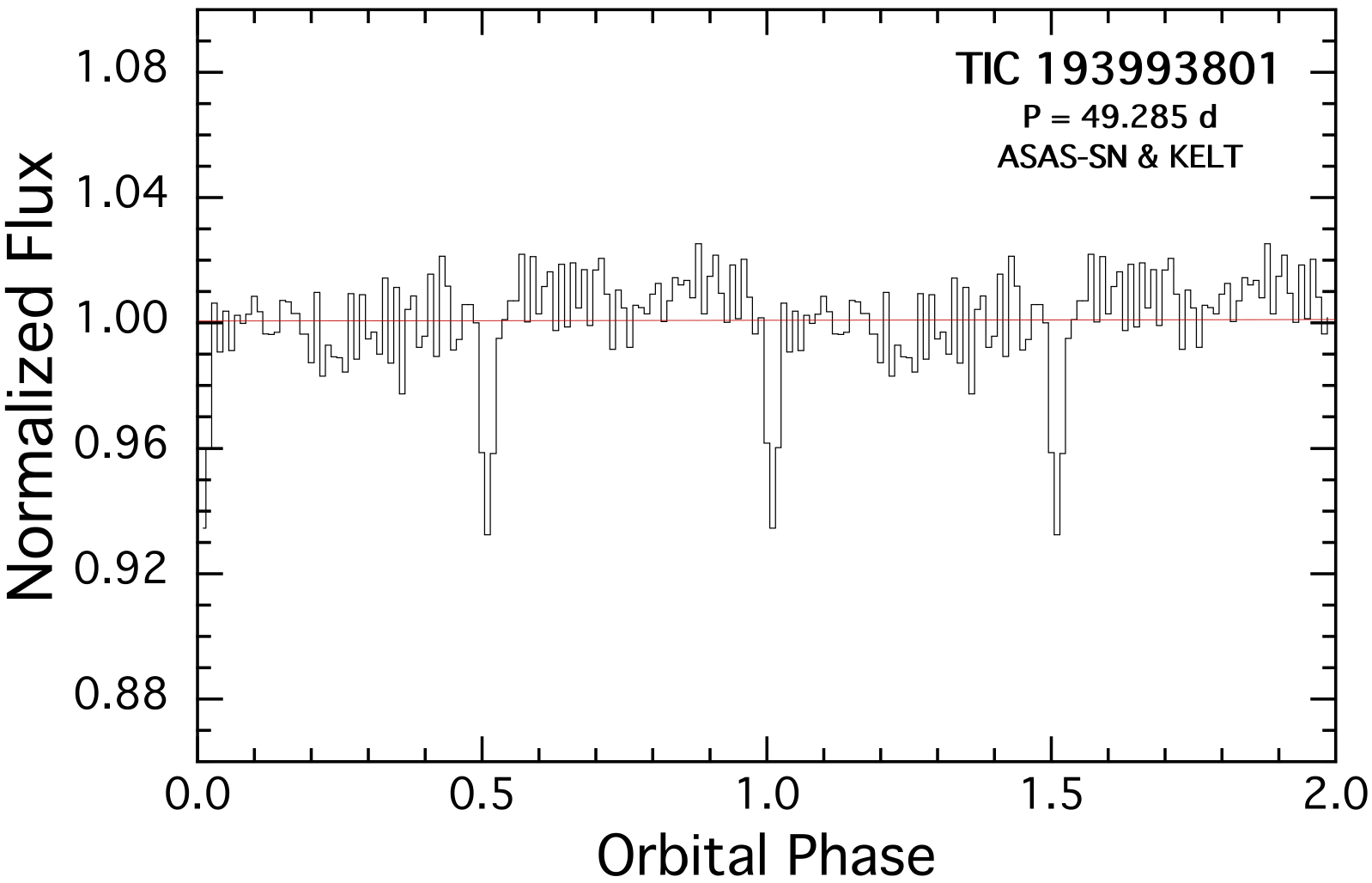}
\caption{The outer orbit of TIC 193993801 as diagnosed by the ASAS-SN and KELT archival photometric data. {\it Top panel}: fold of the data about the period of the eclipsing binary.  {\it Middle panel}: BLS transform of the data after subtracting out the lightcurve of the EB. The highest six peaks in the BLS are all due to the 49.285-day outer orbit and its harmonics. {\it Bottom panel}: fold of the data about the outer orbital period of 49.285 d as determined from the BLS transform of the cleaned data set. One can see that the primary and secondary outer eclipses are separated by close to half the orbital cycle making $e \cos \omega_{\rm out} \lesssim 0.02$.}
\label{fig:193993801}
\end{center}
\end{figure}

\begin{figure}
\vspace{0.0cm}
\begin{center}
\includegraphics[width=0.45 \textwidth]{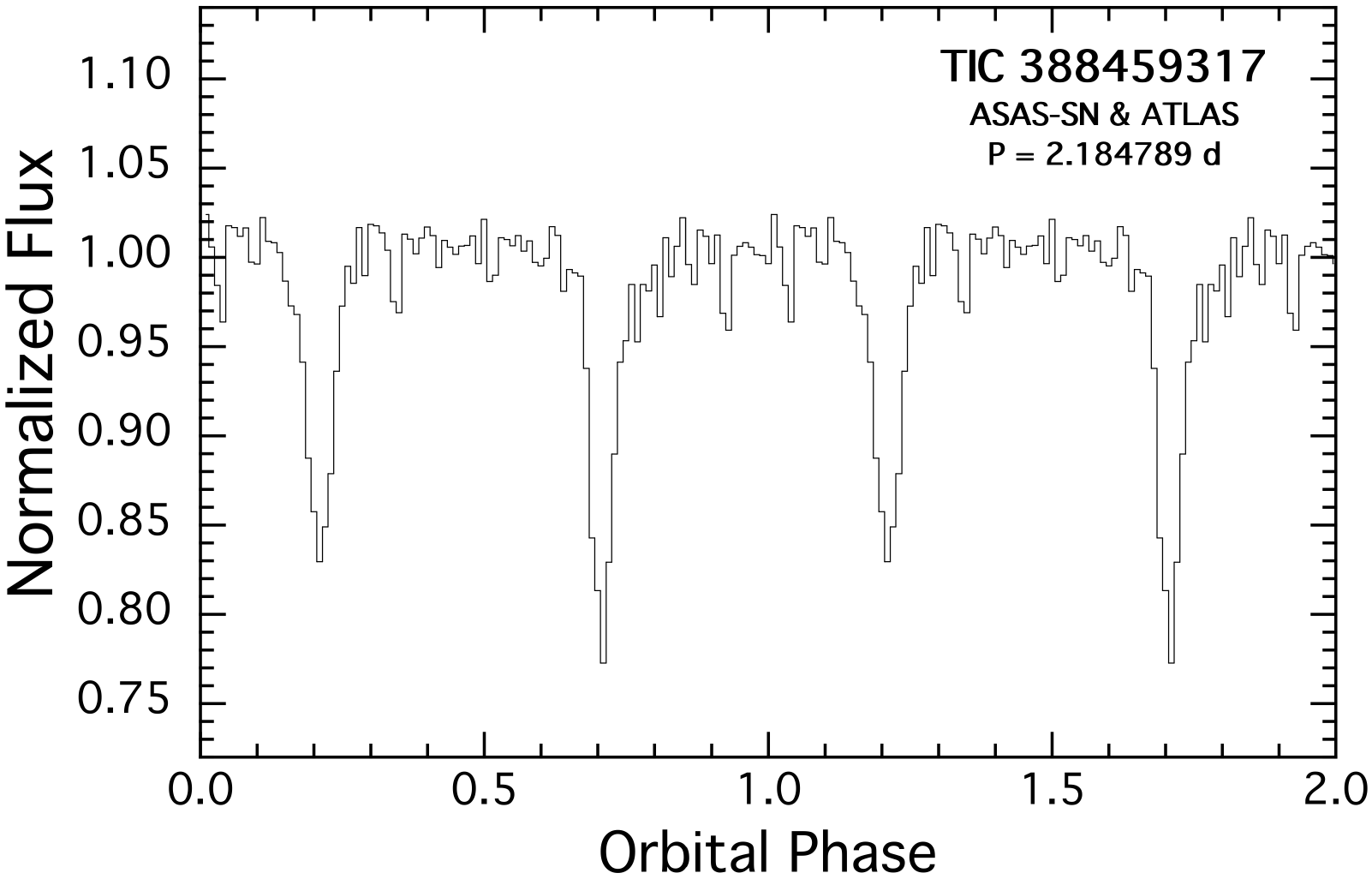} \vglue0.02cm \hglue-0.32cm
\includegraphics[width=0.45 \textwidth]{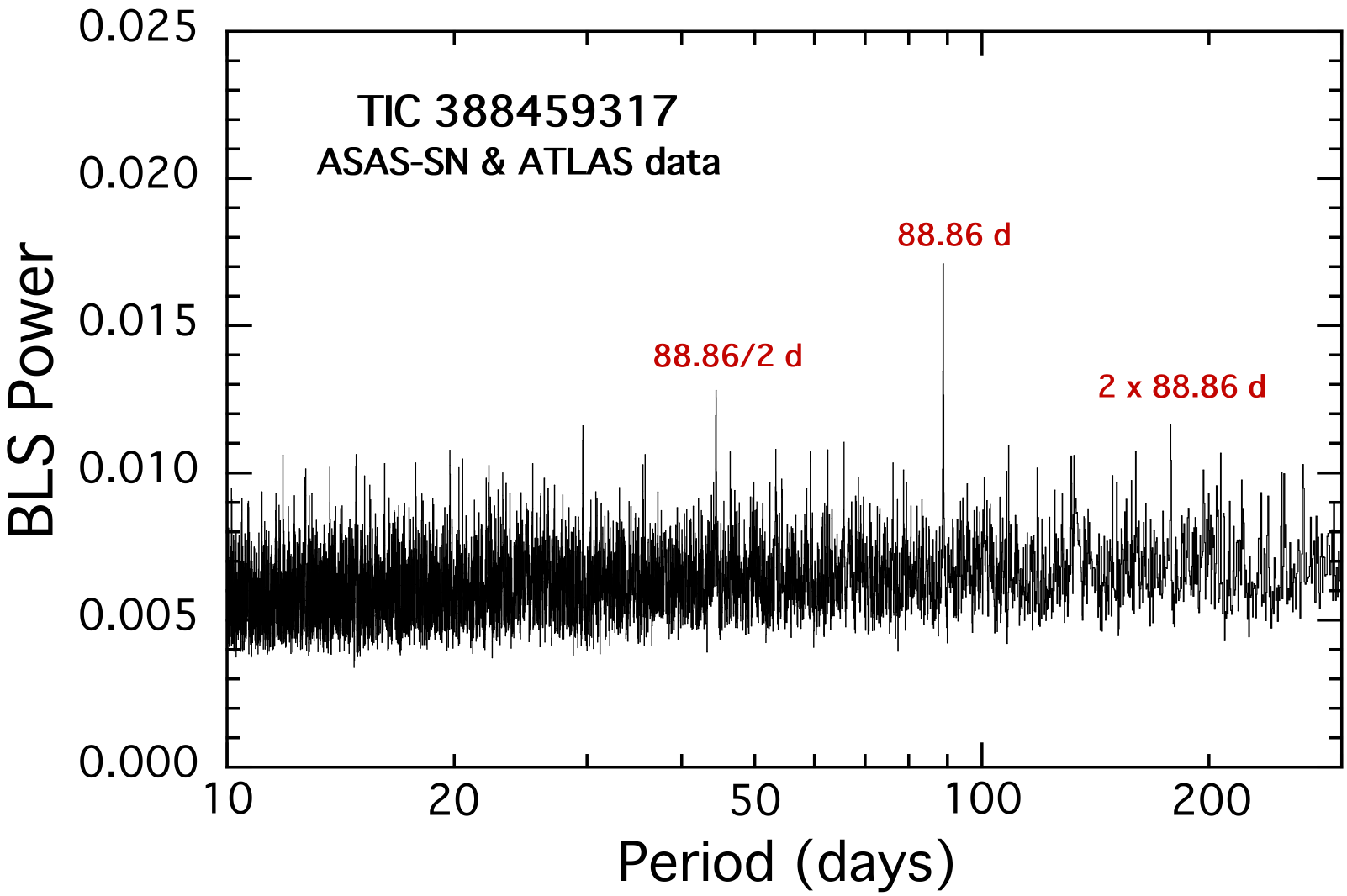}  \vglue0.05cm \hglue0.15cm
\includegraphics[width=0.45 \textwidth]{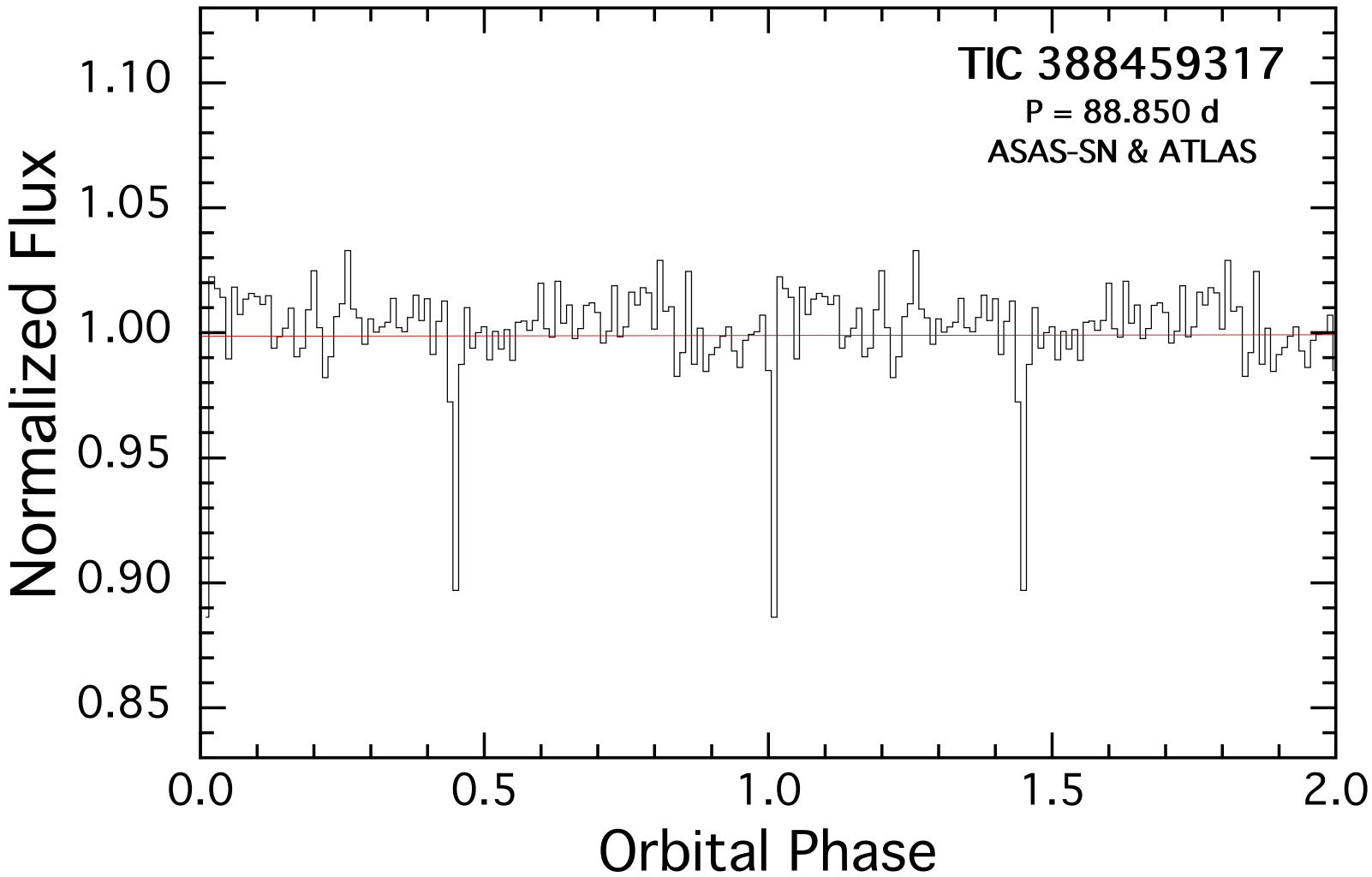}
\caption{The outer orbit of TIC 388459317 as diagnosed by the ASAS-SN and ATLAS archival photometric data. {\it Top panel}: fold of the data about the period of the eclipsing binary.  {\it Middle panel}: BLS transform of the data after subtracting out the lightcurve of the EB.  The highest three peaks in the BLS are all due to the 88.6-day outer orbit and its harmonics.  {\it Bottom panel} : fold of the data about the outer orbital period of 88.85 d as determined from the BLS transform of the cleaned data set. The eccentricity of the outer orbit is self-evident from the folded lightcurve.}
\label{fig:388459317}
\end{center}
\end{figure}

\begin{figure}
\vspace{0.0cm}
\begin{center}
\includegraphics[width=0.45 \textwidth]{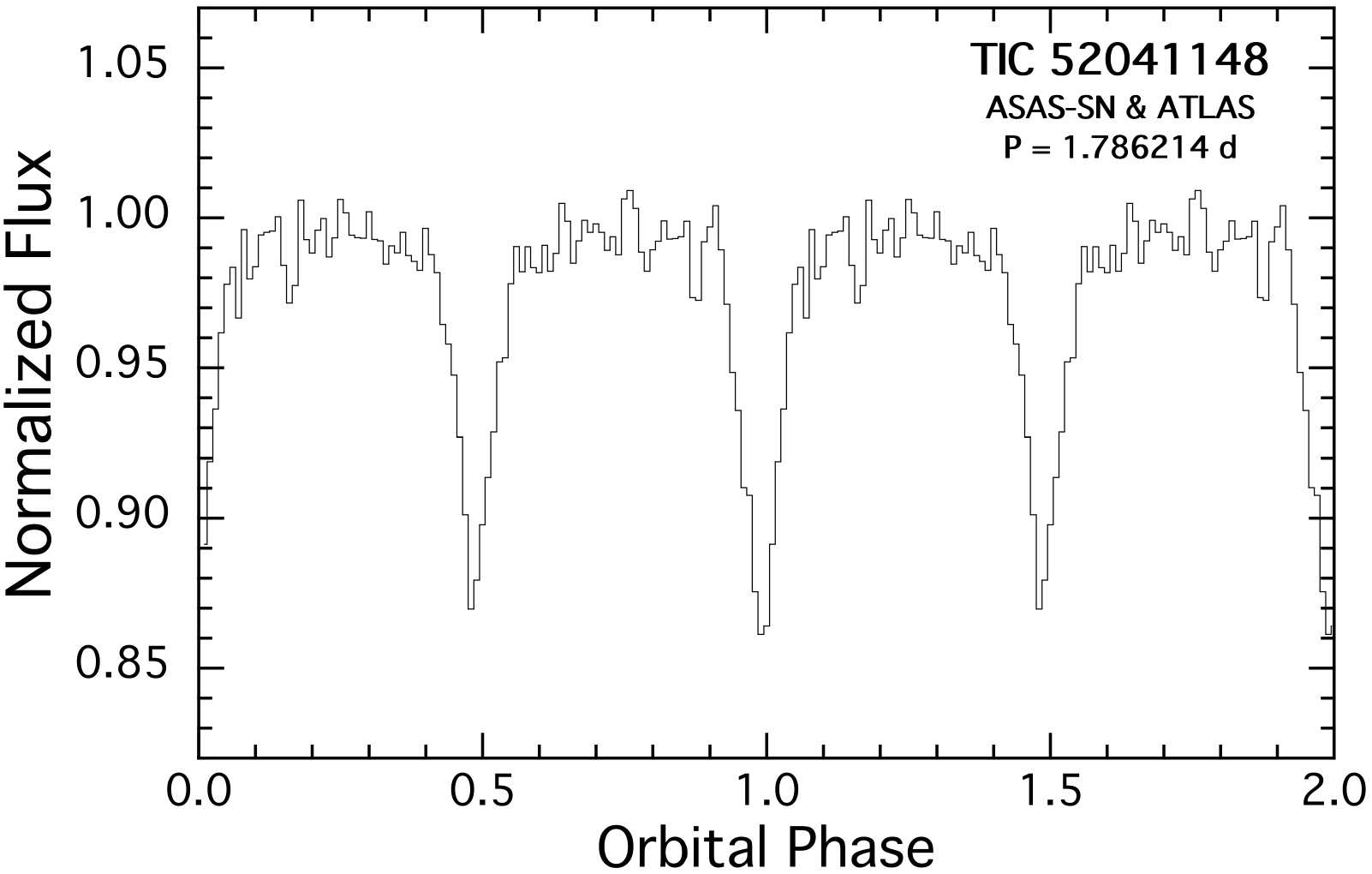} \vglue0.02cm \hglue-0.32cm
\includegraphics[width=0.45 \textwidth]{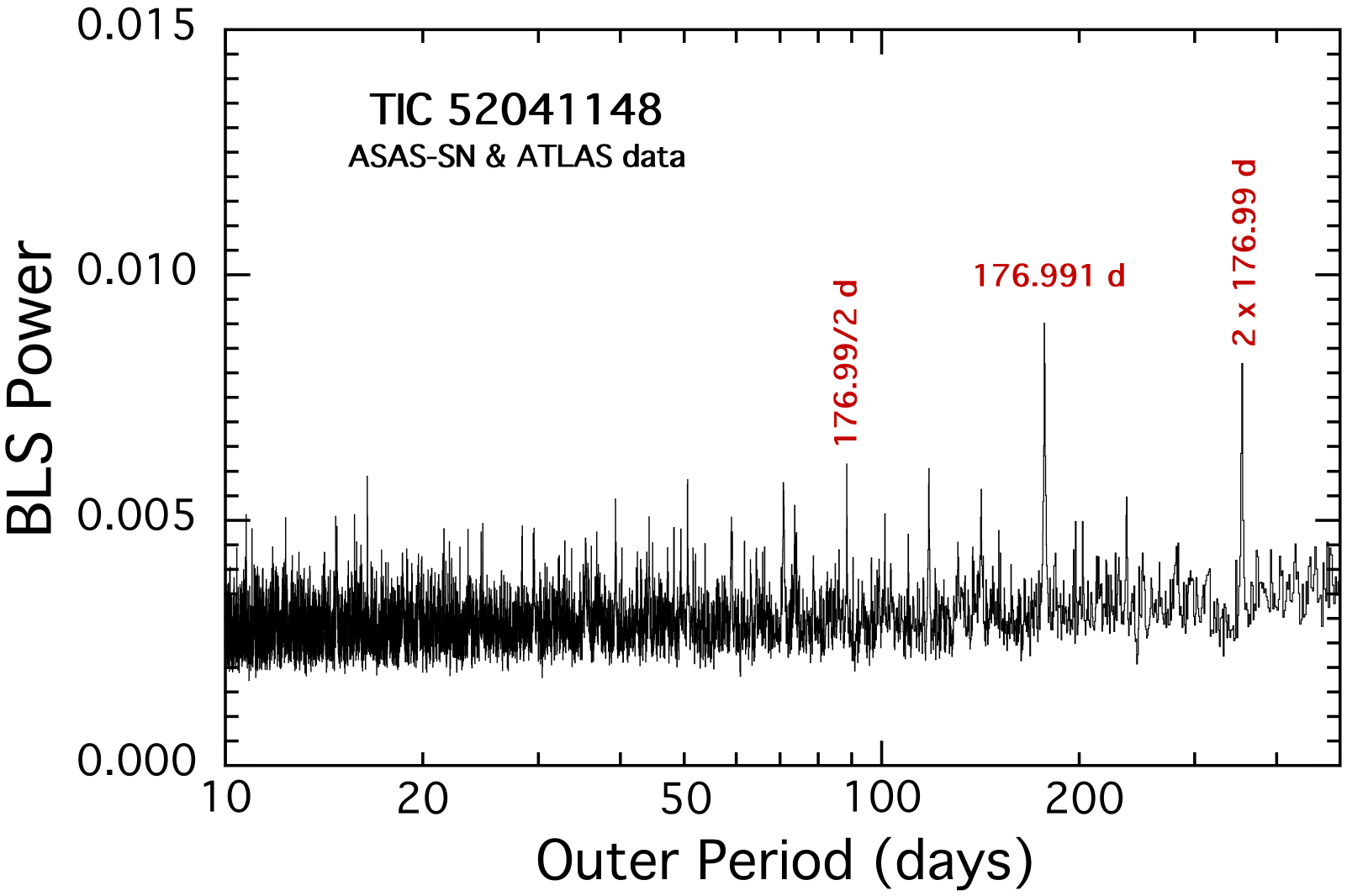} \vglue0.05cm \hglue0.18cm
\includegraphics[width=0.45 \textwidth]{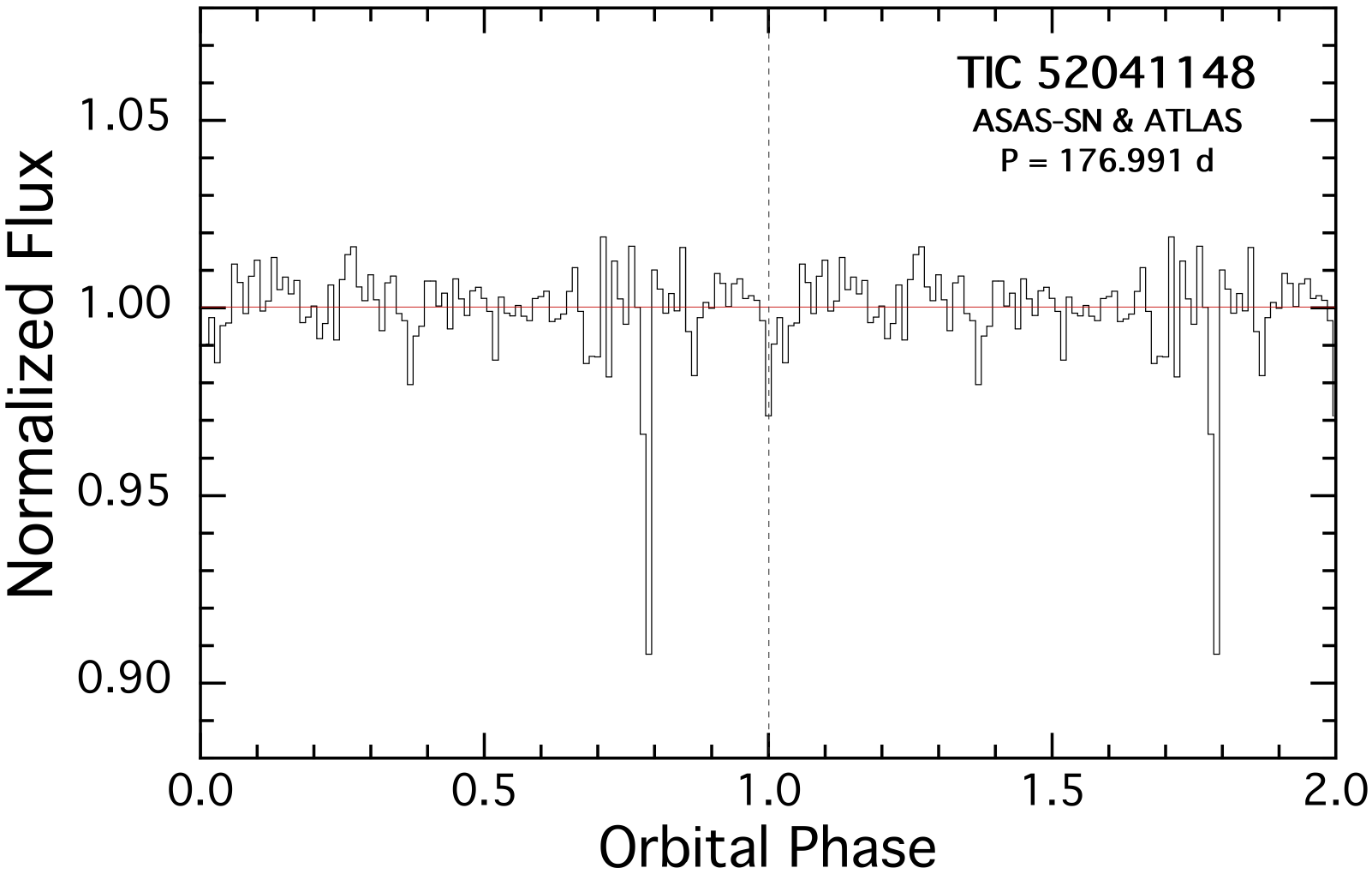}
\caption{The outer orbit of TIC 52041148 as diagnosed by the ASAS-SN and ATLAS archival photometric data. {\it Top panel}: fold of the data about the period of the eclipsing binary.  {\it Middle panel}: BLS transform of the data after subtracting out the lightcurve of the EB. The highest three peaks in the BLS are all due to the 176.99-day outer orbit and its harmonics. {\it Bottom panel} : fold of the data about the outer orbital period of 176.99 d as determined from the BLS transform of the cleaned data set. One can see from the phasing of the primary outer eclipse that it occurs $\sim$0.21 outer orbital phases prior to the secondary outer eclipse which is the one observed by {\it TESS}.}
\label{fig:52041148}
\end{center}
\end{figure}

We illustrate the results of the above procedure with a set of three figures for each of our sources (Figs.~\ref{fig:193993801}, \ref{fig:388459317}, and \ref{fig:52041148}).  In the top panel we show the fold of the archival data for the eclipsing binary in the triple.  In the middle panel we show the BLS transform of the data after the EB lightcurve has been removed via its Fourier components.  Finally, the bottom panel displays the fold of the data about the outer orbital period as deduced from the BLS transform.  In all cases, zero phase for the outer orbit is taken to be the time of one of the third-body eclipses in the {\em TESS} data.

\subsection{Photometric follow up }
\label{sec:follow_up}

Following the discovery and the preliminary characterization of these (and some other) triply eclipsing triple systems, we organized photometric follow-up observational campaigns with the participation of amateur and professional astronomers at different sites in Hungary and the US, as follows:

\textit{Gothard Astrophysical Observatory} (Szombathely, Hungary): Observations were carried out with a newly installed 0.8-m alt-azimuth Ritchey-Chr\'etien-type ASA AZ800 f/7 telescope, equipped with an FLI MicroLine ML 16803 CCD camera with $4096 \times 4096$ pixels and $9\times9~\mu$m pixel size. The field of view of the instrument is $22.6 \times 22.6$ arcmin, and the pixel scale is 0.33 arcsec per pixel. The CCD uses Sloan $g'$, $r'$, $i'$, $z'$ and Clear filters. TIC~193993801 was observed in the Sloan $z'$ filter with an exposure time of 120\,s. The remaining two fainter objects,  TIC~52041148 and TIC~388459317, were observed in white light using the Clear filter and had an exposure time of 60\,s. The photometric observations were reduced using the software package {\sc IRAF}. The reduction included dark and flat corrections, astrometric calibrations, and differential photometry. During the differential photometry several close comparison stars of similar brightness were used.

\textit{Baja Astronomical Observatory of Szeged University} (Baja, Hungary): An identical ASA AZ800 f/7 telescope to that at Gothard Astrophysical Observatory is operated here as well.  It has an FLI ProLine PL23042 back-illuminated CCD camera with $2048 \times 2048$, $15~\mu$m square pixels. This setup provides an $18.9 \times 18.9$~arcmin field of view, with a plate scale of $0.55$ arcseconds per pixel. The Sloan $r'$ and $z'$ filters with exposure times between $10$ and $120$ seconds were used for the triple-star observations (there was no 'Clear' filter available). The usual photometric calibration procedures were performed using site-specific packages written in {\sc IRAF} and {\sc Python}. The reduction included bias, dark and flat corrections, followed by astrometric calibration (using {\tt astrometry.net}) and aperture photometry. The closest few comparison stars were used for obtaining differential magnitudes.

\textit{Piszk\'estet\H o Observatory} (Piszk\'es-tet\H o, Hungary):  We used the 1-m Ritchey-Chr\'etien-Coud\'e telescope located at the observatory and equipped with an SI 1100S CCD. The camera has a 4\,k$\times$4\,k pixel frame with 15 micron pixel-size resulting in an effective field-of-view of 16$\times$16 arcmin and effective resolution of 0.23 arcsec per pixel. We obtained photometric images of TIC~193993801 in Johnson-Cousins R and TIC~388459317 in a Clear filter both with 60\,s exposure time. The reduction of the images was performed with the {\sc FITSH} \citep[][]{2012MNRAS.421.1825P} software package applying bias, dark and flat field corrections, astrometric calibrations and differential photometry using the nearby Gaia DR2 1594781593922650624 and Gaia DR2 2003450996324196480 as comparison stars, respectively.

\textit{Junk Bond Observatory} (Arizona, US): JBO is in southeast Arizona at 1210 meter elevation. It houses a domed, 0.8 meter Ritchey-Chr\'etien at F4.6, robotically controlled with Astronomers Control Panel (ACP) software. The camera is an SBIG STL6303E CCD and the images are unfiltered. The aperture photometry was pipelined through AstroimageJ software with standard darks, flats and bias frames.

\textit{Patterson Observatory} (Arizona, US): Patterson Observatory is in southeast Arizona at an elevation of 1412 meters. It houses a domed, 0.5 meter Ritchey-Chr\'etien at F8, robotically controlled with Astronomers Control Panel (ACP) software. The camera is an SBIG STL-1001e unfiltered. The aperture photometry was pipelined through AstroimageJ software with standard darks, flats and bias frames.

\textit{Hereford Arizona Observatory} (Arizona, US): The Hereford Arizona Observatory consists of an AstroTech 0.41-m 
Ritchey-Chretien telescope located at a Southern Arizona site with an altitude of 1420 m. An SBIG ST-10XME CCD camera was used without filter. Image sets with exposure times of 10 seconds were obtained of TIC 193993801 on three dates in 2020 September and October. Standard bias, dark and flat-field calibrations were performed before photometry measurements were made with a single reference star with an $r'=12.338$ magnitude from the APASS catalog.

\begin{figure*}
\includegraphics[width=0.8\textwidth]{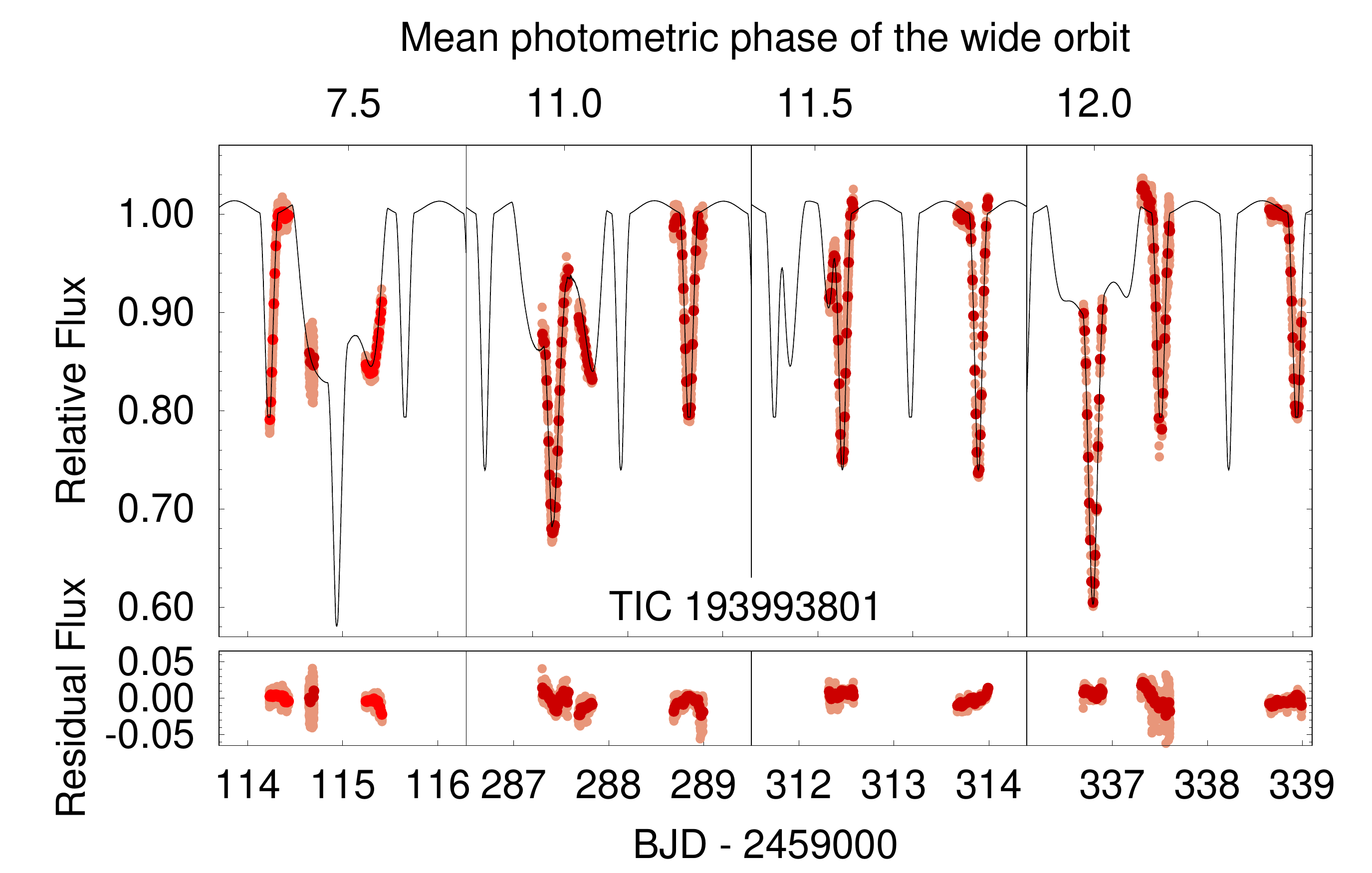}
\includegraphics[width=0.8\textwidth]{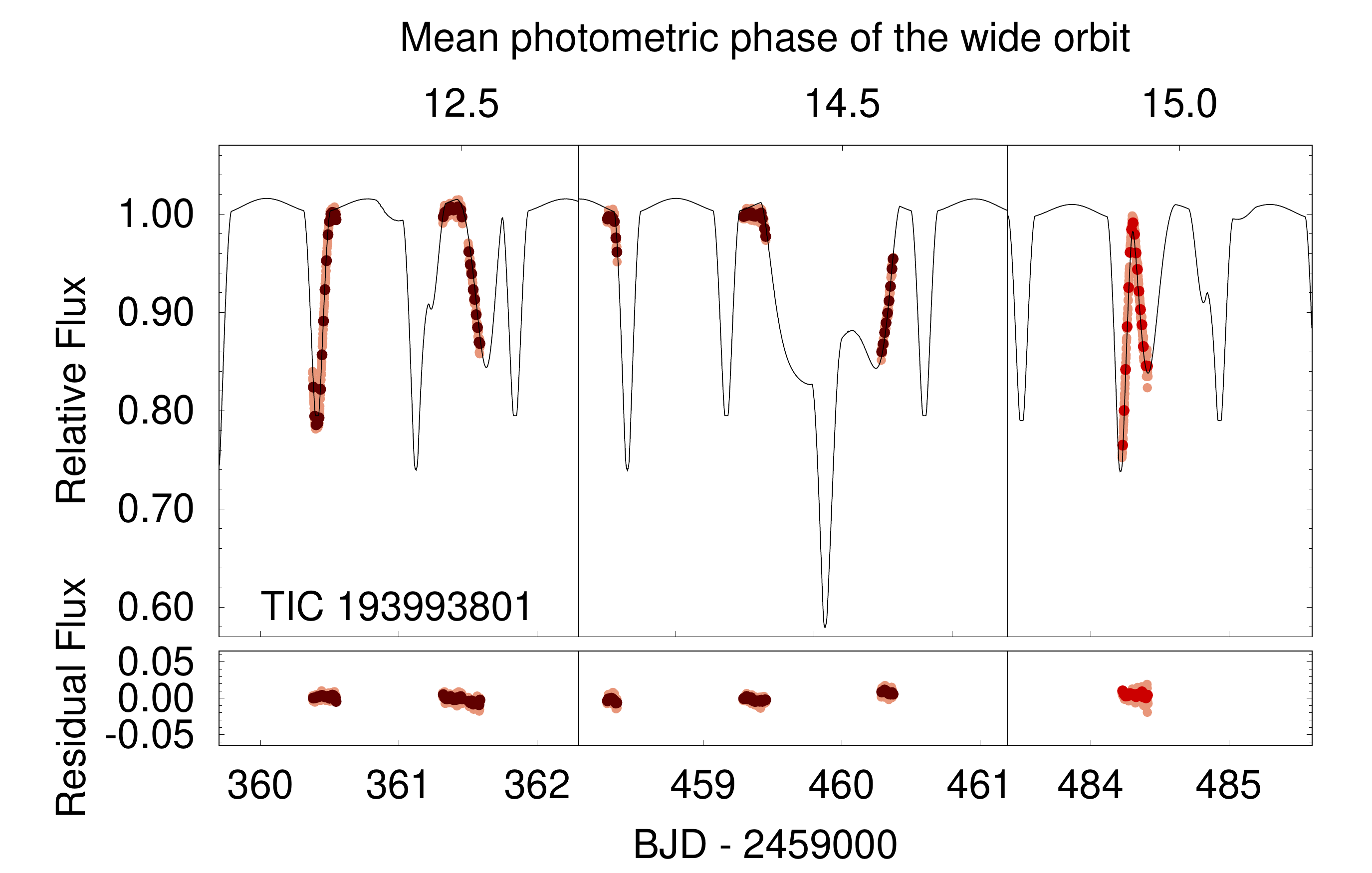}
\caption{Seven third-body eclipses of TIC~193993801 partially observed during our ground-based follow up photometric campaign. Red, claret and dark brown dots refer to observations made with $R_C$, $r'$ and $z'$ filters, respectively. The lighter dots stand for the original observations, while the darker ones show their 900-sec averages which have been used for the photodynamical fits. The black line represents the best-fitting spectro-photodynamical model solution. (Note, the first contact of the last event around BJD~2\,459\,484.3 was observed after finalizing the spectro-photodynamical modelling and, therefore naturally, it was not taken into account for the parameter search.) Residuals against this photodynamical model is also shown below.}
\label{fig:T193993801lcE3ground}
\end{figure*}

During the whole multi-site photometric campaign, TIC 193993801 was observed 42 nights between 18 September 2020 and 11 October 2021. These measurements cover portions of seven third-body eclipses (Fig.~\ref{fig:T193993801lcE3ground}). Moreover, 29 regular eclipses of the inner binary were also observed. The mid-eclipse times for these eclipses are also listed in Table~\ref{tab:T193993801ToM}.  

\begin{figure*}
\includegraphics[width=0.8\textwidth]{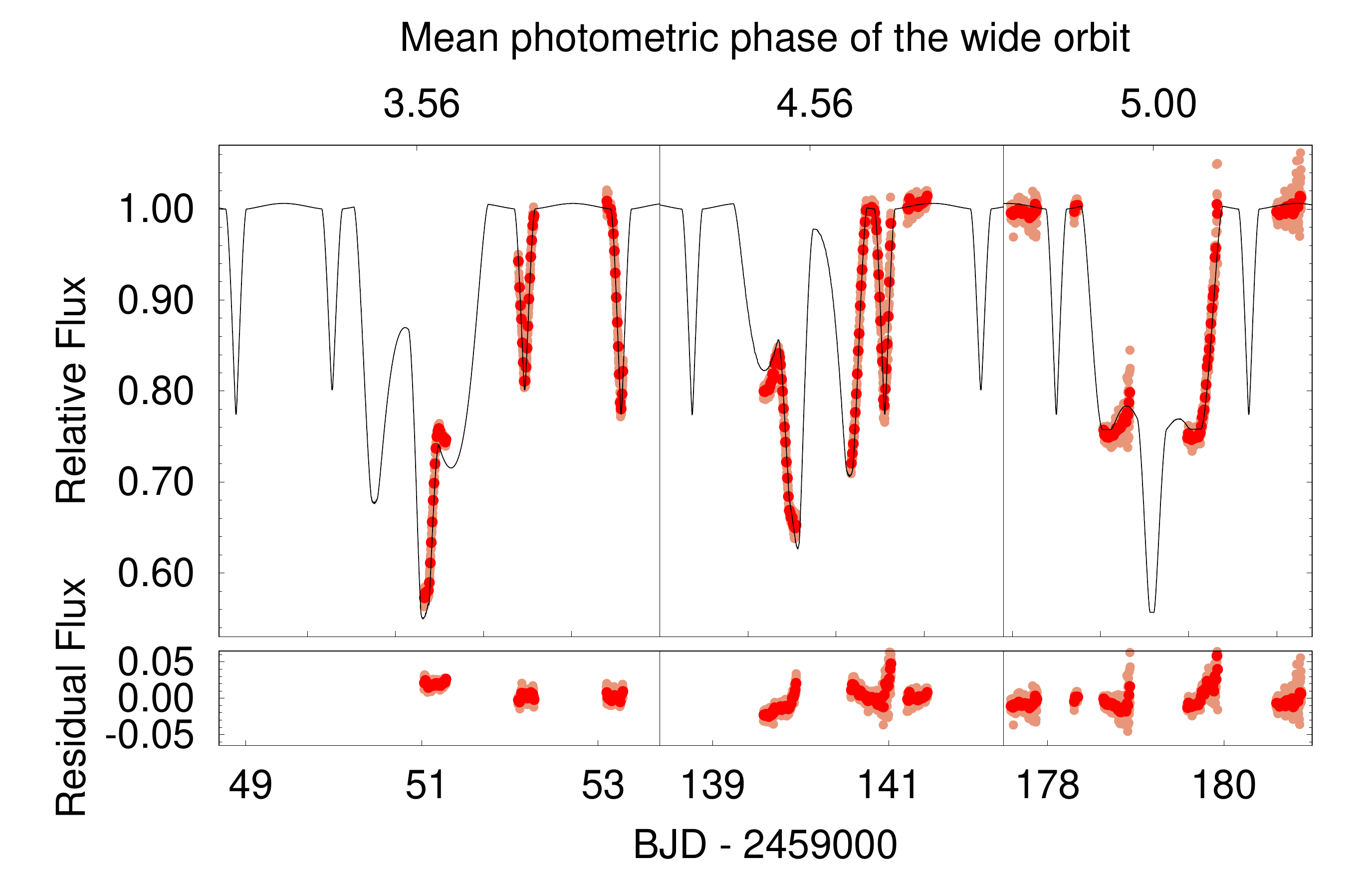}
\caption{Unfiltered ground-based follow up photometric observations of three third-body (i.e., outer) eclipses of TIC~388459317. Note, during the photodynamical analysis the dataset was modelled in the $R_C$-band. The lighter dots stand for the original observations, while the darker ones show their 900-sec averages which have been used for the photodynamical fits. The black line represents the best-fitting spectro-photodynamical model solution. Residuals against this photodynamical model are also shown in the bottom panel.}
\label{fig:T388459317lcE3ground}
\end{figure*}

The photometric campaign observations of TIC 388459317 covered 16 nights between 9 July 2020 and 13 October 2021. Sections of three additional third-body eclipses (Fig.~\ref{fig:T388459317lcE3ground}) were obtained. Additionally, we observed 11 regular, inner binary eclipses, of which the mid-eclipse times can be found in the last rows of Table~\ref{tab:T388459317ToM}.

\begin{figure*}
\includegraphics[width=0.8\textwidth]{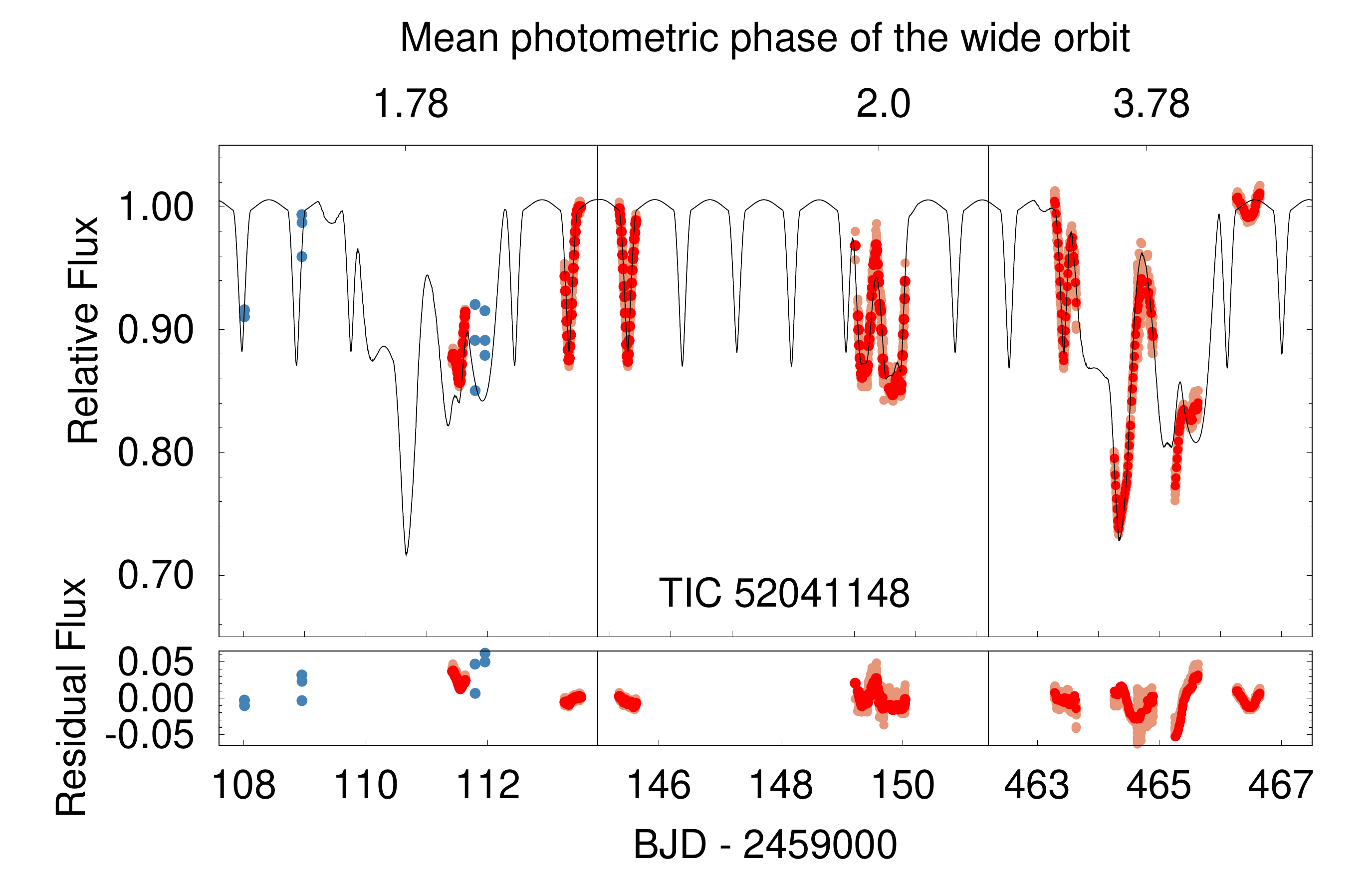}
\caption{Three third-body eclipses of TIC 52041148 together with the ground-based follow up photometric observations that partially cover the events. Note, during the photodynamical analysis the dataset was modelled in $R_C$-band, though different sections of the observations were carried out with different filters, or even without any filter. The lighter red dots stand for the original observations, while the darker ones show their 900-sec averages which have been used for the photodynamical fits. Moreover, in the case of the first event, we plot also the ASAS-SN points (light blue dots), though these latter data were not used for our analysis.  The black line represents the best-fitting spectro-photodynamical model solution.  The substantial discrepancies between the observations and the model during the last two nights of the third event (right panel) are discussed in the text. Residuals against this photodynamical model are also shown in the bottom panel.}
\label{fig:T052041148lcE3ground}
\end{figure*}

Finally, TIC 52041148 was observed on 28 nights during the interval between 11 June 2020 and 24 October 2021. In addition to the sections of three third-body eclipses (Fig.~\ref{fig:T052041148lcE3ground}), we also obtained lightcurves of 16 regular eclipses of the inner binary (see Table~\ref{tab:T052041148ToM} for the mid-eclipse times determined from these observations).

\subsection{Radial-velocity data}
\label{sec:RVs}
Eighteen high-resolution spectra of TIC~193993801 were obtained from March 26 to June 26, 2021. The observations were obtained with a 1.3 m, f/8.36 Nasmyth-Cassegrain telescope equipped with a fibre-fed \'Echelle spectrograph at the Skalnat\'{e} Pleso (SP) Observatory at an altitude of 1783\,m above sea level. Its layout follows the MUSICOS design \citep{MUSICOS}. The spectra were recorded by an Andor iKon-L DZ936N-BV  CCD  camera with a $2048\times2048$ array, 13.5 $\mu$m square pixels, 2.9 e$^-$ read-out noise, and gain close to unity. The spectral range of the instrument is 4250--7375 \AA~ (56 \'echelle orders) with the maximum resolution of $R$ = 38\,000. The wavelength stability is 100-200 m/s. To increase the SNR three consecutive 900-sec exposures were combined. This resulted in the SNR ranging from 10 to 18 in the yellow part of the spectrum. 

The raw spectroscopic data were reduced using {\sc IRAF} package tasks, LINUX shell scripts, and FORTRAN programs as described in \citet{eShel}. In the first step, master dark  frames were produced. In the second step, the photometric calibration of the frames was done using dark and flat-field frames. Bad pixels were cleaned using a bad-pixel mask, and cosmic hits were removed using the program of \citet{pych04}. Order positions were defined by fitting sixth order Chebyshev polynomials to tungsten-lamp and blue LED spectra. In the following step, scattered light was modeled and subtracted. Aperture spectra were then extracted for both object and ThAr lamp and then the resulting 2D spectra were dispersion solved. Finally, 2D spectra were combined to 1D spectra.

One dimensional cross-correlation functions (CCFs) were calculated for each observed spectrum with the iSpec software \citep{blancocuaresmaetal14,blancocuaresma19} using the built-in NARVAL Sun linelist as a template. As was expected from the preliminary model, the system was found to be a triple-lined (SB3) spectroscopic triple star based on the CCFs, with two rotationally broadened components (the inner pair), and a sharp-lined tertiary.  The CCFs were then shifted into the barycentric reference frame and fitted with the sum of three Gaussians in order to derive the RVs of the constituent stars. The RV data identified as the centers of the fitted Gaussians are tabulated in Table~\ref{tab:RVdata}, while a section of them is plotted in Fig.~\ref{fig:T193993801RVwithfit} together with the best-fitting spectro-photodynamical model (see below, in Sect.~\ref{sec:dyn_mod}).

\begin{figure}
\center
\includegraphics[width=0.5\textwidth]{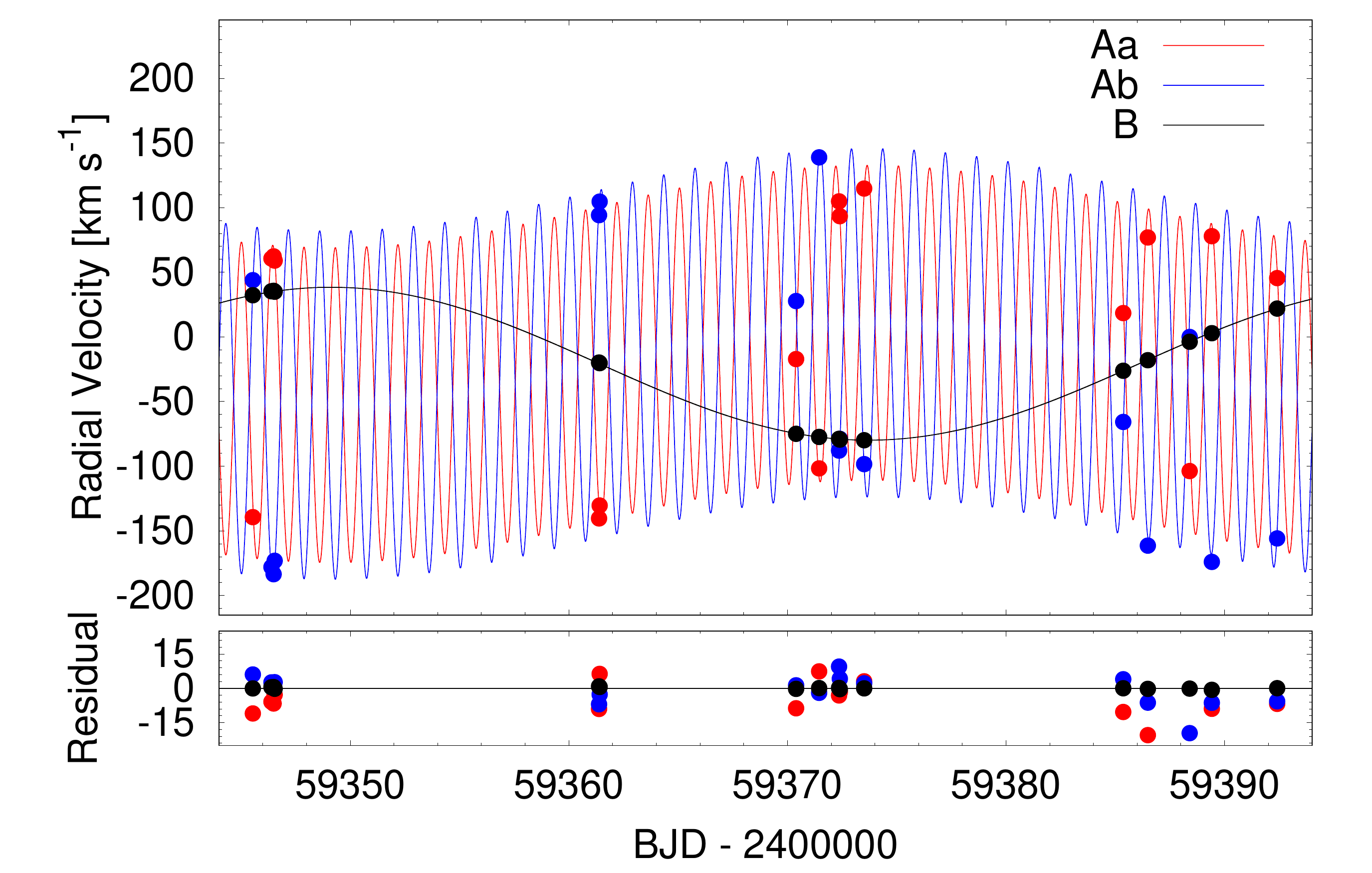}  
\caption{Radial velocity measurements of TIC 193993801 Aa, Ab and B (red, blue and black dots, respectively), together with the best-fit photodynamical solutions (the corresponding smooth curves) between BJDs 2\,459\,344 and 2\,459\,394 (i.e. one complete orbital cycle of the outer orbit). The lower panel shows the residuals.}
\label{fig:T193993801RVwithfit}
\end{figure}  

\begin{table*}
\centering
\caption{Measured radial velocities of the three stellar components of TIC~193993801. The date is given as BJD -- 2\,400\,000, while the RVs and their uncertainties are in km\,s$^{-1}$.}
\label{tab:RVdata}
\begin{tabular}{lrrrrrr}
\hline
\hline
Date & RV$_\mathrm{Aa}$ & $\sigma_\mathrm{Aa}$ & RV$_\mathrm{Ab}$ & $\sigma_\mathrm{Ab}$ & RV$_\mathrm{B}$ & $\sigma_\mathrm{B}$ \\  
\hline
$59\,300.59944$ & $  54.28$ & $ 2.40$ & $-180.11$ & $ 2.21$ & $ 38.02$ & $ 0.17$ \\
$59\,304.53943$ & $ -91.97$ & $16.24$ & $ -25.30$ & $24.12$ & $ 28.29$ & $ 0.17$ \\
$59\,345.54053$ & $-139.38$ & $ 2.32$ & $  43.84$ & $ 8.60$ & $ 32.22$ & $ 0.28$ \\
$59\,346.39172$ & $  60.58$ & $ 3.91$ & $-178.00$ & $ 2.45$ & $ 35.28$ & $ 0.25$ \\
$59\,346.49289$ & $  62.28$ & $ 2.38$ & $-183.50$ & $ 1.83$ & $ 35.62$ & $ 0.18$ \\
$59\,346.54078$ & $  59.01$ & $ 2.60$ & $-173.17$ & $ 1.51$ & $ 34.98$ & $ 0.18$ \\
$59\,361.38238$ & $-140.40$ & $ 2.79$ & $  94.20$ & $ 3.64$ & $-19.85$ & $ 0.26$ \\
$59\,361.41056$ & $-130.39$ & $ 2.85$ & $ 104.60$ & $ 4.79$ & $-20.31$ & $ 0.24$ \\
$59\,370.39586$ & $ -17.16$ & $ 1.42$ & $  27.72$ & $ 4.60$ & $-74.95$ & $ 0.21$ \\
$59\,371.44955$ & $-101.67$ & $ 2.49$ & $ 138.86$ & $ 2.01$ & $-77.35$ & $ 0.18$ \\
$59\,372.36236$ & $ 104.81$ & $ 2.89$ & $ -87.96$ & $ 2.24$ & $-78.82$ & $ 0.22$ \\
$59\,372.39921$ & $  93.37$ & $ 2.21$ & $ -78.81$ & $ 1.93$ & $-79.42$ & $ 0.17$ \\
$59\,373.51221$ & $ 114.74$ & $ 1.46$ & $ -98.41$ & $ 2.65$ & $-79.89$ & $ 0.17$ \\
$59\,385.36677$ & $  18.44$ & $ 5.81$ & $ -65.80$ & $ 1.72$ & $-26.17$ & $ 0.22$ \\
$59\,386.49217$ & $  76.92$ & $ 6.48$ & $-161.45$ & $ 5.47$ & $-17.99$ & $ 0.25$ \\
$59\,388.40531$ & $-103.72$ & $ 6.39$ & $  -0.06$ & $ 8.25$ & $ -3.75$ & $ 0.12$ \\
$59\,389.41875$ & $  77.90$ & $ 4.30$ & $-174.00$ & $ 4.15$ & $  2.86$ & $ 0.22$ \\
$59\,392.40528$ & $  45.50$ & $ 6.09$ & $-155.89$ & $ 6.09$ & $ 21.90$ & $ 0.53$ \\
\hline
\end{tabular}
\end{table*}

High dispersion spectra from the Skalnat\'e Pleso observatory were independently analyzed by the broadening-function (BF) technique \citep{rucinski92}. According to the $J-K = 0.274(28)$ color index, the system seems to be of the F5V spectral type. Several slowly rotating templates of the F spectral type were used in an attempt to deconvolve the spectra. The best match of the object and template spectrum was found for HD~102870 (F8V). For this star the BF integral was close to unity hence the strength of the metallic lines of the object and template is also similar \citep[see][]{rucinski13}. The BFs were extracted from the 4950 \AA \,to 6046 \AA \,spectral region devoid of strong hydrogen Balmer lines. In addition to the wide profiles of the close-binary components, all BFs clearly show a third slowly-rotating component which indicates a variable radial velocity (RV). While the third RV component was very well defined, the RV components of the eclipsing pair were rather noisy.

The BFs were modeled by a multi-component Gaussian fit. In the next step, the Gaussian fit to the third component was subtracted from the BFs. The relative light contribution of the third component, estimated from the multi-profile fit, is about $L_3/(L_1+L_2) = 0.599\pm0.034$. In the following step, the resulting cleaned BFs were modeled by two rotational 
profiles which are more appropriate for stars where the rotation determines the line profiles. This resulted in RVs and rotational velocities of the components, $v \sin i$. While the third component rotates slowly and its $v \sin i < 8\,\mathrm{kms}^{-1}$ (limited by the spectral resolution), the primary and secondary components rotate at $v \sin i = 59.5\pm1.6\,\mathrm{kms}^{-1}$ and $v \sin i = 47.4\pm2.6\,\mathrm{kms}^{-1}$, respectively.  Assuming that the stellar equators are aligned with the orbital planes, and taking the stellar radii (Table~\ref{tab:syntheticfit_TIC193993801}) from the spectro-photodynamical solution discussed below in Sect.~\ref{sec:dyn_mod}, one can calculate the spin periods of the three stars to be $P_\mathrm{rot,Aa}=1\fd44\pm0\fd05$, $P_\mathrm{rot,Ab}=1\fd40\pm0\fd10$, $P_\mathrm{rot,B}>8\fd5$, respectively. These findings indicate that the rotation of the inner pair is synchronized to the orbital period ($P_\mathrm{in}=1\fd43$). (Since the RVs determined via the BF fitting method agree with the CCF results (Table~\ref{tab:RVdata}) to well within the tabulated uncertainties for each RV point, we do not list them separately.)

\section{Spectro-photodynamical modeling}
\label{sec:dyn_mod}

We carried out combined (spectro-)photodynamical analyses with the software package {\sc Lightcurvefactory} \citep[see, e.g.][and further references therein]{borkovitsetal19a,borkovitsetal20a}. This code, amongst other features, contains (i) a built-in numerical integrator to calculate the three-body perturbed coordinates and velocities of the three bodies; (ii) multi-band light curve, ETV and RV curve emulators, and (iii) for the inverse problem, an MCMC-based parameter search routine, which employs an implementation of the generic Metropolis-Hastings algorithm \citep[see e. g.][]{ford05}. The usage of this software package and the philosophy and consecutive steps of the whole process were previously explained in detail for a variety of systems \citep{borkovitsetal18,borkovitsetal19a,borkovitsetal19b,borkovitsetal20a,borkovitsetal20b,borkovitsetal21,mitnyanetal20}.  These included tight and less tight triple systems (either exhibiting extra eclipses or not)  as well as quadruple systems (having hierarchies of either the 2+2 or 2+1+1 types) and, therefore, here we discuss only the points specific to the current studies.

For all three of the triple systems considered in this work, we made two different kinds of analyses. The first of these we refer to as the `astrophysical model-independent solution'. Here we fit simultaneously the (multiple band) lightcurves, the ETV curves deduced from them, and, in the case of TIC 193993801, the RV data as well. This means that we use almost exclusively geometrical and dynamical information coded into the observational data without any a priori astrophysical model assumption.\footnote{We say `almost' because, even in this case, we use some weak astrophysical model dependence through the setting and interpolation of limb darkening and gravitational darkening coefficients. These parameters, however, have only minor influences on the results.} As a consequence, accurate results from this type of analysis (generally, precisions of $1-2\%$ or better in the fundamental stellar parameters, e.g., in masses and radii) can be used as strong constraints for stellar evolutionary models \citep[see, e.g][and further references therein]{torresetal2010,southworth20,maxtedetal20}.

For such an accurate analysis of an EB, however, one needs RV data for both components as these provide the information about the masses of the components. In the absence of this knowledge, one cannot convert the dimensionless quantities -- such as the fractional radii of the stars, even if precisely determined from the lightcurve solution -- into absolute, physical ones. Moreover, a further fundamental question in such an accurate analysis is the precision of the effective temperatures of the binary members.  Recently \citet{milleretal20} have developed a new method to determine the temperatures of eclipsing binary stars with a precision better than $1\%$, via a careful analysis of the system's net spectral energy distribution (SED) while taking into account the components' flux ratio, as `quasi observables', that can be measured from high quality multi-band occultation lightcurves of some totally eclipsing binaries. Unfortunately, however, due to the significant light contributions of the third stellar components in our systems, as well as the lack of flat-bottomed binary eclipses and the strongly varying profiles of the third-body eclipses, this latter rigorous method for precise SED analysis cannot be applied directly to our triples.  Moreover, amongst these three triply eclipsing triple systems RV data are available only for TIC 193993801. Thus, in the cases of the current triples, and especially for TIC 388459317 and 52041148, we require some other means to deduce these important parameters and, therefore, be able to characterize the system's true physical nature and evolutionary status.

In the case of a dynamically interacting, not too distant third stellar component, however, its gravitational perturbations to the Keplerian motion of the inner EB carries information about the inner and outer mass ratios of the system.  And, in special cases, even the masses of the individual stellar components \citep[see e.~g.][]{borkovitsetal15} can be deduced in this way. Unfortunately, however, in most cases the necessary precision cannot be reached via perturbations induced by the third star, and RV measurements are required.  Therefore, we follow a different strategy. We introduce into the analysis some a priori knowledge about stellar astrophysics and evolution with the use of \texttt{PARSEC} isochrones and evolutionary tracks \citep{PARSEC}. We use tabulated three dimensional grids of \texttt{PARSEC} isochrones that contain stellar temperatures, radii, surface gravities, luminosities, and magnitudes in different passbands of several photometric systems for [age, metallicity, initial stellar mass] triplets.  Then, allowing these latter three parameters to be freely adjustable variables, the stellar temperature, radius, and actual passband magnitude are calculated through trilinear interpolations from the grid points and these values are used to generate synthetic lightcurves, and SED that can be compared to their observational counterparts.  This process is described in detail in \citet{borkovitsetal20a}.

Regarding the technical details of the two different kind of analyses, in the case of the stellar evolution model independent runs, the freely adjusted (i.e., trial) parameters were as follows:
\begin{itemize}
\item[(i)] Five plus 0--4 lightcurve related parameters: the temperature ratios of $T_\mathrm{Ab}/T_\mathrm{Aa}$ and $T_\mathrm{B}/T_{Aa}$; the durations of the primary eclipses of the inner pair $(\Delta t_\mathrm{pri})$; the ratios of the radii of the inner stars $(R_\mathrm{Ab}/R_\mathrm{Aa})$; the fractional radius of the third component ($r_\mathrm{B}=R_\mathrm{B}/a_\mathrm{out}$) and, the contaminated extra light in different passbands ($\ell_4$). Here we tabulate also the effective temperature of the primary of the inner binary ($T_\mathrm{Aa}$) which was also adjusted in the MCMC process with a uniform prior. The initial value of this parameter in the initial runs was set according to the catalog values (given in Table~\ref{tbl:mags}), but later, after getting the results of the preliminary model-dependent runs, it was set according to these results.
\item[(ii)] Three of six orbital-element related parameters of the inner, and six parameters of the outer orbits, i.e. the components of the eccentricity vectors of the two orbits $(e\sin\omega)_\mathrm{in,out}$, $(e\cos\omega)_\mathrm{in,out}$, the inclinations relative to the plane of the sky ($i_\mathrm{in,out}$), and moreover, three other parameters for the outer orbit, including the period ($P_\mathrm{out}$), time of the first (inferior or superior) conjunction of the third component observed in the \textit{TESS} data ($\mathcal{T}_\mathrm{out}^\mathrm{inf,sup})$ and finally, the longitude of the node relative to the inner binary's node ($\Omega_\mathrm{out}$) .
\item[(iii)] Three mass-related parameters: the mass of the primary of the inner binary ($m_\mathrm{Aa}$), and the mass ratios of the two (inner and outer) binaries ($q_\mathrm{in,out}$). 
\end{itemize} 

In the case of the model dependent runs, the lightcurve-related parameters (i above), apart from the extra light ($\ell_4$), were no longer adjusted as all three radii and temperatures are now calculated from the \texttt{PARSEC} tables. On the other hand, the following new adjustable parameters are introduced: the metallicity of the system ([$M/H$]), the (logarithmic) age of the three stars ($\log\tau$), the interstellar extinction $E(B-V)$ toward the given triple, and its distance.  Here additional notes about the `age' and the 'distance' are in order. First, regarding the 'age' parameter, our former experience has shown that it is better to allow the ages of the three stars to be adjusted separately instead of demanding strictly equal ages. This issue was partly discussed in \citet{rowdenetal20} and \citet{borkovitsetal21}, and we will return to this question in the discussion below. Turning now to the distance of the system, one can argue that the accurate trigonometric distances calculated from the Gaia parallax measurements \citep{bailer-jonesetal21} should be used as Gaussian priors to penalize the model solutions. However, unfortunately, neither DR2 nor the recently released EDR3 Gaia parallaxes have been corrected for the binary or multiple natures of the targets and, therefore the published parallaxes and determined distances are not necessarily accurate or, even reliable for our systems. For example, in the case of TIC~193993801 the EDR3 and DR2 parallaxes differ by about 5 of their mutual $\sigma$'s. Therefore, we decided not to utilize the Gaia distances. Instead, we constrained the distance by minimizing the $\chi^2_\mathrm{SED}$ value a posteriori, at the end of each trial step.

Some other parameters were also constrained instead of being adjusted or fixed during our analyses. Thus, the orbital period of the inner binary ($P_\mathrm{in}$) and its orbital phase (through the time of an arbitrary primary eclipse or, more strictly, the time of the inferior conjunction of the secondary star -- $\mathcal{T}^\mathrm{inf}_\mathrm{in}$) were constrained internally through the ETV curves in both kinds of analyses. Moreover, in the case of TIC~193993801 the systemic radial velocity ($\gamma$) was also constrained internally via minimization of the $\chi^2_\mathrm{RV}$ contribution a posteriori in each trial step.

Regarding the atmospheric parameters of the stars under analysis, we handled them in a similar manner as in our previous works. Therefore, we used the logarithmic limb-darkening law \citep{klinglesmithsobieski70} for which the passband-dependent linear and non-linear coefficients were interpolated in each trial step with the use of the tables from the original version of the {\tt Phoebe} software \citep{Phoebe}. We set the gravity darkening coefficients for all late type stars to $\beta=0.32$ in accordance with the classic model of \citet{lucy67} valid for convective stars and keep them as constant. In the case of TIC~388459317, however, the analysis of the net SED has revealed that the system consists of hotter stars, having radiative envelopes. Therefore, we set $\beta=1.0$ for the radiative components of TIC~388459317.  This parameter, however, has only minor significance, as the stars under the present investigations are close to spheroids.

While preparing the observational data for analysis, in order to save computational time, we dropped out the out-of-eclipse sections of the 30-min cadence \textit{TESS} lightcurves, retaining only the $\pm0\fp15$ phase-domain regions around the binary eclipses themselves.  However, during sections of the data containing the third-body (i.e., `outer') eclipses, we kept the data for an entire binary period both before and after the first and last contacts of the given third-body eclipse. 

Regarding the data obtained during the follow-up campaigns, these data mostly cover times around predicted third-body eclipses and several regular eclipses as well. These data are quite inhomogeneous. They are, however, extremely important for constraining the dynamical properties of the systems very precisely. Independent of the original exposure times, we binned these data into 900 sec and normalized the fluxes from one data set to another. For the two fainter targets, most of the measurements were carried out without a passband filter. For the analysis, however, for practical reasons, we consider them as Cousins-R ($R_C$) data. For TIC~193993801, the bulk of the observations were carried out with two similar 80-cm telescope with $r'$ and $z'$ filters. These data were analysed accordingly. The remaining, inhomogeneous data set were also considered to be taken in $R_C$. We emphasize again, however, that these data have only a minor influence on the stellar properties, e.g., the flux (or temperature) ratios, and constrains almost exclusively the dynamical parameters through the shapes and timings of the third-body eclipses. 

In addition to the photometric lightcurves, ETV data (Tables~\ref{tab:T193993801ToM} -- \ref{tab:T052041148ToM}), and RV data in the case of TIC~193993801 (Table~\ref{tab:RVdata}), were also simultaneously included into both kinds of analyses. Moreover, in the case of the second type of analysis, using SED fitting and \texttt{PARSEC} tables, the observed passband magnitudes (Table~\ref{tbl:mags}) were also fitted against the code-generated synthetic SEDs. Note, however, during the SED fitting, in order to avoid the over-dominance of the very precise Gaia magnitudes the uncertainties for each passband magnitude were not allowed to be less than 0.03~mag.

\begin{figure}
\begin{center}
\includegraphics[width=0.50 \textwidth]{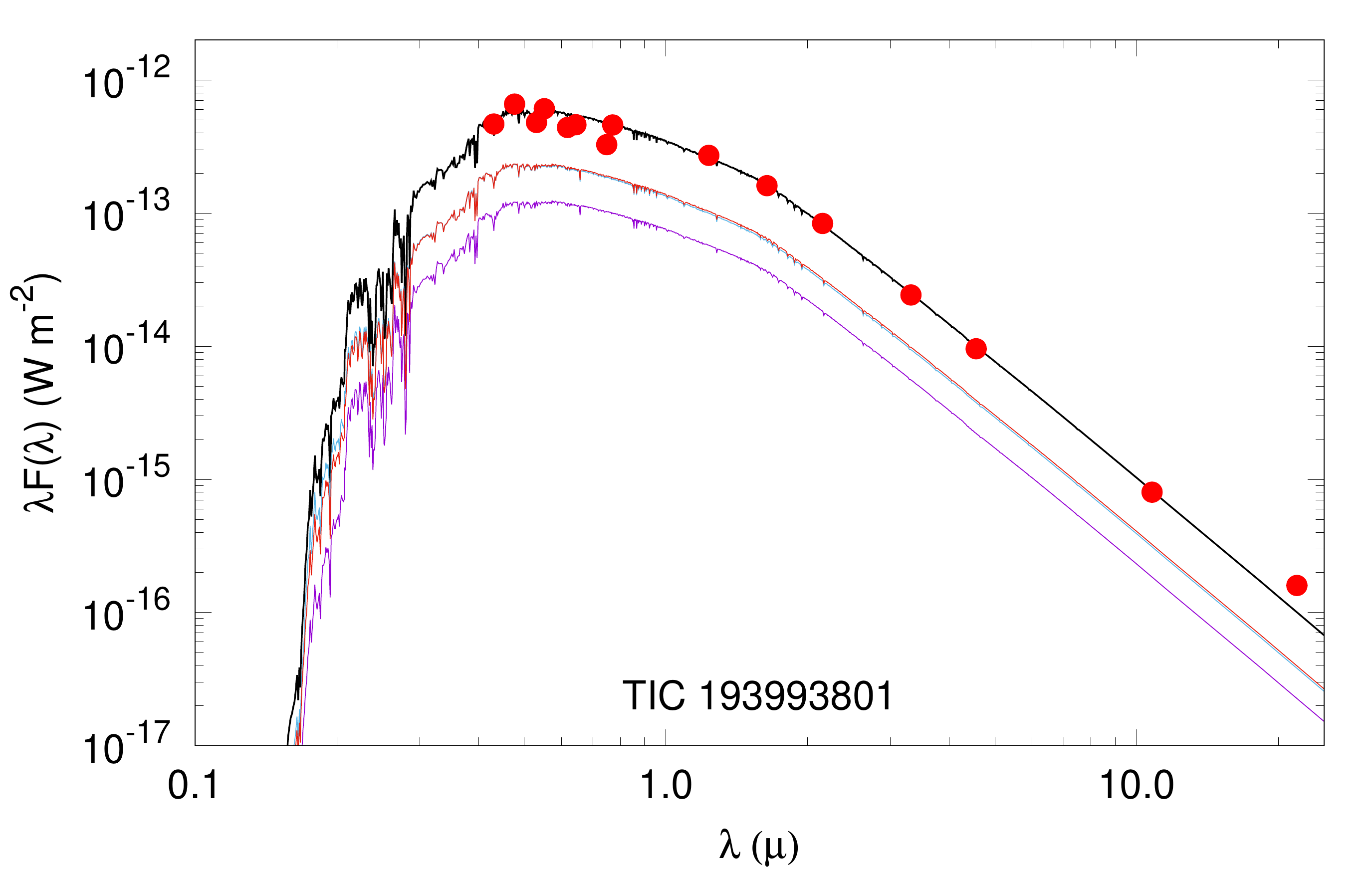}
\includegraphics[width=0.50 \textwidth]{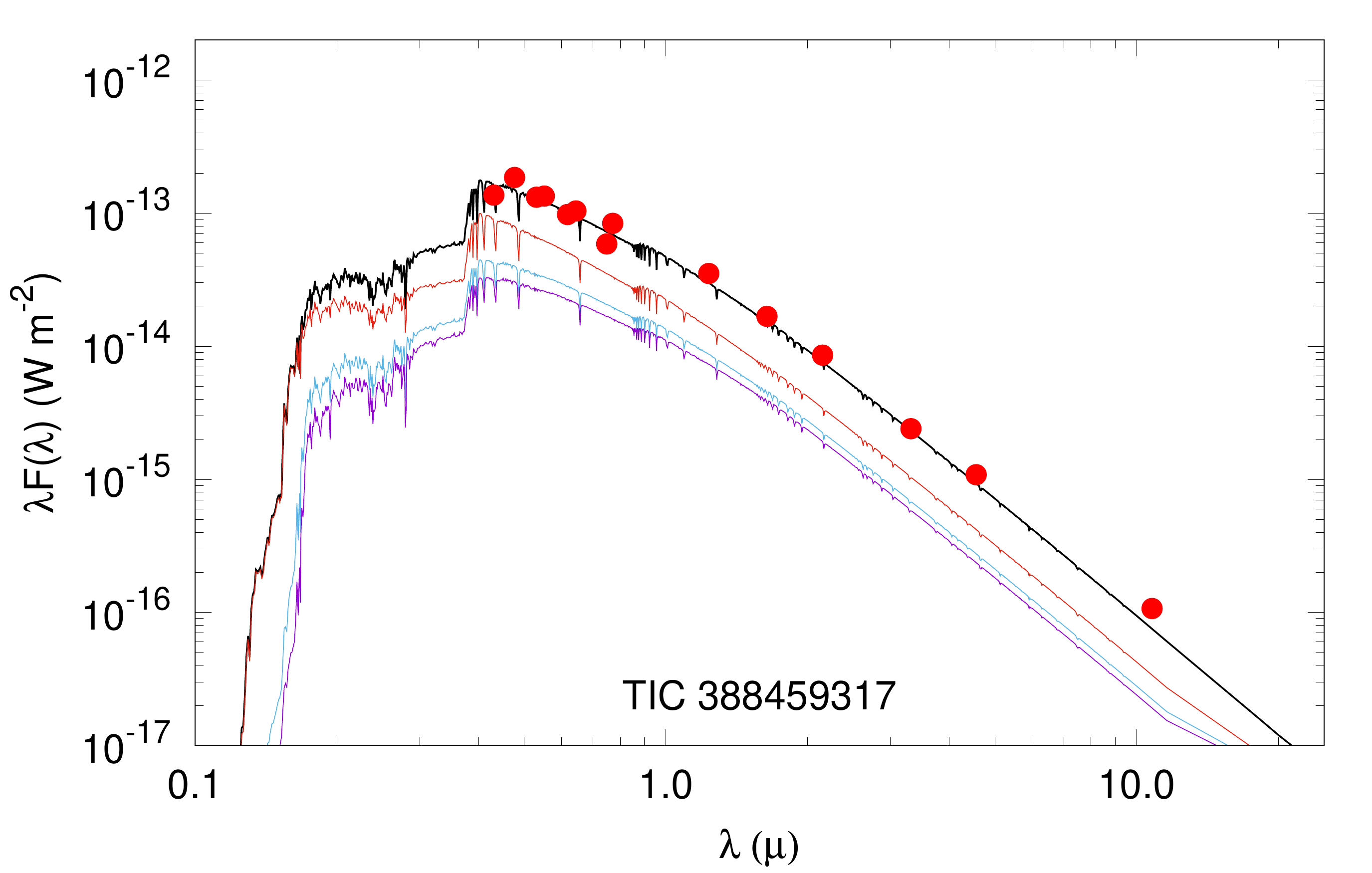}
\includegraphics[width=0.50 \textwidth]{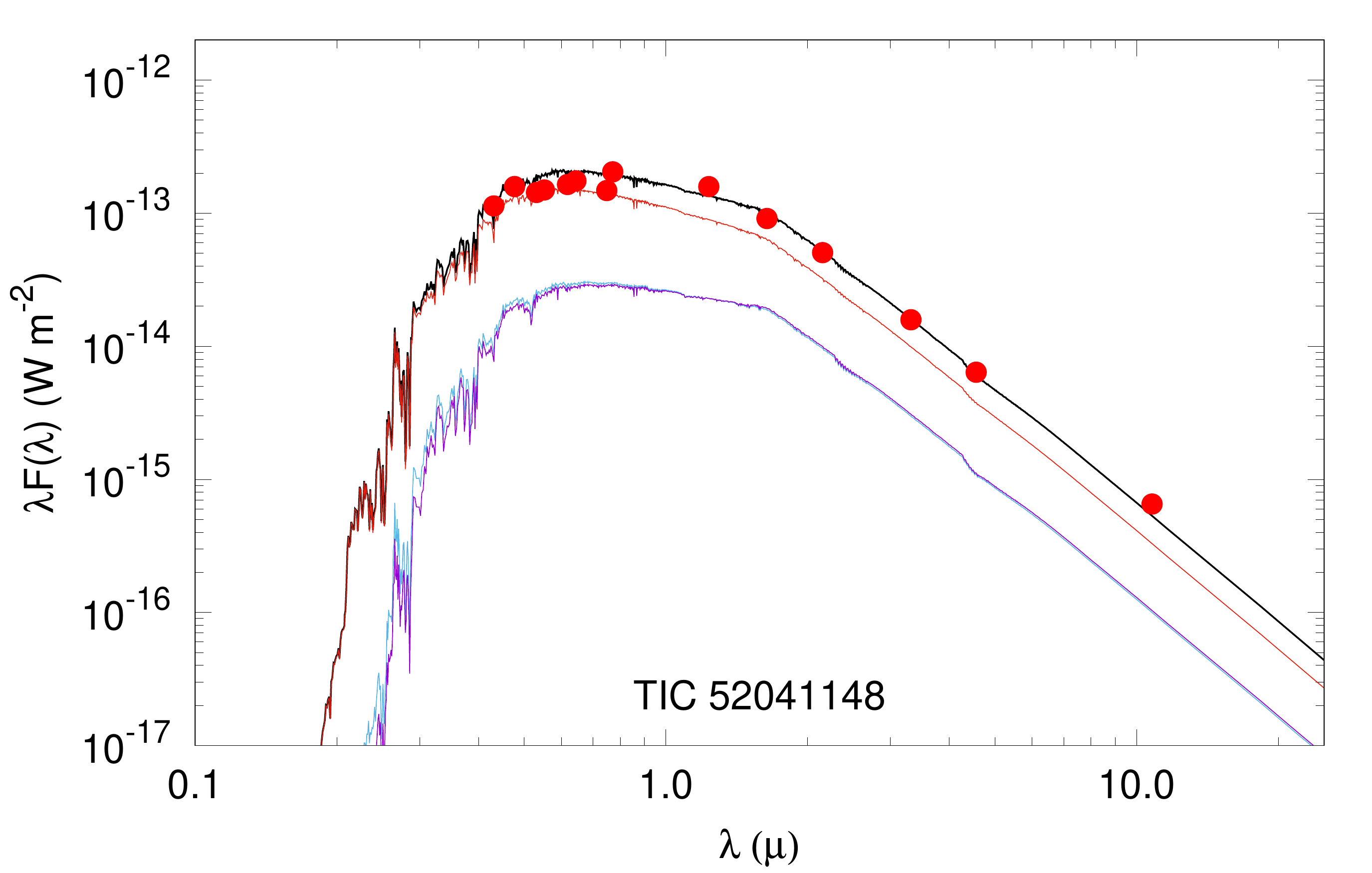}
\caption{The summed SED of the stars of TICs 193993801 (upper panel), 388459317 (middle panel) and 52041148 (lower panel) in the flux domain. The dereddened observed magnitudes are converted into the flux domain (red filled circles), and overplotted with the quasi-continuous summed SEDs for the multiple star systems (thick black line). These SEDs are computed from the \citet{castellikurucz04} ATLAS9 stellar atmospheres models (\url{http://wwwuser.oats.inaf.it/castelli/grids/gridp00k2odfnew/fp00k2tab.html}). The separate SEDs of the component stars are also shown with thin green, black and purple lines, respectively. } 
\label{fig:sedfit} 
\end{center}
\end{figure}  



\begin{table*}
 \centering
\caption{Orbital and astrophysical parameters of TIC\,193993801 derived from the joint photodynamical lightcurve, RV and ETV solution with and without the involvement of the stellar energy distribution and \texttt{PARSEC} isochrone fitting (MIR and MDR models, respectively). Note, the instantaneous, osculating orbital elements (including the mutual inclination of the orbital planes -- $i_\mathrm{m}$) are given for epoch $t_0=2\,458\,738.5$ (BJD).}
 \label{tab:syntheticfit_TIC193993801}
\begin{tabular}{@{}lllllll}
\hline
 & \multicolumn{3}{c}{without SED+\texttt{PARSEC} (MIR)} & \multicolumn{3}{c}{with SED+\texttt{PARSEC} (MDR)} \\ 
\hline
\multicolumn{7}{c}{orbital elements} \\
\hline
   & \multicolumn{6}{c}{subsystem}  \\
   & \multicolumn{2}{c}{Aa--Ab} & A--B & \multicolumn{2}{c}{Aa--Ab} & A--B \\
  $P$ [days] & \multicolumn{2}{c}{$1.430848_{-0.000014}^{+0.000014}$} & $49.4020_{-0.0010}^{+0.0009}$ & \multicolumn{2}{c}{$1.430841_{-0.000014}^{+0.000014}$} & $49.4021_{-0.0008}^{+0.0008}$ \\
  $a$ [R$_\odot$] & \multicolumn{2}{c}{$7.250_{-0.026}^{+0.028}$} & $88.76_{-0.48}^{+0.51}$ & \multicolumn{2}{c}{$7.244_{-0.023}^{+0.022}$} & $88.56_{-0.37}^{+0.39}$ \\
  $e$ & \multicolumn{2}{c}{$0.00081_{-0.00008}^{+0.00008}$} & $0.0025_{-0.0009}^{+0.0020}$ &  \multicolumn{2}{c}{$0.00081_{-0.00008}^{+0.00007}$} & $0.0029_{-0.0013}^{+0.0024}$ \\
  $\omega$ [deg]& \multicolumn{2}{c}{$147_{-13}^{+12}$} & $65_{-24}^{+12}$ & \multicolumn{2}{c}{$149_{-11}^{+12}$} & $67_{-23}^{+12}$ \\ 
  $i$ [deg] & \multicolumn{2}{c}{$88.60_{-0.64}^{+0.46}$} & $88.996_{-0.043}^{+0.047}$ & \multicolumn{2}{c}{$88.93_{-0.47}^{+0.36}$} & $88.982_{-0.032}^{+0.042}$ \\
  $\mathcal{T}_0^\mathrm{inf}$ [BJD - 2400000]& \multicolumn{2}{c}{$58\,739.93959_{-0.00005}^{+0.00006}$} & $58\,745.4792_{-0.0037}^{+0.0039}$ & \multicolumn{2}{c}{$58\,739.93958_{-0.00005}^{+0.00005}$} & $58\,745.4768_{-0.0037}^{+0.0037}$ \\
  $\Omega$ [deg] & \multicolumn{2}{c}{$0.0$} & $0.18_{-0.34}^{+0.33}$ & \multicolumn{2}{c}{$0.0$} & $-0.03_{-0.34}^{+0.32}$ \\
  $i_\mathrm{m}$ [deg] & \multicolumn{3}{c}{$0.47_{-0.26}^{+0.55}$}& \multicolumn{3}{c}{$0.42_{-0.21}^{+0.30}$} \\
  \hline
  mass ratio $[q=m_\mathrm{sec}/m_\mathrm{pri}]$ & \multicolumn{2}{c}{$0.910_{-0.010}^{+0.011}$} & $0.539_{-0.008}^{+0.010}$ & \multicolumn{2}{c}{$0.866_{-0.005}^{+0.006}$} & $0.533_{-0.006}^{+0.006}$ \\
  $\gamma$ [km\,s$^{-1}$] & \multicolumn{3}{c}{$-20.85_{-0.04}^{+0.04}$} & \multicolumn{3}{c}{$-20.83_{-0.04}^{+0.05}$} \\
  \hline  
\multicolumn{7}{c}{stellar parameters} \\
\hline
   & Aa & Ab &  B & Aa & Ab &  B \\
  \hline
 \multicolumn{7}{c}{Relative quantities} \\
  \hline
 fractional radius [$R/a$] & $0.2334_{-0.0015}^{+0.0015}$ & $0.1799_{-0.0025}^{+0.0020}$ & $0.0187_{-0.0007}^{+0.0006}$ & $0.2318_{-0.0014}^{+0.0014}$ & $0.1811_{-0.0026}^{+0.0023}$ & $0.0190_{-0.0005}^{+0.0005}$ \\
 temperature relative to $(T_\mathrm{eff})_\mathrm{Aa}$ & $1$ & $0.9637_{-0.0025}^{+0.0026}$ & $0.9881_{-0.0103}^{+0.0099}$ & $1$ & $0.9671_{-0.0026}^{+0.0027}$ & $0.9957_{-0.0103}^{+0.0112}$ \\
 fractional flux [in $TESS$-band] & $0.3959_{-0.0085}^{+0.0100}$ & $0.2092_{-0.0021}^{+0.0021}$ & $0.3724_{-0.0211}^{+0.0173}$ & $0.3854_{-0.0075}^{+0.0086}$ & $0.2117_{-0.0019}^{+0.0020}$ & $0.3865_{-0.0154}^{+0.0124}$ \\
 fractional flux [in $R_C$-band] & $0.3985_{-0.0175}^{+0.0169}$ & $0.2080_{-0.0062}^{+0.0059}$ & $0.3741_{-0.0201}^{+0.0147}$ & $0.3811_{-0.0115}^{+0.0111}$ & $0.2068_{-0.0044}^{+0.0041}$ & $0.3826_{-0.0147}^{+0.0119}$ \\
 fractional flux [in $r'$-band] & $0.4012_{-0.0111}^{+0.0125}$ & $0.2068_{-0.0036}^{+0.0039}$ & $0.3699_{-0.0208}^{+0.0164}$ & $0.3880_{-0.0086}^{+0.0095}$ & $0.2088_{-0.0033}^{+0.0032}$ & $0.3837_{-0.0162}^{+0.0127}$ \\
 fractional flux [in $z'$-band] & $0.3901_{-0.0102}^{+0.0122}$ & $0.2114_{-0.0038}^{+0.0032}$ & $0.3717_{-0.0222}^{+0.0167}$ & $0.3797_{-0.0092}^{+0.0095}$ & $0.2130_{-0.0036}^{+0.0035}$ & $0.3831_{-0.0147}^{+0.0115}$ \\
 \hline
 \multicolumn{7}{c}{Physical Quantities} \\
  \hline 
 $m$ [M$_\odot$] & $1.306_{-0.013}^{+0.016}$ & $1.189_{-0.017}^{+0.018}$ & $1.344_{-0.034}^{+0.039}$ & $1.334_{-0.014}^{+0.012}$ & $1.155_{-0.011}^{+0.012}$ & $1.327_{-0.026}^{+0.028}$ \\
 $R$ [R$_\odot$] & $1.692_{-0.013}^{+0.014}$ & $1.304_{-0.018}^{+0.015}$ & $1.663_{-0.061}^{+0.054}$ & $1.679_{-0.011}^{+0.011}$ & $1.312_{-0.021}^{+0.019}$ & $1.682_{-0.049}^{+0.043}$ \\
 $T_\mathrm{eff}$ [K]& $6317_{-31}^{+54}$ & $6092_{-37}^{+48}$ & $6247_{-71}^{+72}$ & $6410_{-138}^{+100}$ & $6199_{-134}^{+99}$ & $6378_{-108}^{+106}$ \\
 $L_\mathrm{bol}$ [L$_\odot$] & $4.108_{-0.114}^{+0.139}$ & $2.104_{-0.086}^{+0.093}$ & $3.789_{-0.305}^{+0.258}$ & $4.274_{-0.338}^{+0.265}$ & $2.269_{-0.146}^{+0.148}$ & $4.162_{-0.236}^{+0.334}$ \\
 $M_\mathrm{bol}$ & $3.21_{-0.04}^{+0.03}$ & $3.93_{-0.05}^{+0.05}$ & $3.29_{-0.07}^{+0.09}$ & $3.19_{-0.07}^{+0.09}$ & $3.88_{-0.07}^{+0.07}$ & $3.22_{-0.08}^{+0.06}$ \\
 $M_V           $ & $3.21_{-0.04}^{+0.03}$ & $3.97_{-0.05}^{+0.05}$ & $3.31_{-0.07}^{+0.10}$ & $3.18_{-0.06}^{+0.09}$ & $3.88_{-0.07}^{+0.08}$ & $3.21_{-0.08}^{+0.06}$ \\
 $\log g$ [dex]   & $4.098_{-0.006}^{+0.006}$ & $4.245_{-0.010}^{+0.012}$ & $4.126_{-0.027}^{+0.034}$ & $4.112_{-0.005}^{+0.005}$ & $4.263_{-0.010}^{+0.012}$ & $4.109_{-0.020}^{+0.023}$ \\
 \hline
$\log$(age) [dex] & \multicolumn{3}{c}{$-$} & $9.422_{-0.022}^{+0.020}$ & $9.571_{-0.053}^{+0.040}$ & $9.435_{-0.035}^{+0.029}$ \\
$[M/H]$ [dex]     & \multicolumn{3}{c}{$-$} & \multicolumn{3}{c}{$0.039_{-0.072}^{+0.131}$} \\
$E(B-V)$ [mag]    & \multicolumn{3}{c}{$-$} & \multicolumn{3}{c}{$0.030_{-0.024}^{+0.022}$} \\ 
extra light $\ell_4$ [in $TESS$-band] & \multicolumn{3}{c}{$0.022_{-0.014}^{+0.017}$} & \multicolumn{3}{c}{$0.015_{-0.011}^{+0.016}$} \\
extra light $\ell_4$ [in $R_C$-band] & \multicolumn{3}{c}{$0.014_{-0.010}^{+0.019}$} & \multicolumn{3}{c}{$0.030_{-0.019}^{+0.022}$} \\
extra light $\ell_4$ [in $r'$-band] & \multicolumn{3}{c}{$0.020_{-0.013}^{+0.021}$} & \multicolumn{3}{c}{$0.017_{-0.012}^{+0.020}$} \\
extra light $\ell_4$ [in $z'$-band] & \multicolumn{3}{c}{$0.028_{-0.020}^{+0.021}$} & \multicolumn{3}{c}{$0.024_{-0.016}^{+0.018}$} \\
$(M_V)_\mathrm{tot}$ & \multicolumn{3}{c}{$2.26_{-0.05}^{+0.05}$} & \multicolumn{3}{c}{$2.19_{-0.07}^{+0.07}$} \\
distance [pc]     &\multicolumn{3}{c}{$-$} & \multicolumn{3}{c}{$668.5_{-9.0}^{+8.9}$} \\
\hline
\end{tabular}

\end{table*}

\begin{table*}
 \centering
 \caption{The same results for TIC~193993801 as in Table~\ref{tab:syntheticfit_TIC193993801} but without the use of the RV data (MIN and MDN models, respectively).}
 \label{tab:syntheticfit_TIC193993801noRV}
\begin{tabular}{@{}lllllll}
\hline
 & \multicolumn{3}{c}{without SED+\texttt{PARSEC} (MIN)} & \multicolumn{3}{c}{with SED+\texttt{PARSEC} (MDN)} \\ 
\hline
\multicolumn{7}{c}{orbital elements} \\
\hline
   & \multicolumn{6}{c}{subsystem}  \\
   & \multicolumn{2}{c}{Aa--Ab} & A--B & \multicolumn{2}{c}{Aa--Ab} & A--B \\
  $P$ [days] & \multicolumn{2}{c}{$1.430813_{-0.000042}^{+0.000031}$} & $49.3999_{-0.0022}^{+0.0018}$ & \multicolumn{2}{c}{$1.430848_{-0.000014}^{+0.000015}$} & $49.4026_{-0.0007}^{+0.0006}$ \\
  $a$ [R$_\odot$] & \multicolumn{2}{c}{$7.422_{-0.288}^{+0.244}$} & $91.54_{-3.20}^{+2.60}$ & \multicolumn{2}{c}{$7.417_{-0.087}^{+0.042}$} & $90.48_{-1.02}^{+0.56}$ \\
  $e$ & \multicolumn{2}{c}{$0.00084_{-0.00008}^{+0.00008}$} & $0.0034_{-0.0017}^{+0.0029}$ &  \multicolumn{2}{c}{$0.00081_{-0.00007}^{+0.00008}$} & $0.0025_{-0.0010}^{+0.0019}$ \\
  $\omega$ [deg]& \multicolumn{2}{c}{$153_{-12}^{+12}$} & $73_{-24}^{+34}$ & \multicolumn{2}{c}{$148_{-13}^{+13}$} & $75_{-29}^{+43}$ \\ 
  $i$ [deg] & \multicolumn{2}{c}{$88.76_{-0.40}^{+0.32}$} & $88.987_{-0.040}^{+0.038}$ & \multicolumn{2}{c}{$88.50_{-0.37}^{+0.38}$} & $89.013_{-0.027}^{+0.029}$ \\
  $\mathcal{T}_0^\mathrm{inf}$ [BJD - 2400000]& \multicolumn{2}{c}{$58\,739.93948_{-0.00008}^{+0.00008}$} & $58\,745.4795_{-0.0038}^{+0.0039}$ & \multicolumn{2}{c}{$58\,739.93957_{-0.00006}^{+0.00005}$} & $58\,745.4791_{-0.0039}^{+0.0038}$ \\
  $\Omega$ [deg] & \multicolumn{2}{c}{$0.0$} & $0.11_{-0.30}^{+0.28}$ & \multicolumn{2}{c}{$0.0$} & $0.02_{-0.24}^{+0.28}$ \\
  $i_\mathrm{m}$ [deg] & \multicolumn{3}{c}{$0.42_{-0.20}^{+0.30}$}& \multicolumn{3}{c}{$0.58_{-0.27}^{+0.34}$} \\
  \hline
  mass ratio $[q=m_\mathrm{sec}/m_\mathrm{pri}]$ & \multicolumn{2}{c}{$0.909_{-0.019}^{+0.023}$} & $0.576_{-0.032}^{+0.041}$ & \multicolumn{2}{c}{$0.879_{-0.005}^{+0.005}$} & $0.524_{-0.004}^{+0.004}$ \\
  \hline  
\multicolumn{7}{c}{stellar parameters} \\
\hline
   & Aa & Ab &  B & Aa & Ab &  B \\
  \hline
 \multicolumn{7}{c}{Relative quantities} \\
  \hline
 fractional radius [$R/a$] & $0.2332_{-0.0013}^{+0.0013}$ & $0.1800_{-0.0023}^{+0.0021}$ & $0.0188_{-0.0005}^{+0.0006}$ & $0.2333_{-0.0014}^{+0.0015}$ & $0.1786_{-0.0024}^{+0.0025}$ & $0.0184_{-0.0004}^{+0.0004}$ \\
 temperature relative to $(T_\mathrm{eff})_\mathrm{Aa}$ & $1$ & $0.9642_{-0.0024}^{+0.0025}$ & $0.9858_{-0.0109}^{+0.0115}$ & $1$ & $0.9646_{-0.0026}^{+0.0030}$ & $0.9975_{-0.0188}^{+0.0138}$ \\
 fractional flux [in $TESS$-band] & $0.3939_{-0.0082}^{+0.0088}$ & $0.2088_{-0.0018}^{+0.0019}$ & $0.3796_{-0.013}^{+0.012}$ & $0.3994_{-0.0091}^{+0.0103}$ & $0.2096_{-0.0019}^{+0.0019}$ & $0.3715_{-0.0143}^{+0.0123}$ \\
 fractional flux [in $R_C$-band] & $0.3925_{-0.0123}^{+0.0102}$ & $0.2058_{-0.0049}^{+0.0036}$ & $0.3766_{-0.0115}^{+0.0130}$ & $0.4010_{-0.0091}^{+0.0110}$ & $0.2079_{-0.0037}^{+0.0038}$ & $0.3749_{-0.0122}^{+0.0105}$ \\
 fractional flux [in $r'$-band] & $0.3990_{-0.0108}^{+0.0093}$ & $0.2063_{-0.0035}^{+0.0034}$ & $0.3757_{-0.0127}^{+0.0125}$ & $0.4052_{-0.0104}^{+0.0117}$ & $0.2077_{-0.0031}^{+0.0030}$ & $0.3724_{-0.0142}^{+0.0105}$ \\
 fractional flux [in $z'$-band] & $0.3902_{-0.0088}^{+0.0095}$ & $0.2119_{-0.0033}^{+0.0034}$ & $0.3802_{-0.0108}^{+0.0112}$ & $0.3928_{-0.0116}^{+0.0128}$ & $0.2108_{-0.0035}^{+0.0037}$ & $0.3674_{-0.0115}^{+0.0100}$ \\
 \hline
 \multicolumn{7}{c}{Physical Quantities} \\
  \hline 
 $m$ [M$_\odot$] & $1.400_{-0.160}^{+0.143}$ & $1.274_{-0.138}^{+0.127}$ & $1.533_{-0.126}^{+0.134}$ & $1.423_{-0.051}^{+0.023}$ & $1.249_{-0.042}^{+0.022}$ & $1.396_{-0.044}^{+0.030}$ \\
 $R$ [R$_\odot$] & $1.730_{-0.067}^{+0.058}$ & $1.330_{-0.042}^{+0.047}$ & $1.724_{-0.066}^{+0.059}$ & $1.728_{-0.022}^{+0.017}$ & $1.323_{-0.026}^{+0.026}$ & $1.663_{-0.042}^{+0.038}$ \\
 $T_\mathrm{eff}$ [K]& $6285_{-42}^{+57}$ & $6061_{-46}^{+54}$ & $6205_{-100}^{+86}$ & $6547_{-96}^{+195}$ & $6326_{-109}^{+175}$ & $6528_{-105}^{+197}$ \\
 $L_\mathrm{bol}$ [L$_\odot$] & $4.223_{-0.524}^{+0.281}$ & $2.165_{-0.241}^{+0.167}$ & $3.962_{-0.484}^{+0.352}$ & $4.983_{-0.388}^{+0.473}$ & $2.549_{-0.231}^{+0.231}$ & $4.610_{-0.453}^{+0.452}$ \\
 $M_\mathrm{bol}$ & $3.18_{-0.07}^{+0.14}$ & $3.90_{-0.08}^{+0.13}$ & $3.25_{-0.09}^{+0.14}$ & $3.03_{-0.10}^{+0.09}$ & $3.75_{-0.09}^{+0.10}$ & $3.11_{-0.10}^{+0.11}$ \\
 $M_V           $ & $3.19_{-0.07}^{+0.15}$ & $3.94_{-0.08}^{+0.14}$ & $3.26_{-0.10}^{+0.15}$ & $2.99_{-0.09}^{+0.09}$ & $3.74_{-0.09}^{+0.11}$ & $3.08_{-0.10}^{+0.11}$ \\
 $\log g$ [dex]   & $4.108_{-0.018}^{+0.016}$ & $4.293_{-0.022}^{+0.018}$ & $4.153_{-0.034}^{+0.031}$ & $4.113_{-0.006}^{+0.006}$ & $4.288_{-0.009}^{+0.009}$ & $4.139_{-0.013}^{+0.014}$ \\
 \hline
$\log$(age) [dex] & \multicolumn{3}{c}{$-$} & \multicolumn{3}{c}{$9.305_{-0.029}^{+0.030}$} \\
$[M/H]$ [dex]     & \multicolumn{3}{c}{$-$} & \multicolumn{3}{c}{$0.174_{-0.234}^{+0.111}$} \\
$E(B-V)$ [mag]    & \multicolumn{3}{c}{$-$} & \multicolumn{3}{c}{$0.059_{-0.020}^{+0.024}$} \\ 
extra light $\ell_4$ [in $TESS$-band] & \multicolumn{3}{c}{$0.016_{-0.011}^{+0.014}$} & \multicolumn{3}{c}{$0.018_{-0.017}^{+0.012}$} \\
extra light $\ell_4$ [in $R_C$-band] & \multicolumn{3}{c}{$0.022_{-0.015}^{+0.023}$} & \multicolumn{3}{c}{$0.014_{-0.009}^{+0.014}$} \\
extra light $\ell_4$ [in $r'$-band] & \multicolumn{3}{c}{$0.015_{-0.010}^{+0.022}$} & \multicolumn{3}{c}{$0.012_{-0.009}^{+0.015}$} \\
extra light $\ell_4$ [in $z'$-band] & \multicolumn{3}{c}{$0.015_{-0.010}^{+0.015}$} & \multicolumn{3}{c}{$0.030_{-0.019}^{+0.015}$} \\
$(M_V)_\mathrm{tot}$ & \multicolumn{3}{c}{$2.22_{-0.08}^{+0.15}$} & \multicolumn{3}{c}{$2.03_{-0.09}^{+0.10}$} \\
distance [pc]     &\multicolumn{3}{c}{$-$} & \multicolumn{3}{c}{$682.4_{-12.0}^{+10.2}$} \\
\hline
\end{tabular}

\end{table*}

The median values of the orbital and physical parameters of the three triple systems, derived from the MCMC posteriors and their $1\sigma$ uncertainties are tabulated in Tables~\ref{tab:syntheticfit_TIC193993801}--\ref{tab:syntheticfit_TIC052041148}. Furthermore, the observed vs.~model lightcurves are plotted in Figs.~\ref{fig:T193993801lcswithfit}--\ref{fig:T052041148lcswithfit}, \ref{fig:T193993801lcE3ground}, \ref{fig:T388459317lcE3ground}, and \ref{fig:T052041148lcE3ground}.  The observed vs.~model ETV, RV curves and summed SEDs are shown in Figs.~\ref{fig:ETVswithfit}, \ref{fig:T193993801RVwithfit}, and \ref{fig:sedfit}, respectively.

\begin{table*}
 \centering
\caption{The derived system parameters for TIC~388459317 according to the MDN model (i.e., with the use of the model-dependent SED and \texttt{PARSEC} isochrones, but without the availability of RV data). Note, that the instantaneous, osculating orbital elements, are given for epoch $t_0=2\,458\,730.0$ (BJD).}
 \label{tab: syntheticfit_TIC388459317}
\begin{tabular}{@{}llll}
\hline
\multicolumn{4}{c}{orbital elements} \\
\hline
   & \multicolumn{3}{c}{subsystem}  \\
   & \multicolumn{2}{c}{Aa--Ab} & A--B \\
  $P$ [days] & \multicolumn{2}{c}{$2.184868_{-0.000012}^{+0.000013}$} & $89.0312_{-0.0070}^{+0.0076}$ \\
  $a$ [R$_\odot$] & \multicolumn{2}{c}{$10.81_{-0.43}^{+0.21}$} & $150.2_{-6.8}^{+3.2}$ \\
  $e$ & \multicolumn{2}{c}{$0.00364_{-0.00053}^{+0.00126}$} & $0.1034_{-0.0020}^{+0.0021}$ \\
  $\omega$ [deg] & \multicolumn{2}{c}{$280_{-5}^{+6}$} & $155_{-3}^{+3}$ \\ 
  $i$ [deg] & \multicolumn{2}{c}{$89.54_{-0.43}^{+0.70}$} & $90.02_{-0.21}^{+0.15}$ \\
  $\mathcal{T}_0^\mathrm{inf/sup}$ [BJD - 2400000] & \multicolumn{2}{c}{$58\,738.95606_{-0.00020}^{+0.00019}$} & $58\,784.7073_{-0.0071}^{+0.0077}$ \\
  $\tau$ [BJD - 2400000] & \multicolumn{2}{c}{$58\,736.852_{-0.038}^{+0.029}$} & $58\,709.317_{-0.593}^{+0.374}$ \\
  $\Omega$ [deg] & \multicolumn{2}{c}{$0.0$} & $-0.42_{-2.23}^{+1.20}$ \\
  $i_\mathrm{m}$ [deg] & \multicolumn{3}{c}{$1.29_{-0.75}^{+1.50}$} \\
  \hline
  mass ratio $[q=m_\mathrm{sec}/m_\mathrm{pri}]$ & \multicolumn{2}{c}{$0.951_{-0.020}^{+0.019}$} & $0.617_{-0.026}^{+0.013}$  \\
  \hline  
\multicolumn{4}{c}{stellar parameters} \\
\hline
   & Aa & Ab &  B \\
  \hline
 \multicolumn{4}{c}{Relative quantities} \\
  \hline
 fractional radius [$R/a$] & $0.1703_{-0.0042}^{+0.0042}$ & $0.1612_{-0.0053}^{+0.0043}$  & $0.0162_{-0.0006}^{+0.0007}$ \\
 temperature relative to $(T_\mathrm{eff})_\mathrm{Aa}$ & $1$ & $0.9695_{-0.0110}^{+0.0121}$ & $1.0917_{-0.0619}^{+0.0310}$ \\
 fractional flux [in $TESS$-band] & $0.2515_{-0.0111}^{+0.0148}$ & $0.2078_{-0.0089}^{+0.0095}$ & $0.5394_{-0.0119}^{+0.0110}$ \\
 fractional flux [in $R_C$-band] & $0.2477_{-0.0121}^{+0.0167}$ & $0.2007_{-0.0092}^{+0.0105}$ & $0.5499_{-0.0139}^{+0.0126}$ \\
 \hline
 \multicolumn{4}{c}{Physical Quantities} \\
  \hline 
 $m$ [M$_\odot$] & $1.822_{-0.204}^{+0.097}$ & $1.728_{-0.199}^{+0.114}$ & $2.191_{-0.341}^{+0.161}$ \\
 $R$ [R$_\odot$] & $1.839_{-0.087}^{+0.066}$ & $1.742_{-0.123}^{+0.079}$ & $2.422_{-0.103}^{+0.104}$ \\
 $T_\mathrm{eff}$ [K] & $7892_{-275}^{+233}$ & $7639_{-244}^{+243}$ & $8595_{-648}^{+412}$ \\
 $L_\mathrm{bol}$ [L$_\odot$] & $11.79_{-2.48}^{+2.15}$ & $9.17_{-1.88}^{+2.06}$ & $28.63_{-8.32}^{+7.24}$ \\
 $M_\mathrm{bol}$ & $2.09_{-0.18}^{+0.26}$ & $2.36_{-0.22}^{+0.25}$ & $1.13_{-0.25}^{+0.37}$ \\
 $M_V           $ & $2.03_{-0.18}^{+0.26}$ & $2.30_{-0.23}^{+0.27}$ & $1.11_{-0.20}^{+0.34}$ \\
 $\log g$ [dex]   & $4.165_{-0.019}^{+0.017}$ & $4.192_{-0.015}^{+0.014}$ & $4.000_{-0.050}^{+0.046}$ \\
 \hline
$\log$(age) [dex] & \multicolumn{3}{c}{$8.73_{-0.12}^{+0.28}$} \\
$[M/H]$ [dex]     & \multicolumn{3}{c}{$0.20_{-0.28}^{+0.19}$} \\
$E(B-V)$ [mag]    & \multicolumn{3}{c}{$0.739_{-0.069}^{+0.042}$} \\ 
$(M_V)_\mathrm{tot}$ & \multicolumn{3}{c}{$0.49_{-0.19}^{+0.30}$} \\
distance [pc]     &\multicolumn{3}{c}{$3021_{-169}^{+146}$} \\
\hline
\end{tabular}

\end{table*}

\begin{table*}
 \centering
 \caption{The derived system parameters for TIC~52041148 according to the MDN model (i.e., with the use of the model-dependent SED and \texttt{PARSEC} isochrones, but without the availability of RV data). Note, that the instantaneous, osculating orbital elements, are given for epoch $t_0=2\,458\,790.5$ (BJD).}
 \label{tab:syntheticfit_TIC052041148}
\begin{tabular}{@{}llll}
\hline
\multicolumn{4}{c}{orbital elements} \\
\hline
   & \multicolumn{3}{c}{subsystem}  \\
   & \multicolumn{2}{c}{Aa--Ab} & A--B  \\
  $P$ [days] & \multicolumn{2}{c}{$1.78721_{-0.00015}^{+0.00016}$} & $177.862_{-0.014}^{+0.020}$ \\
  $a$ [R$_\odot$] & \multicolumn{2}{c}{$9.04_{-0.16}^{+0.09}$} & $239.7_{-2.5}^{+1.9}$ \\
  $e$ & \multicolumn{2}{c}{$0.00187_{-0.00024}^{+0.00038}$} & $0.6204_{-0.0046}^{+0.0046}$ \\
  $\omega$ [deg] & \multicolumn{2}{c}{$283_{-147}^{+160}$} & $228.7_{-0.6}^{+0.6}$ \\ 
  $i$ [deg]  & \multicolumn{2}{c}{$87.61_{-0.45}^{+0.89}$} & $89.46_{-0.05}^{+1.15}$ \\
  $\mathcal{T}_0^\mathrm{inf}$ [BJD - 2400000] & \multicolumn{2}{c}{$58\,792.71293_{-0.00035}^{+0.00022}$} & $58\,795.4691_{-0.0043}^{+0.0041}$ \\
  $\tau$ [BJD - 2400000] & \multicolumn{2}{c}{$58\,791.306_{-0.217}^{+0.100}$} & $58\,791.458_{-0.139}^{+0.139}$ \\
  $\Omega$ [deg] & \multicolumn{2}{c}{$0.0$} & $0.29_{-2.10}^{+1.22}$ \\
  $i_\mathrm{m}$ [deg] & \multicolumn{3}{c}{$2.56_{-0.80}^{+1.12}$} \\
  \hline
  mass ratio $[q=m_\mathrm{sec}/m_\mathrm{pri}]$ & \multicolumn{2}{c}{$0.916_{-0.011}^{+0.012}$} & $0.889_{-0.024}^{+0.032}$ \\
  \hline  
\multicolumn{4}{c}{stellar parameters} \\
\hline
   & Aa & Ab &  B \\
  \hline
 \multicolumn{4}{c}{Relative quantities} \\
  \hline
 fractional radius [$R/a$] & $0.2365_{-0.0029}^{+0.0030}$ & $0.2466_{-0.0026}^{+0.0025}$  & $0.0160_{-0.0002}^{+0.0002}$ \\
 temperature relative to $(T_\mathrm{eff})_\mathrm{Aa}$ & $1$ & $0.9781_{-0.0024}^{+0.0022}$ & $1.1086_{-0.0073}^{+0.0095}$ \\
 fractional flux [in $TESS$-band] & $0.1408_{-0.0026}^{+0.0027}$ & $0.1384_{-0.0036}^{+0.0036}$ & $0.7052_{-0.0177}^{+0.0114}$ \\
 fractional flux [in $R_C$-band] & $0.1332_{-0.0032}^{+0.0038}$ & $0.1289_{-0.0039}^{+0.0043}$ & $0.7205_{-0.0184}^{+0.0103}$ \\
 \hline
 \multicolumn{4}{c}{Physical Quantities} \\
  \hline 
 $m$ [M$_\odot$] & $1.616_{-0.073}^{+0.043}$ & $1.481_{-0.082}^{+0.050}$ & $2.743_{-0.040}^{+0.040}$ \\
 $R$ [R$_\odot$] & $2.136_{-0.041}^{+0.033}$ & $2.225_{-0.042}^{+0.038}$ & $3.821_{-0.070}^{+0.083}$ \\
 $T_\mathrm{eff}$ [K] & $4857_{-43}^{+37}$ & $4751_{-42}^{+32}$ & $5388_{-53}^{+51}$ \\
 $L_\mathrm{bol}$ [L$_\odot$] & $2.281_{-0.154}^{+0.132}$ & $2.269_{-0.159}^{+0.132}$ & $11.085_{-0.794}^{+0.808}$ \\
 $M_\mathrm{bol}$ & $3.87_{-0.06}^{+0.08}$ & $3.88_{-0.06}^{+0.08}$ & $2.16_{-0.08}^{+0.08}$ \\
 $M_V           $ & $4.23_{-0.08}^{+0.10}$ & $4.29_{-0.08}^{+0.11}$ & $2.29_{-0.09}^{+0.10}$ \\
 $\log g$ [dex]   & $3.985_{-0.013}^{+0.010}$ & $3.910_{-0.010}^{+0.011}$ & $3.710_{-0.013}^{+0.011}$ \\
 \hline
$\log$(age) [dex] & $6.46_{-0.03}^{+0.02}$ & $6.33_{-0.02}^{+0.02}$ & $6.25_{-0.04}^{+0.03}$ \\
$[M/H]$ [dex]     & \multicolumn{3}{c}{$0.35_{-0.06}^{+0.08}$} \\
$E(B-V)$ [mag]    & \multicolumn{3}{c}{$0.464_{-0.020}^{+0.018}$} \\ 
extra light $\ell_4$ [in $TESS$-band] & \multicolumn{3}{c}{$0.015_{-0.011}^{+0.018}$} \\
extra light $\ell_4$ [in $R_C$-band] & \multicolumn{3}{c}{$0.016_{-0.011}^{+0.021}$} \\
$(M_V)_\mathrm{tot}$ & \multicolumn{3}{c}{$2.04_{-0.08}^{+0.04}$} \\
distance [pc]     & \multicolumn{3}{c}{$1357_{-29}^{+30}$} \\
\hline
\end{tabular}

\end{table*}

\section{Discussion}
\label{sec:discussion}

\subsection{TIC 193993801} 

\subsubsection{Use as a `calibration system' with and without RVs}
We use this triple system as a benchmark to test how well the use of theoretical \texttt{PARSEC} isochrones and SED fitting can serve as a partial substitute for RV data, and at what level we can trust the results of the former model-dependent analysis.  This is instructive since we do have RV information for TIC 193993801 while we do not for TIC 388459317 or TIC 52041148. Therefore, we carried out four different kinds of analyses for this system, namely both model-dependent and model-independent runs with and without the use of the RV dataset. The results of the model-independent and model-dependent runs with the use of the RV data (hereafter `MIR' and `MDR', respectively) are tabulated in columns 2--4 and 5--7 of Table~\ref{tab:syntheticfit_TIC193993801}, while their non-RV data counterparts (hereafter `MIN' and `MDN', respectively) can be found in Table~\ref{tab:syntheticfit_TIC193993801noRV}. First we compare and discuss the results of these models, and then will discuss the astrophysical and orbital results for TIC~193993801 itself.  

\begin{figure}
\begin{center}
\includegraphics[width=0.99 \columnwidth]{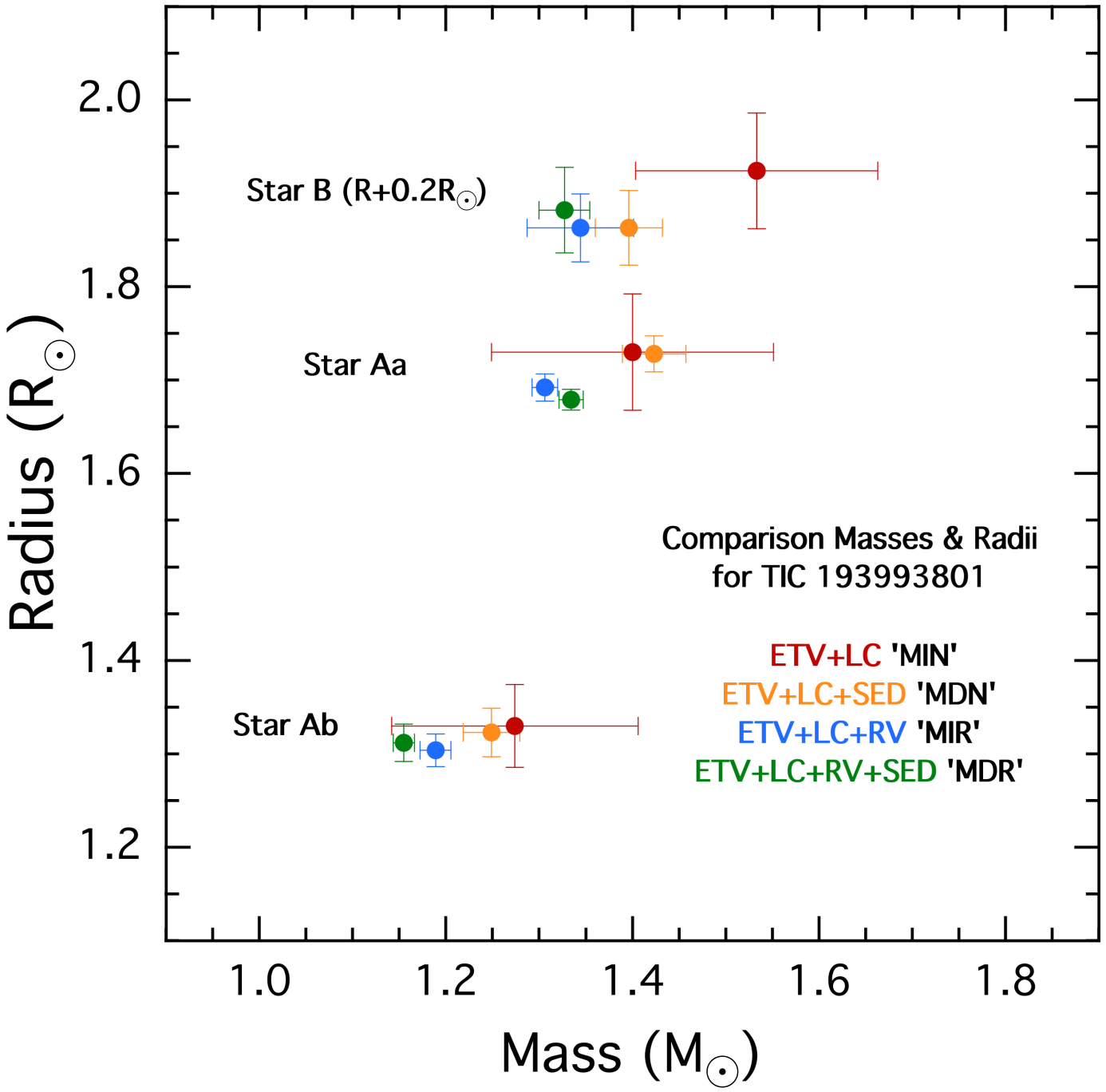}
\caption{Comparison of the masses and radii for the three stars in TIC 193993801 as determined by four different model fits. (The radii for star B have been shifted upward by 0.2 R$_\odot$ to avoid overlap with the points for star Aa.) The red points are for the model-independent analysis of the ETV and light curves, but without the use of the RVs (`MIN'); the orange points result when we add the SED analysis plus the use of stellar isochrone models (`MDN'); blue points are from the `MIR' analysis which are model-independent with the use of RVs; and green points are from the `full' model using all the information available (`MDR').} 
\label{fig:comparison} 
\end{center}
\end{figure}  

The most important question is, how the dynamical and astrophysical parameters that can be directly obtained and/or inferred from the MIR and MDN solutions compare to each other. One can see from a comparison of columns 2--4 in Table~\ref{tab:syntheticfit_TIC193993801} and columns 5--7 in Table~\ref{tab:syntheticfit_TIC193993801noRV} that the majority of the parameters, including practically all the directly measurable orbital, atmospheric (temperature ratios), and geometric (i.e. fractional radii of the stars) parameters agree well within their $1\sigma$ uncertainties. The masses and the other mass-related parameters, however, show larger discrepancies. The masses in the MDN solution were found to be systematically larger. Namely, $\Delta m_\mathrm{Aa}\approx0.12\,\mathrm{M}_\odot$ (that is about $\sim9\%$ of the MIR mass or, $7\sigma_\mathrm{MIR}$ and, $2.5\sigma_\mathrm{MDN}$), $\Delta m_\mathrm{Ab}\approx0.06\,\mathrm{M}_\odot$ ($\sim5\%$, $3.5\sigma_\mathrm{MIR}$, $1.5\sigma_\mathrm{MDN}$) and, $\Delta m_\mathrm{B}\approx0.05\,\mathrm{M}_\odot$ ($\sim4\%$, $1.5\sigma_\mathrm{MIR}$, $1.5\sigma_\mathrm{MDN}$). These comparisons may be easier to visualize graphically, as we show in Fig.~\ref{fig:comparison}.  Here we plot the radius and mass for all three stars in TIC 19399380 for each of the three different types of solutions, i.e., MIN, MDN, MIR, and MDR, in roughly increasing order of accuracy.  

Regarding the mass ratios, in the case of the inner binary ($q_\mathrm{in}$), the MDN value was found to be lower with $\Delta q_\mathrm{in}=-0.031$, i.e., almost $3\sigma_\mathrm{MIR}$ from the MIR solution, while for the outer mass ratio ($q_\mathrm{out}$), the MDN value is lower with $\Delta q_\mathrm{out}=0.015$ ($1.5\sigma_\mathrm{MIR}$). In this context one naturally should consider the stellar radii, as well. As the fractional radii agree in both solutions within $1\sigma$, and the orbital semi-major axes ($a_\mathrm{in,out}$) scales with $m_\mathrm{A,AB}^{1/3}$ the discrepancy in the physical stellar sizes remains below $2.1\%$ for all the three stars (see Fig.~\ref{fig:comparison} for a graphical comparison).

In order to investigate the origin of the tendency for higher masses to be deduced from the MDN solution, we include the MDR (i.e. full) solution in our discussion. In this model, the RV dataset with the implied lower masses forces the \texttt{PARSEC}+SED fitting section of the whole process to find some evolutionary tracks that belong to less massive stars and consistent with the overall solution. However, we were then unable to find a satisfactory coeval solution. Therefore, we allowed the ages of all three stars to vary independently.  In such a manner we were able to find an MDR (i.e., model-dependent `full') solution with masses consistent with the base MIR solution results within $1-3\%$ ($m_\mathrm{Aa}^\mathrm{MIR}=1.31\pm0.02\,\mathrm{M}_\odot$ vs $m_\mathrm{Aa}^\mathrm{MDR}=1.33\pm0.01\,\mathrm{M}_\odot$; $m_\mathrm{Ab}^\mathrm{MIR}=1.19\pm0.02\,\mathrm{M}_\odot$ vs $m_\mathrm{Ab}^\mathrm{MDR}=1.16\pm0.01\,\mathrm{M}_\odot$; $m_\mathrm{B}^\mathrm{MIR}=1.34\pm0.04\,\mathrm{M}_\odot$ vs $m_\mathrm{B}^\mathrm{MDR}=1.33\pm0.03\,\mathrm{M}_\odot$).  Note that the total mass of the inner binary, however, was found to be almost equal in the two cases ($m_\mathrm{A}^\mathrm{MIR}=1.495,\mathrm{M}_\odot$ vs $m_\mathrm{A}^\mathrm{MDR}=1.489\,\mathrm{M}_\odot$). The mass of the third stellar component in the two solutions differs by less than $1\sigma$ or, about $1.3\%$. The better agreement in the masses of the wide binary (and, accordingly the outer mass ratio $\Delta q_\mathrm{out}=-0.006$) can be explained by the fact that the strongly rotationally broadened spectral lines of the inner binary stars result in several kms$^{-1}$ uncertainties in the determination of their RV values, which are 10-15 times larger uncertainties than those of the RVs of the third stellar component.  Therefore, the RV solution is mostly dominated by the orbital solution of the outer orbit. The cost for this less massive, MIR-consistent MDR solution was, however, that while the two similar mass stars Aa and B were found to be essentially coeval (well within $1\sigma$; $\tau_\mathrm{Aa,B}=2.7\pm0.2$\,Gyr), for the less massive inner secondary star Ab we obtained a significantly older age ($\tau_\mathrm{Ab}=3.7\pm0.4$\,Gyr), i.e., the discrepancy is $\sim3\sigma$.

Comparing the two different kinds of model-dependent solutions (i.e., the coeval MDN and non-coeval MDR models), we find that the latter led to an older ($\tau^\mathrm{MDR}=2.7-3.7$\,Gyr vs $\tau^\mathrm{MDN}=2.0\pm0.2$\,Gyr) and less metal-rich solution ($[M/H]^\mathrm{MDR}=0.0\pm0.2$\,[dex] vs $[M/H]^\mathrm{MDN}=0.2\pm0.2$\,[dex]).  This is a natural consequence of the fact that, for a given stellar mass, a star with lower metallicity is hotter than a more metal-rich one.

Finally, we turn to the MIN model, i.e., where only the lightcurves and the ETV curves were utilized in the analysis. In general, the lightcurve of a detached eclipsing binary is not sensitive to either the masses or to the mass ratio of the two components. In our case, however, the presence of the third star on a tight orbit and, especially the third-body eclipses, should carry information about both the inner and outer mass ratios, and even on the individual masses as well. Note, that the source of this extra information is not restricted to only the timing variations of the inner and third-body eclipses.  In fact, the shapes of the third body eclipses also carry substantial information on the global system geometry and, therefore, on both mass ratios, as was discussed in the Appendix~A of \citet{borkovitsetal13}. The \textit{TESS} observations of TIC~193993801 cover continuously one full and another half outer orbit, three completely observed third-body eclipses and, moreover, during our ground-based follow up photometric campaign, we were able to measure at least sections of seven other third-body eclipses. It is therefore a relevant question to ask whether these data make it possible to obtain reliable fundamental stellar parameters in a model-independent way, even in the absence of RV data and, if so, what precision can be achieved. This latter, MIN solution is tabulated in columns 2--4 of Table~\ref{tab:syntheticfit_TIC193993801noRV}. Comparing the masses and their statistical $1\sigma$ uncertainties to the base solution MIR, one can see that the $\sigma_\mathrm{MIN}$ uncertainties are about one order of magnitude larger than those of the MIR solutions (and also of the other two, MDR and MDN) solutions, and they represent $\sim$$10-12\%$ of the masses themselves. Moreover, the deviation of the median masses (and other parameters) of the MIN solution from the `true' MIR results in most cases are well within the corresponding $1\sigma_\mathrm{MIN}$ uncertainties.

These results, in our interpretation, suggest that in the absence of reliable RV measurements, the MDN model can be used as a tolerably good substitute. However, the MDN solutions lead to lower accuracies of $5-10\%$ instead of the $1-2\%$ or better precision that can be reached in the case of high quality RV data. Moreover, careful application of the \texttt{PARSEC} isochrones and the correspondingly generated theoretical SEDs, i.e., the inclusion of some a priori stellar astrophysical knowledge in addition to the RV data (the MDR model), neither alters the model independent (MIR) results nor does it reduce or increase their uncertainties.  Rather, the MDR results can help to improve the information content of the entire solution, making it possible to find stellar temperatures, evolutionary states and some other parameters (see below) of the stars under investigation. In addition, the MDR results may help to eliminate some possibly false solutions from the MIR results, which might arise from combinations of the fitted parameters that lead to astrophysically unrealistic stars.

\subsubsection{Results for TIC 193993801}
Turning now to the discussion of our findings for TIC~193993801, both models (MIR and MDR) reveal that this triple system consists of two very similar, mid F-type stars (components Aa and B), and a slightly less massive late F-type component (Ab). The stars in both solutions are slightly evolved (i.e., a bit oversized for their masses). The discrepancies between the MDR and the MDN solutions, and also the non-existence of satisfactory coeval solutions in the MDR fit may imply some small systematic deviations from the single-star evolutionary tracks that were used. We may speculate that these discrepancies might have arisen from the rapid rotation of the members of the inner binary or, from some prior mass transfer phase between the close binary pair, though the latter seems to be unlikely.

The MDR model has yielded a metallicity $[M/H]$ and interstellar extinction ($E(B-V)$) that are in good agreement with the catalog values (Table~\ref{tbl:mags}). The photometric distance of TIC~193993801 was found to be $d_\mathrm{phot}=669\pm9$\,pc which is in perfect agreement with the Gaia EDR3 derived geometric distance of $d_\mathrm{geo}^\mathrm{EDR3}=676\pm26$\,pc \citep{bailer-jonesetal21}.

Regarding the dynamical properties of TIC~193993801, it has the fourth shortest outer period known for a triply eclipsing triple star (see Fig.~\ref{fig:triples}), with its outer anomalistic period of $P_\mathrm{out}=49\fd402\pm0\fd001$.\footnote{In what follows we use the numerical results of the MIR model solution (columns 2 -- 4 of Table~\ref{tab:syntheticfit_TIC193993801}).} On the other hand, given the short period of the inner binary ($P_\mathrm{in}=1\fd43081\pm0\fd00004$) the outer-to-inner period ratio is $P_\mathrm{out}/P_\mathrm{in}\approx34.5$, which implies only moderate third-body perturbations to the Keplerian motion. Moreover, similar to the majority of previously investigated triply eclipsing systems, this one is also found to be very flat, with $i_\mathrm{mut}=0\fdg47_{-0\fdg26}^{+0\fdg55}$.  Furthermore, both the inner and outer orbits are almost circular ($e_\mathrm{in}=0.0008\pm0.0001$; $e_\mathrm{out}=0.003\pm0.002$), where the latter is certainly atypical, even for such tight triple systems.  For example, the even more compact triply eclipsing triples KOI-126 ($P_\mathrm{out}=33\fd92$) and HD~144548 ($P_\mathrm{out}=33\fd95$) have outer eccentricities $e_\mathrm{out}=0.31$ \citep{carter11,yenawineetal21} and $e_\mathrm{out}=0.27$ \citep{alonso15}, respectively. On the other hand, however, as a rare counterexample, we note that the much longer outer period ($P_\mathrm{out}=235\fd55$) triply eclipsing triple star TIC~278825952 was also found to be doubly circular and flat \citep{mitnyanetal20}.

Finally, given the small eccentricities and the flatness of the system, we conclude that the present configuration of this triple system is dynamically stable and, moreover, we do not expect significant, measurable non-Keplerian variations in the orbital elements and, therefore in the system observables. In accordance with the analytical findings of \citet{borkovitsetal03,borkovitsetal15}, the ETV curve is fully dominated by the light-travel time effect (since for a doubly circular and flat configuration the most prominent quadrupole perturbation terms in the ETV curve disappear), and the same holds for the RV measurements. Moreover, on longer timescales, in the present system neither apsidal motion nor orbital plane precession is expected.

\subsection{TIC 388459317}
\begin{figure}
\begin{center}
\includegraphics[width=0.99 \columnwidth]{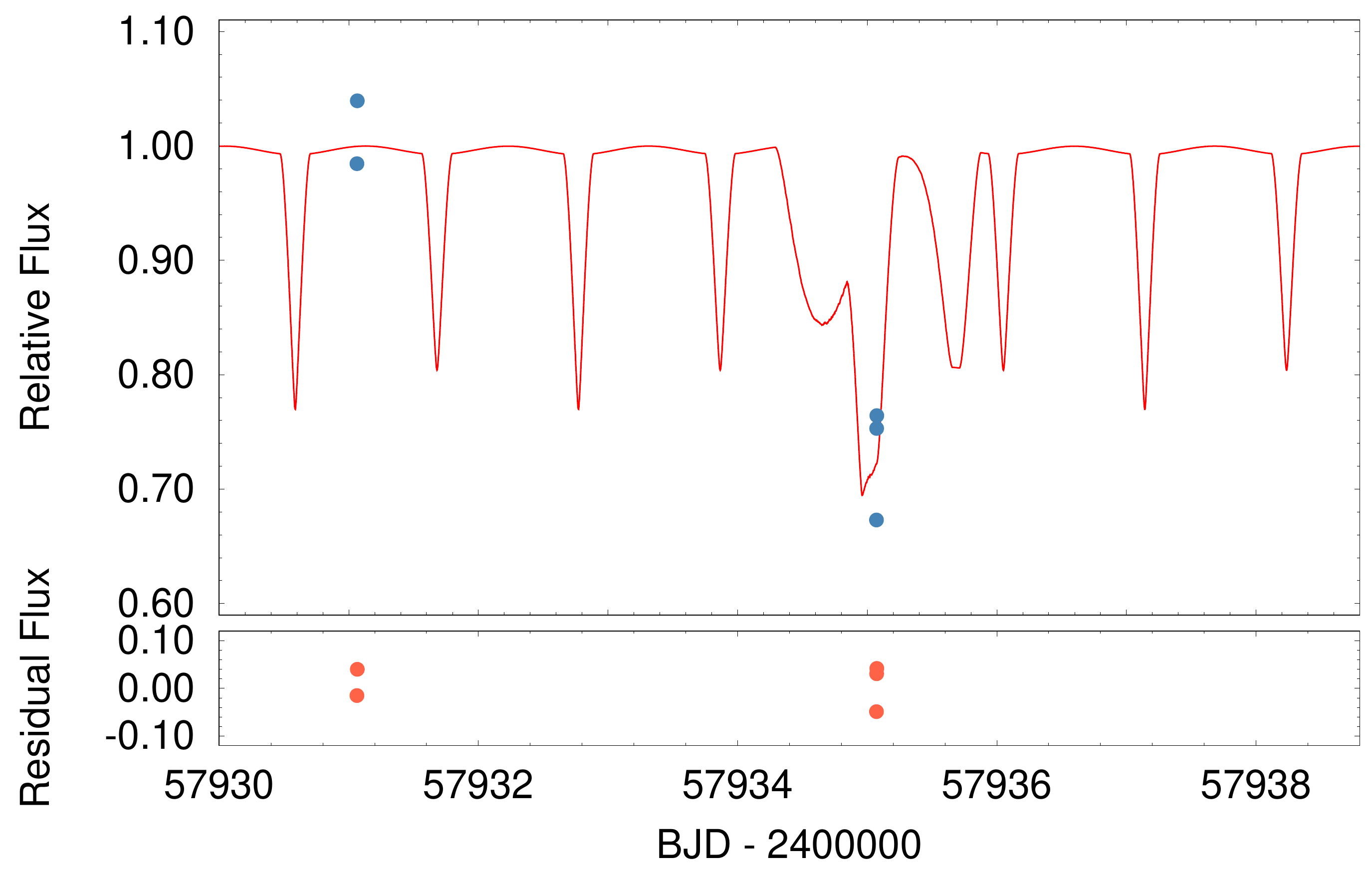}
\includegraphics[width=0.99 \columnwidth]{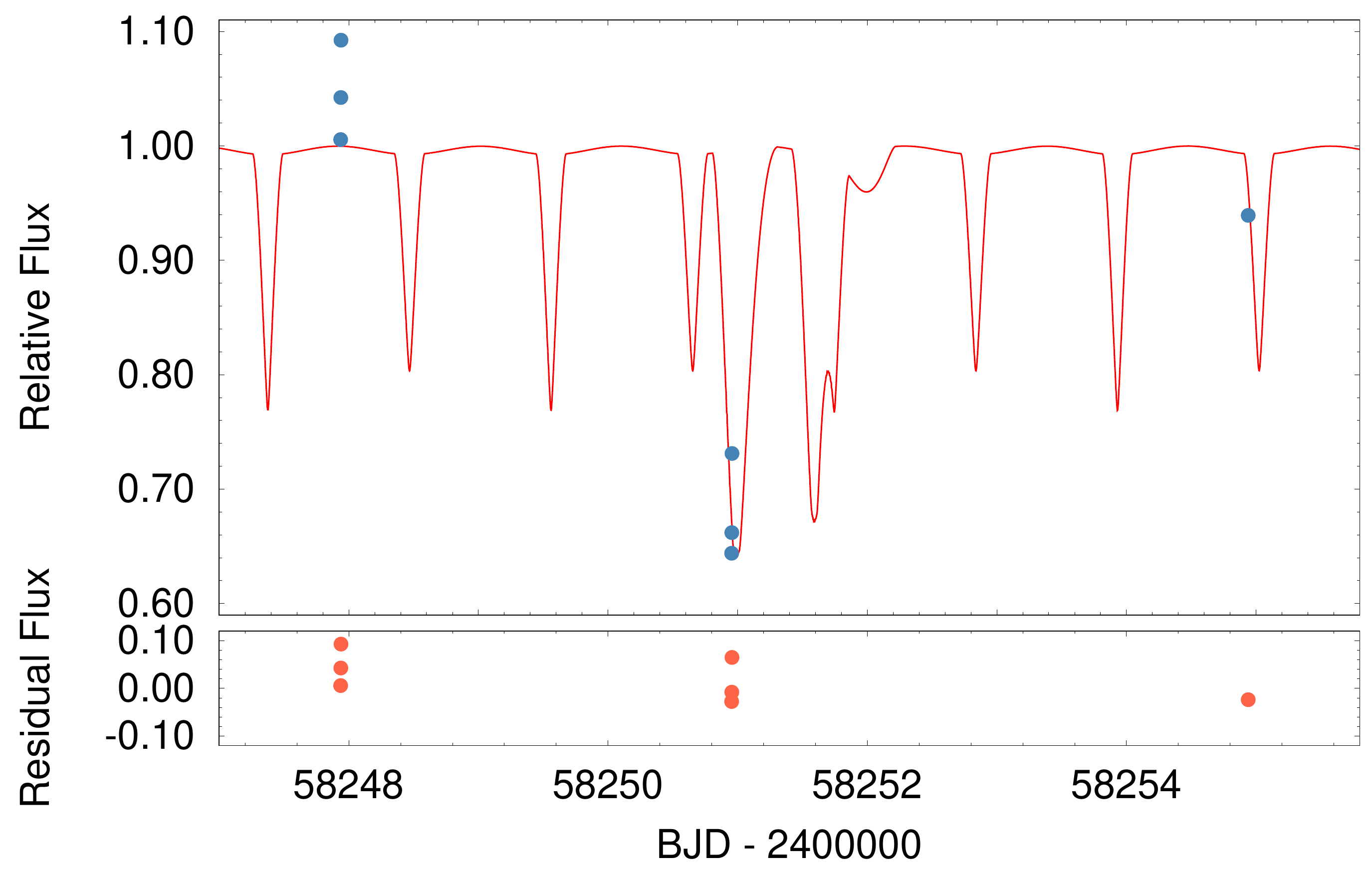}
\includegraphics[width=0.99 \columnwidth]{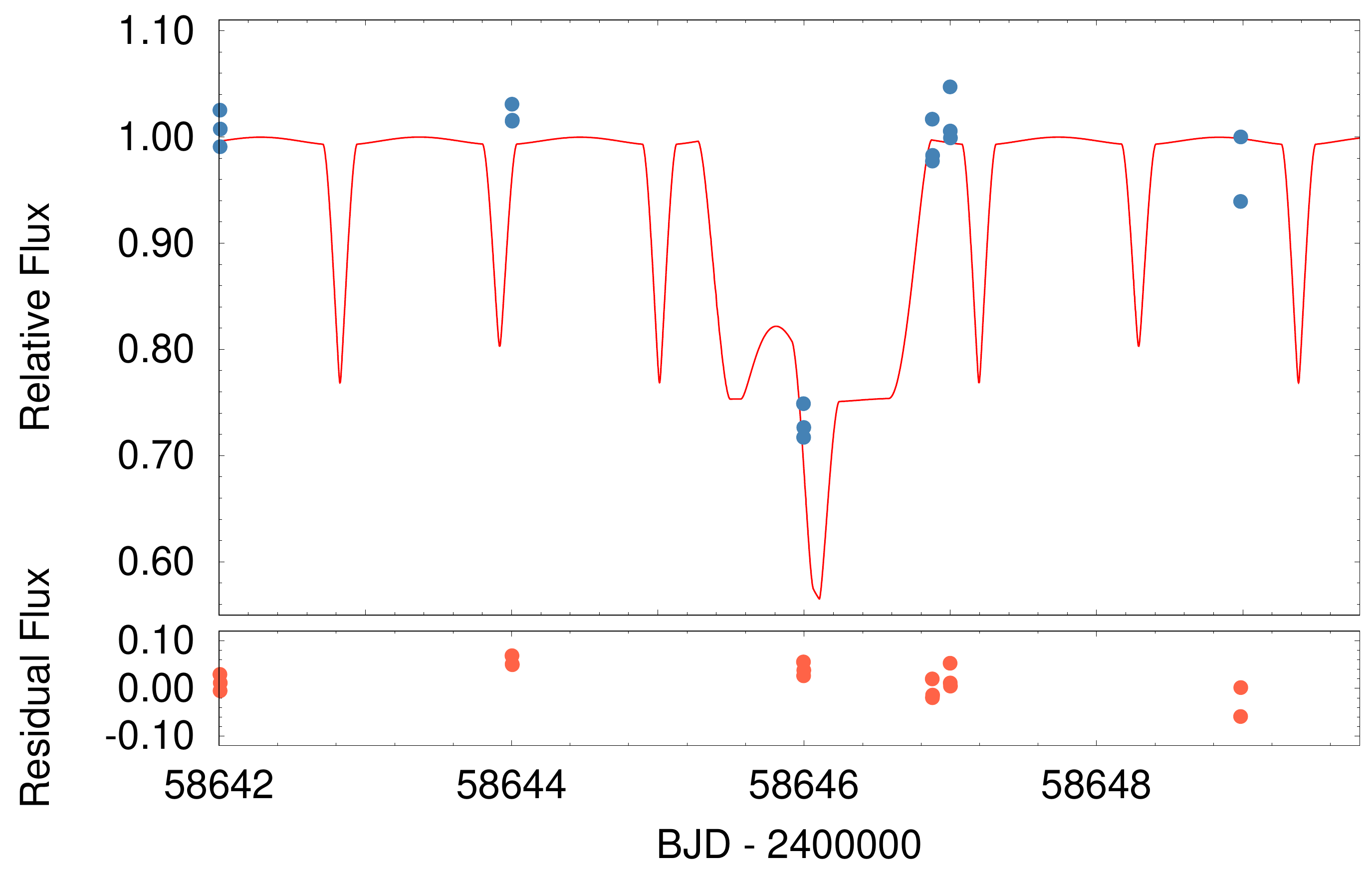}
\caption{Sections of the spectro-photodynamical lightcurve solution of TIC~388459317 (red lines) integrated back to the times of the ASAS-SN observations (blue dots). As one can see, despite the large scatter, the lower-flux data points are in agreement with the back-projected times of third-body eclipses in the past.} 
\label{fig:T388459317ASNlcback} 
\end{center}
\end{figure}

.  The \textit{TESS} observations of TIC~388549317 have shown only a single third-body eclipse and, moreover, the ETV curve generated from the satellite data was found to be featureless. Therefore, the determination of even the outer orbital period was a great challenge. However, having a strong clue about the outer period from the analysis of the archival ATLAS and ASAS-SN data (see Sect.~\ref{sec:archival}), and observing segments of three additional third-body eclipses during our photometric campaign, made it possible to carry out a full photodynamical analysis of this target. In the absence of RV data for this target we utilized only the MIN and MDN modeling in our analysis (the results of the latter tabulated in Table~\ref{tab: syntheticfit_TIC388459317}). The orbital and the other dimensionless lightcurve parameters (fractional radii, temperature ratios) obtained from the two solutions agree fairly well, mostly within their mutual $1\sigma$ uncertainties. The solutions reveal a slightly less tight ($P_\mathrm{in}=2\fd18487\pm0\fd00001$; $P_\mathrm{out}=89\fd031\pm0\fd007$; $P_\mathrm{out}/P_\mathrm{in}\approx40.7$) triple star system with almost circular inner and moderately eccentric outer orbits ($e_\mathrm{in}=0.004\pm0.001$ and $e_\mathrm{out}=0.103\pm0.002$, respectively. This system was also found to be almost coplanar with $i_\mathrm{mut}=1\fdg3_{-0\fdg8}^{+1\fdg5}$. In order to check the consistency of the orbital elements of the outer orbit that we obtained, we calculated synthetic lightcurves for the earlier times of the ASAS-SN observations. Some illustrative sections of this back-integration can be seen in the panels of Fig.~\ref{fig:T388459317ASNlcback}.

Turning to the astrophysical parameters for TIC 388459317, the SED fitting has revealed clearly that this triple comprises three radiative, A-type stars. In this case we were able to find a satisfactory coeval ($\tau\approx540_{-130}^{^+500}$\,Myr) solution, though with a somewhat large uncertainty in age. The stellar masses in this system were found with relatively large $\approx10-15\%$ uncertainties. We explain this with the short and featureless ETV curve observed with \textit{TESS} which therefore does not contain any information about even the outer mass ratio. Therefore, the masses and their ratios are constrained almost exclusively by the theoretical stellar isochrones. What is certain, however, is that the third distant star is by far the more massive and luminous in this triple, while the inner binary is formed by two rather similar stars ($q_\mathrm{in}=0.95\pm0.02$). Our solution gives a photometric distance of $d_\mathrm{phot}=3010\pm150$\,pc which is smaller by $\approx4\sigma$ than the geometric distance of $d^\mathrm{EDR3}_\mathrm{geo}=4042\pm268$\,pc \citep{bailer-jonesetal21} that is calculated from the Gaia EDR3 parallax. This discrepancy, again, may arise from the way that the triple nature of this target affects the Gaia measurements.

\subsection{TIC 52041148} 
TIC 52041148 was found to be the most enigmatic of our three triple star systems. First, Gaia DR2 gives an effective temperature of $T^\mathrm{Gaia}_\mathrm{eff}=4400_{-150}^{+600}$\,K, which clearly contradicts the TIC8.2 value of $T^\mathrm{TIC}_\mathrm{eff}=6994\pm126$\,K.  Moreover, the Gaia measurements imply a very distant system with $d_\mathrm{geo}^\mathrm{EDR3}=5931\pm407$\,pc \citep{bailer-jonesetal21}. This corresponds to an unreddened distance modulus of $\mu=13.9$\,mag, which suggests an $M_\mathrm{V}\lesssim0.2$ for the triple system. This brightness would imply that if the Gaia temperature were correct, then at least one star in TIC~52041148 should be a giant. In contrast to this, however, a quick perusal of the eclipsing lightcurve clearly reveals that neither star in the binary can be as large as is implied by the inferred $M_V$. Simply put, in an $\sim$$1.79$-day period EB with narrow eclipses and small ellipsoidal variations there is no place for a giant star. Quantitatively, one can see in Table~\ref{tab:syntheticfit_TIC052041148} that the surface gravities of both stars Aa and Ab are $\log~g\approx4.0$. The only other possibility is to assume that it is the third star which is a giant. This hypothesis, however, again can be refuted easily, as follows.  First, the depths of the regular binary eclipses are about $\sim$$12-13$\%. Therefore, the inner binary contributes at least 25\% of the total system flux\footnote{We arrive at the same conclusion if assume that during the flat, $\sim28\%$-deep, middle part of the \textit{TESS}-observed third body eclipse at BJD 2\,458\,795 (see upper panel in Fig.~\ref{fig:T052041148lcswithfit}) both binary stars were totally eclipsed by the tertiary.}. Second, both the archival ground based survey observations (see Sect.~\ref{sec:archival} and especially Fig.~\ref{fig:52041148}), and our follow up third-body eclipse measurements (Fig.~\ref{fig:T052041148lcE3ground}), show clearly that those third-body eclipses in which the third star is (partially) eclipsed by the inner binary are deeper than the other type of third-body eclipses (i.e., the secondary outer eclipses). And, since these deeper primary third-body eclipses occur closer to the apastron of the outer orbit, i.e., when the blocked surface area might be even smaller than in the case of the secondary third-body eclipses, it follows that the third component is the hottest star in the system. In that case, a hotter giant tertiary star would certainly produce much more than $\sim75$\% of the total flux of the triple system, which is a clear contradiction to what is observed.  Therefore, the tertiary star also cannot be a giant.

On the other hand, if we simply accept the TIC catalog value for the composite effective temperature, three similar subgiant stars with $T_\mathrm{eff}\sim7000$\,K and radii of  $R\sim4.0\,\mathrm{R}_\odot$ could have produced the inferred system brightness of $M_\mathrm{V}\sim0.2$\,mag. According to our MDN solution (Table~\ref{tab:syntheticfit_TIC052041148}), however, the temperatures of the three stars are around 5\,000\,K, i.e., in between the Gaia and TIC catalog values, but much closer to the Gaia DR2 value. Combining these temperatures with the above discussed surface gravity values, we can immediately conclude that the stars (i) cannot be located on the main sequence and, (ii) the Gaia parallax derived distance cannot be realistic.  Then, in order to decide between the pre- and post-MS scenarios, we have to turn to the dynamics of the triple system through its ETV curve (Fig.~\ref{fig:ETVswithfit}, bottom panel). The strongly asymmetric ETV curve indicates a highly eccentric outer orbit. The photodynamical solution reveals that the LTTE and the dynamical effects contribute to the ETV curve in nearly equal amounts.  Such a combination might appear to be auspicious in the sense that the LTTE part constrains the mass function ($m_\mathrm{A}q_\mathrm{out}^3\sin^3i_\mathrm{out}/(1+q_\mathrm{out})^2$), while the dynamical part gives $q_\mathrm{out}$ itself \citep{borkovitsetal15}.  In principle, one could then obtain reasonable constraints on the individual masses, themselves. Unfortunately, however, while the \textit{TESS} observations cover two consecutive upper, steeper extrema of the ETV curve, none of the lower extrema were observed. Therefore, we cannot determine the amplitude of the ETV curve precisely enough for a reasonably robust mass determination.

In spite of the issues discussed above, however, we are able deduce reasonably accurate values for the masses and the mass ratio of the outer binary. The photodynamical solutions reveal that the third star is substantially more massive than the inner binary components and, moreover, the stars themselves are intermediate mass stars in the regime of $\sim1.5-4\,\mathrm{M}_\odot$. In the case of the MIN solution the three masses are found to be $m_\mathrm{Aa}^\mathrm{MIN}=2.1_{-0.3}^{+0.8}\,\mathrm{M}_\odot$, $m_\mathrm{Ab}^\mathrm{MIN}=1.8_{-0.2}^{+0.5}\,\mathrm{M}_\odot$ and $m_\mathrm{B}^\mathrm{MIN}=3.8_{-0.9}^{+0.6}\,\mathrm{M}_\odot$.  At this point it is also interesting to note that, while according to the unconstrained MIN solution the tertiary component B is heavier from Aa by a factor of $\approx1.8$, the ratio of their temperatures was found to be only $T^\mathrm{MIN}_\mathrm{eff,B}/T^\mathrm{MIN}_\mathrm{eff,Aa}=1.06\pm0.03$.

Keeping in mind these constraints, and assuming that there was no former mass exchange amongst the three stars, we were not able to find any acceptable post-MS solutions. Oppositely, however, we obtain satisfactory pre-MS solutions (though we had to allow the ages of all the three stars to be adjusted independently). The results obtained from our pre-MS MDN model are tabulated in Table~\ref{tab:syntheticfit_TIC052041148}.

As one can see, the MDN masses, i.e., those that were constrained with the \texttt{PARSEC} evolutionary tracks are lower by about $0.5-1\,\mathrm{M}_\odot$ as compared to those based on the MIN solutions (deduced from only lightcurves and ETV curves). Moreover, the outer mass ratio is also somewhat smaller for the MDN model ($q_\mathrm{out}^\mathrm{MDN}=0.89\pm0.03$ vs $q_\mathrm{out}^\mathrm{MIN}=0.97\pm0.05$).  In this regard, we note that the $\chi^2_\mathrm{ETV}$ values for the MDN solutions are about 20\% higher than for the MIN solutions. We may say that this is `the cost for an astrophysically realistic solution'. On the other hand, we stress again that the most important parameter of the ETV curve, namely its amplitude, is relatively poorly determined due to the absence of coverage of the lower extremum and, therefore, the inferred masses and their ratios are necessarily less robust.

The MDN solution prefers a very young ($\tau=2.3\pm0.5$\,Myr) system with enhanced stellar metallicity ($[M/H]=0.35\pm0.1$).  The distance of the triple would be $d=1357\pm30$\,pc, which is however, strongly discrepant with the measured Gaia parallaxes. Regarding the fact that the semimajor axis of the outer orbit is close to 1\,au, and the period near half a year, we may speculate that these conditions might play havoc with the Gaia parallaxes.

Regarding the dynamical properties of the system, the outer eccentricity of $e_\mathrm{out}=0.620\pm0.005$ is unusually high (though, not without precedent) for a $P_\mathrm{out}=177$-day orbit. For example, in the survey of 222 tight triple stars in the original \textit{Kepler}-field, investigated by \citet{borkovitsetal16}, there are 14  triples with shorter outer period than half a year and, amongst them, there is only one triple (KIC~6531485) with similar, but a slightly smaller outer eccentricity of $\sim0.57$. All the other outer eccentricities remain well below 0.4. Note, however, as one can see on the right panel of Fig.~\ref{fig:T052041148lcE3ground} the best-fit models do not precisely predict the flux variation of the second part of the last observed primary third-body eclipse. This suggests that our results are not fully perfect. With very small manual adjustments to some outer orbital parameters (e.g. changes in the \textit{third} decimal digit of $e\cos\omega_\mathrm{out}$, and/or the second decimal digit of $e\sin\omega_\mathrm{out}$, some tenth of degrees in the two inclinations $i_\mathrm{in,out}$ and, $1\degr-2\degr$ in $\Omega_\mathrm{out}$) we are able to find such solutions that describe much better these observations.  However, with these slightly tweaked parameters the corresponding model strongly contradicts our (and also the ASAS-SN) flux measurements of the one-year (i.e. two orbital cycles) earlier primary third-body eclipse (left panel of Fig.~\ref{fig:T052041148lcE3ground}). Moreover, all those MCMC runs which were initiated with the use of such manually adjusted initial parameters also converged to very similar solutions as the best-fit one, which is plotted in the panels of Fig.~\ref{fig:T052041148lcE3ground}. Then, as a final trial, we left out of the analysis the BJD~2\,459\,111 data, i.e., when a short section of the former, above mentioned primary third-body eclipse was observed (left panel of Fig.~\ref{fig:T052041148lcE3ground}), but the new runs again converged to the former solutions. 

From the fact that the two observed primary outer eclipses cannot be modeled satisfactorily with the same initial set of $e\cos\omega_\mathrm{out}$, $e\sin\omega_\mathrm{out}$ parameters, we may suppose that the origin of this slight discrepancy should be in the non fully satisfactory modeling of the apsidal advance (with a theoretical period of $U=700\pm20$\,yr) of the highly eccentric outer orbit. In this regard, we note that the theoretical dynamical (third-body) and general relativistic advance rates for the MDN model configurations we find are $\Delta\omega_\mathrm{3b}=900\pm20$\,arcsec/cycle and $\Delta\omega_\mathrm{GR}=0.33\pm0.01$\,arcsec/cycle, i.e., the latter is negligible. Thus, we conclude that for a more precise analysis we need further observations, especially during the future primary third-body eclipses. On the other hand, we stress again that very fine adjustments (e.g., in the third decimal place of the outer eccentricity or at the $\sim1\sigma$ level of the other outer orbital elements) did allow us to model much better the questionable lightcurve section (though, the modeling of the other parts of the lightcurve became worse).  From these considerations,  we are therefore convinced that our present findings give a satisfactory approximation of the true system parameters.

TIC~52041148 was also found to be a nearly coplanar system with $i_\mathrm{mut}=2\fdg56_{-0\fdg80}^{+1\fdg12}$.  We also made runs with nearly coplanar, but retrograde configurations, but found only much weaker solutions. Therefore, we conclude with some certainty that the revolutions of the inner and outer orbits are prograde.

Finally, note that TIC~52041148 is especially interesting in the sense that, due to its large outer eccentricity and young age, this triple system may shed some light on the formation of compact hierarchies \citep[see the recent review of][and references therein]{tokovinin21}.

\section{Conclusions}
\label{sec:conclusions}

In this paper we report the discovery and the spectro-photodynamical analyses of three triply eclipsing triple systems found in the northern sky during the second year of observations with the \textit{TESS} space telescope. All three triple-star targets were observed during only two or three \textit{TESS} sectors, yielding precise but relatively short space-borne photometric data trains.  However, when combined with archival ground-based survey photometric measurements, as well as recent targeted follow-up photometric and spectroscopic observations, we were able to carry out detailed spectro-photodynamical analyses and obtain reasonably accurate orbital and stellar parameters for the three triple systems.

For TIC 193993801, which was found to be a triple lined spectroscopic triple system (SB3), besides photometric data, we were able to carry out high resolution spectroscopic observations and obtain RV data for all three of the stars. Therefore, for this system we were able to determine dynamical masses with an accuracy of $\sim1$\% for the inner binary members, and $\sim3$\% for the third stellar component. Moreover, as the stellar radii of the inner binary were also found with an accuracy of $\sim1-1.5$\%, this binary should join the group of those detached EBs for which the parameters are known precisely enough to constrain stellar evolutionary investigations \citep{DEBcat}.

Moreover, we used this system to test how well the use of theoretical \texttt{PARSEC} isochrones and SED fitting can serve as a partial substitute for RV data, and at what level we can trust the results of the model-dependent analysis.  Our results suggest that, in the absence of reliable RV measurements, a photodynamical model with the combination of theoretical \texttt{PARSEC} isochrones and SED fitting can be used as a tolerably good substitute. However, such solutions lead to lower accuracies of $5-10\%$ instead of the $1-2\%$ or better precision that can be reached in the case where high quality RV data are available.

We found that all three of the investigated triple systems are coplanar within $1-3\degr$. The three inner orbits were found to be almost circular, as is expected for such short-period binaries. The eccentricities of the outer orbits, however, ranged over more than two orders of magnitude from $0.003$ (TIC~193993801) to $0.62$ (TIC~52041148). Moreover, a further similarity in the three systems is that each inner pair is comprised of similar mass stars (the inner mass ratios are between 0.88 and 0.96). For two of the three systems the distant tertiary component is the most massive, and even for the third system (TIC~193993801), the mass of the tertiary is very close to the primary of the inner pair.

In order to improve our solutions, especially in the case of TIC~52041148, future observations, especially of third-body eclipses, would be very welcome. TIC~193993801 will be observed with the \textit{TESS} spacecraft again in Sectors 49--51. Unfortunately, however, neither TIC~388459317, nor TIC~52041148 is scheduled for further observations in Cycle 4. Therefore, for the planning of future ground-based photometric follow-up observations, we tabulate ephemerides\footnote{Note, that the periods given in Tables~\ref{tab:syntheticfit_TIC193993801}--\ref{tab:syntheticfit_TIC052041148} are instantaneous, osculating anomalistic periods and therefore, cannot be used for the prediction of the times of future eclipses. See \citet{kostovetal21}, Sect.~5 for a detailed explanation.} for the regular and third-body eclipses of all three triple systems in Table~\ref{tab:ephemerides}.  Such follow-up observations would be extraordinarily important in the case of TIC~52041148 where the outer orbit exhibits measurable apsidal motion with a period of $U\sim700$\,yr which strongly affects the locations of the third-body events relative to each other.

\begin{table}
\centering 
\caption{Derived ephemerides for the three triple systems to be used for planning future observations.}
 \label{tab:ephemerides}
 \begin{tabular}{llll}
 \hline 
TIC ID               & 193993801       & 388459317  & 52041148 \\
\hline
&\multicolumn{3}{c}{Inner binary} \\
\hline
$P$  & 1.431298 &  2.18478   & 1.7862310\\
$\mathcal{T}_0$  & 58\,739.94053 & 58\,738.954 & 58\,792.711\\
$\mathcal{A}_\mathrm{ETV}$  & 0.001 & 0.004 & 0.008 \\
$D$  & 0.187 & 0.226 & 0.273 \\
\hline
&\multicolumn{3}{c}{Wide binary (third body eclipses)} \\
\hline
$P^\mathrm{inf}$ & 49.2777 & 88.860 & 177.07 \\
$\mathcal{T}_0^\mathrm{inf}$  &  58\,745.4775 &  58\,734.9005 & 58\,795.4630 \\
$D^\mathrm{inf}$ & 1.09 & 1.87 &  1.31 \\
$P^\mathrm{sup}$ & 49.2777 & 88.860 &  176.84 \\
$\mathcal{T}_0^\mathrm{sup}$ &  58\,770.0469 &  58\,784.6070 &  58\,933.6056 \\
$D^\mathrm{sup}$ & 1.11 & 1.69 &  3.29 \\
\hline
\end{tabular}

\textit{Notes.} (a) For the inner pairs: $P$, $\mathcal{T}_0$, $\mathcal{A}_\mathrm{ETV}$, $D$ are the period, reference time of a primary minimum, half-amplitude of the ETV curve, and the full duration of an eclipse, respectively. $\mathcal{T}_0$ is given in BJD -- 2\,450\,000, while the other quantities are in days. As all three inner eccentricities are very small and, hence, the shifts of the secondary eclipses relative to phase 0.5 are negligible (quantitatively, they are smaller than the half amplitude of the cyclic ETV curves and, much smaller than the full durations of the individual eclipses), the same reference times and periods can be used to predict the times of the secondary eclipses. (b) For the outer orbits we give separate reference times for the third body eclipses around the inferior and superior conjunctions of the tertiary component. In the case of the strongly eccentric outer orbit in TIC~52041148 the apsidal motion of the outer orbit is significant (see text for details) and therefore, as usual in such cases, we give separate periods for the inferior and superior third-body eclipses. The eclipse durations, $D$, of the third-body eclipses do not give the extent of any specific third body events.  Rather $D$ represents the time difference corresponding to the very first and last moments around a given third-body conjunction when the first/last contact of a third-body event may occur).

\end{table}

\section*{Data availability}

The \textit{TESS} data underlying this article were accessed from MAST (Barbara A. Mikulski Archive for Space Telescopes) Portal (\url{https://mast.stsci.edu/portal/Mashup/Clients/Mast/Portal.html}). The ASAS-SN archival photometric data were accessed from \url{https://asas-sn.osu.edu/}. The ATLAS archival photometric data were accessed from \url{https://fallingstar-data.com/forcedphot/queue/}. A part of the data were derived from sources in public domain as given in the respective footnotes. The derived data generated in this research and the code used for the photodynamical analysis will be shared on reasonable request to the corresponding author.

\section*{Acknowledgments}

RK and TP acknowledge support from the Slovak Research and Development Agency -- the contract No. APVV-20-0148. VBK is thankful for support from NASA grants 80NSSC21K0351. ZG was supported by the VEGA grant of the Slovak Academy of Sciences number 2/0031/18, by an ESA PRODEX grant under contracting with the ELTE University, by the GINOP number 2.3.2-15-2016-00003 of the Hungarian National Research, Development and Innovation Office, and by the City of Szombathely under agreement number 67.177-21/2016. AP acknowledge the financial support of the Hungarian National Research, Development and Innovation Office -- NKFIH Grant K-138962.

The operation of the BRC80 robotic telescope of Baja Astronomical Observatory has been supported by the project ``Transient Astrophysical Objects'' GINOP 2.3.2-15-2016-00033 of the National Research, Development and Innovation Office (NKFIH), Hungary, funded by the European Union.

This project has been supported by the Lend\"ulet grant LP2012-31 of the Hungarian Academy of Sciences.

This paper includes data collected by the \textit{TESS} mission. Funding for the \textit{TESS} mission is provided by the NASA Science Mission directorate. Some of the data presented in this paper were obtained from the Mikulski Archive for Space Telescopes (MAST). STScI is operated by the Association of Universities for Research in Astronomy, Inc., under NASA contract NAS5-26555. Support for MAST for non-HST data is provided by the NASA Office of Space Science via grant NNX09AF08G and by other grants and contracts.

We have made extensive use of the All-Sky Automated Survey for Supernovae archival photometric data.  See \citet{shappee14} and \citet{kochanek17} for details of the ASAS-SN survey.

We also acknowledge use of the photometric archival data from the Asteroid Terrestrial-impact Last Alert System (ATLAS) project.  See \citet{tonry18} and \citet{heinze18} for specifics of the ATLAS survey.

This work has made use  of data  from the European  Space Agency (ESA)  mission {\it Gaia}\footnote{\url{https://www.cosmos.esa.int/gaia}},  processed  by  the {\it   Gaia}   Data   Processing   and  Analysis   Consortium   (DPAC)\footnote{\url{https://www.cosmos.esa.int/web/gaia/dpac/consortium}}.  Funding for the DPAC  has been provided  by national  institutions, in  particular the institutions participating in the {\it Gaia} Multilateral Agreement.

This publication makes use of data products from the Wide-field Infrared Survey Explorer, which is a joint project of the University of California, Los Angeles, and the Jet Propulsion Laboratory/California Institute of Technology, funded by the National Aeronautics and Space Administration. 

This publication makes use of data products from the Two Micron All Sky Survey, which is a joint project of the University of Massachusetts and the Infrared Processing and Analysis Center/California Institute of Technology, funded by the National Aeronautics and Space Administration and the National Science Foundation.

We  used the  Simbad  service  operated by  the  Centre des  Donn\'ees Stellaires (Strasbourg,  France) and the ESO  Science Archive Facility services (data  obtained under request number 396301).






\end{document}